\documentclass[journal]{IEEEtran}
\usepackage{graphicx}
\usepackage{array}
\usepackage{mdwmath}
\usepackage{mdwtab}
\usepackage{eqparbox}
\usepackage{algorithmic}
\usepackage[cmex10]{amsmath}
\usepackage{setspace}
\usepackage{algorithm,caption}
\usepackage{empheq}
\usepackage{cases}
\usepackage{soul}
\usepackage{color}
\usepackage{cite}
\usepackage{amssymb}
\usepackage{epstopdf}
\usepackage{amsmath}
\usepackage{ntheorem}
\usepackage{dsfont}
\usepackage{bbm}
\usepackage{subfig}
\usepackage{stfloats}
\usepackage{mathrsfs}
\usepackage{bm}
\usepackage{makecell}
\allowdisplaybreaks
\ifCLASSINFOpdf
\else
\fi
\begin{document}
\renewcommand{\theenumi}{\roman{enumi}}
\renewcommand{\algorithmiccomment}[1]{\hfill $\triangleright$ #1}
%
% paper title
% can use linebreaks \\ within to get better formatting as desired
\title{Mini-Slot-Assisted Short Packet URLLC: Differential or Coherent Detection?}
%Adaptive Differential and Coherent Detection for Mini-Slot-Assisted Short Packet URLLC}
%{Toward efficient, ultra-reliable, and low-latency short package communications with low overhead}
%{Toward efficient short package URLLC with low overheads}
%{Adaptive Differential and Coherent Detection for Short Packet OFDM URLLC with {3GPP}-Inspired Mini-Slot}
%{Adaptive Differential and Coherent Detection for Mini-Slot-Assisted URLLC with {3GPP}-Inspired Short Packet OFDM}
%{Adaptive Differential and Coherent Detection for {3GPP}-Inspired Short Packet OFDM URLLC with Mini-Slot}
%{Differential Modulation and Beamforming for Finite Blocklength URLLC with mmWave massive MIMO}
% author names and IEEE memberships
% note positions of commas and nonbreaking spaces ( ~ ) LaTeX will not break
% a structure at a ~ so this keeps an author's name from being broken across
% two lines.
% use \thanks{} to gain access to the first footnote area
% a separate \thanks must be used for each paragraph as LaTeX2e's \thanks
% was not built to handle multiple paragraphs

\author{Canjian~Zheng, Fu-Chun~Zheng,~\IEEEmembership{Senior Member,~IEEE}, Jingjing~Luo,~\IEEEmembership{Member,~IEEE}, Pengcheng~Zhu,~\IEEEmembership{Member,~IEEE}, Xiaohu~You,~\IEEEmembership{Fellow,~IEEE} and Daquan~Feng,~\IEEEmembership{Member,~IEEE}}
\maketitle
%\doublespace

%% The paper headers
%\markboth{Journal of \LaTeX\ Class Files,~Vol.~6, No.~1, January~2007}%
%{Shell \MakeLowercase{\textit{et al.}}: Bare Demo of IEEEtran.cls for Journals}

\begin{abstract}
One of the primary challenges in short packet ultra-reliable and low-latency communications (URLLC) is to achieve reliable channel estimation and data detection while minimizing the impact on latency performance. Given the small packet size in mini-slot-assisted URLLC, relying solely on pilot-based coherent detection is almost impossible to meet the seemingly contradictory requirements of high channel estimation accuracy, high reliability, low training overhead, and low latency.
In this paper, we explore differential modulation both in the frequency domain and in the time domain, and propose adopting an adaptive approach that integrates both differential and coherent detection to achieve mini-slot-assisted short packet URLLC, striking a balance among training overhead, system performance, and computational complexity.
Specifically, differential (especially in the frequency domain) and coherent detection schemes can be dynamically activated based on application scenarios, channel statistics, information payloads, mini-slot deployment options, and service requirements.
Furthermore, we derive the block error rate (BLER) for pilot-based, frequency domain, and time domain differential OFDM using non-asymptotic information-theoretic bounds.
Simulation results validate the feasibility and effectiveness of adaptive differential and coherent detection.
\end{abstract}

% Note that keywords are not normally used for peerreview papers.
\begin{IEEEkeywords}
URLLC, short packet transmission, frequency domain differential modulation, time domain differential modulation, mini-slot, OFDM, coherent detection and differential detection.
%GF access, NOMA, stochastic geometry, large-scale multi-cell networks, URLLC.
%URLLC, open-loop communications, clustered user distribution, multi-cell association.
\end{IEEEkeywords}
%
%
%\IEEEpeerreviewmaketitle

% Note that keywords are not normally used for peerreview papers.
%\begin{IEEEkeywords}
%Device-to-Device communications, spectrum sharing, maximum weighted bipartite matching.
%\end{IEEEkeywords}

\IEEEpeerreviewmaketitle

\section{Introduction}
\subsection{Background}
Ultra-reliable and low-latency communications (URLLC) is a primary use case in the 5th generation (5G) wireless networks, facilitating various mission-critical applications such as factory automation, remote surgery, tactile internet, and autonomous driving.
To support these applications, the 3rd generation partnership project (3GPP) has specified typical URLLC requirements of 1 ms physical layer latency and $99.999\%$ (5-nines) reliability for 5G networks\cite{3GPP261,3GPP824}.
%an end-to-end (E2E) latency of less than 1 ms and a block error rate (BLER) of no more than $10^{-5}$ for 5G networks
As emerging services place increasingly stringent demands on the network performance, it is further expected that the physical layer latency and reliability requirements for enhanced URLLC in the forthcoming B5G and 6G networks could reach 0.1 ms levels and $99.99999\%$ (7-nines), respectively\cite{Wang2023On,You2021Towards,she2021PIEEE}.
%However, such stringent performance requirements in URLLC can hardly be supported in traditional slot-based data transmission.
%To this end, the feature of mini-slot has been introduced by 3GPP for URLLC.
%, where one slot occupies 14 OFDM symbols.
%as low as sub-millisecond and $99.99999\%$ (7-nines), respectively \cite{Wang2023On,You2021Towards,she2021PIEEE}.
%To this end, exploiting mini-slot-assisted finite blocklength (FBL) formats is crucial to satisfy the seemingly contradictory requirements of low latency and high reliability in URLLC \cite{Popovski2019Wireless}.

%To satisfy such stringent performance requirements in URLLC, 3GPP has introduced the feature of mini-slot where a group of 2, 4 or 7 instead 14 orthogonal frequency-division multiplexing (OFDM) symbols is used as a basic unit for data scheduling and resource allocation
%To satisfy such stringent performance requirements in URLLC, 3GPP has introduced the feature of mini-slot which occupies 2, 4 or 7 orthogonal frequency-division multiplexing (OFDM) symbols\cite{3GPP912}.
To satisfy such stringent performance requirements in URLLC, 3GPP has introduced the novel feature of mini-slot for 5G networks\cite{3GPP912}.
%More explicitly, the mini-slot generally employs a grouping of relatively short orthogonal frequency-division multiplexing (OFDM) symbols, such as 2, as a basic unit for data scheduling and resource allocation.
Specifically, the mini-slot employs a group of 2, 4, or 7 orthogonal frequency-division multiplexing (OFDM) symbols as a basic unit for data scheduling and resource allocation.
In contrast, the smallest scheduling unit for the other two primary use cases in 5G networks (i.e., enhanced mobile broadband (eMBB) and massive machine-type communications (mMTC)) is a conventional slot occupying 12 or 14 OFDM symbols\cite{3GPP211}.
Unlike the conventional slot, where packets arriving within a slot must wait until the end of the slot for transmission,
%only beginning at the starting point of a slot,
the mini-slot-assisted data transmission with a finer time domain partitioning can start at any OFDM symbol without waiting for a slot boundary\cite{Feng2019toward}.
%As a result, the mini-slot feature is particularly suitable for short packets containing tens to hundreds of bytes, enabling rapid delivery and responsiveness of low-latency packets.
Therefore, the mini-slot feature is particularly suitable for short sized packets containing tens to hundreds of bytes to enable rapid data delivery and transmission responsiveness.

\subsection{Motivations}
%coherent detection of mini-slot-assisted URLLC requires known pilots symbols which are also called demodulation reference signals (DMRS) in 5G networks.
%mini-slot-assisted URLLC requires known pilots symbols for coherent detection. In 5G networks, pilot symbols for channel estimation are also called demodulation reference signals (DMRS).
%mini-slot-assisted URLLC requires known pilots symbols, also called demodulation reference signals (DMRS) in 5G networks, for coherent detection.
In mini-slot-assisted short packet transmission (SPT), using pilot symbols, also known as demodulation reference signals in 5G networks, to acquire accurate channel state information (CSI) is essential for coherent detection.
%acquiring accurate channel state information (CSI) by known pilots symbols, also called demodulation reference signals (DMRS) in 5G networks, is essential for coherent detection.
%However, the reference signal overhead for channel estimation in mini-slot-assisted SPT can consume a non-negligible portion of the valuable power, bandwidth and time resource.
However, the pilot overhead for channel estimation can consume a considerable portion of the valuable time, bandwidth and power resources in mini-slot-assisted SPT\cite{Durisi2016Toward}.
% amount
%especially for ultra short sized OFDM units or .
%%In the conventional slot-based long packet transmission, the data payload size is much larger than the reference signal size so that the channel estimation overheads are generally assumed to be negligible.
%%This assumption, however, does not apply to mini-slot-assisted SPT in which the reference signal size may well be comparable to the data payload size.
%, especially in short packet systems or high mobility scenarios.
Figs.1(a-d) illustrate the pilot patterns specified in 3GPP Release 18 for coherent detected short packet URLLC utilizing 2, 4 and 7 OFDM symbol units\footnote{In order to support different service scenarios and use cases in mini-slot-assisted SPT, 3GPP Release 18 defines a fixed and suboptimal deployment of pilot symbols.
%3GPP Release 18 defines a fixed and suboptimal number of pilot symbols to support different service scenarios and use cases in mini-slot-assisted short packet URLLC
%deploys a fixed and suboptimal number of pilot symbols to support different service scenarios and use cases.
%3GPP Release 18 defines a fixed and suboptimal number of pilot symbols to be deployed in mini-slot-assisted SPT
%According to 3GPP Release 18, a fixed and suboptimal number of pilot symbols is deployed for supporting different service scenarios and use cases in mini-slot-assisted short packet URLLC
%number of pilots
%employs fix and even suboptimal pilot symbol configurations
Specifically, to allow immediate channel estimation and coherent detection as soon as data symbols are received, pilots are always placed at equally spaced subcarriers of the first OFDM symbol in the mini-slot.
Furthermore, for high mobility scenarios, additional pilots can be deployed in the middle of the mini-slot of greater than 4 OFDM symbols to improve the channel estimation performance, as shown in Fig.1(d).}.
As shown in Fig.1(a),
%up to $25\%$ of the total packet are allocated to DMRS for estimating the channel of both the low mobility and high mobility scenarios with only 2 OFDM unit, which is prohibitive for short packet URLLC.
%regardless of whether low mobility or high mobility scenarios
%regardless of mobilities
up to $25\%$ of the total packet of a 2 OFDM symbol unit is allocated to pilot symbols for channel estimation regardless of mobility scenarios, which is expensive for short packet URLLC and impairs significantly the latency performance.
%Such a high portion also incur extra latency for
%Figure 1(a) illustrates that up to $25\%$ of the total packet resources are allocated to DMRS for channel estimation in scenarios of both low and high mobility using a 2 OFDM unit configuration, which is excessive for short packet URLLC.
% in channel estimation of a 2 OFDM unit in the low mobility and high mobility scenarios
%transmit a 2 OFDM unit in both the low mobility and high mobility scenarios
On the other hand, reducing the pilot percentage in mini-slot-assisted SPT may lead to a deterioration in channel estimation accuracy, severely degrading the performance of coherent detection, especially in high mobility scenarios where a fading channel can fluctuate rapidly.
%with rapidly fluctuating channels.
%This degradation becomes even more significant.

\begin{figure*}[!htp]
\centering
\includegraphics [width=0.9\textwidth]{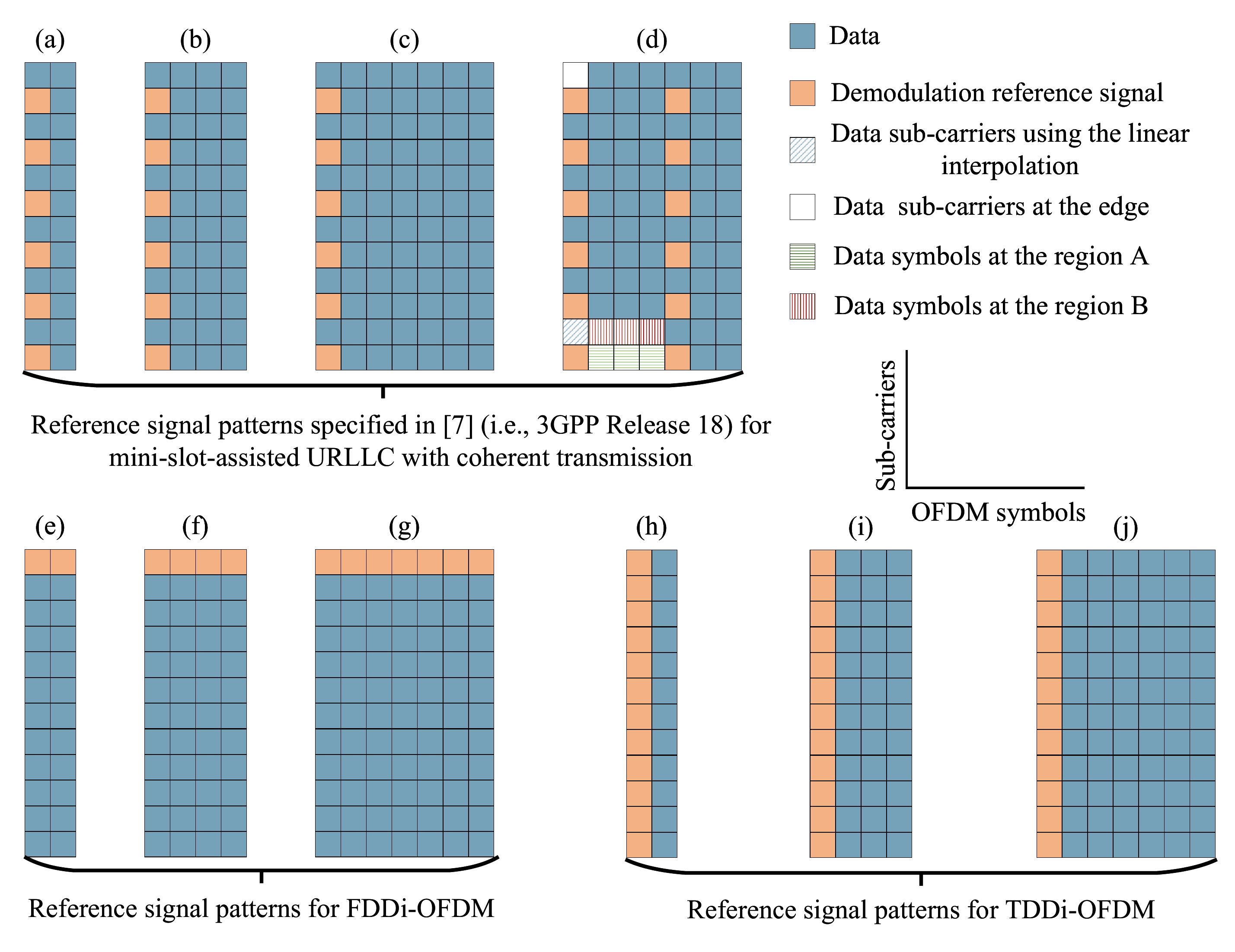}
%\vspace{-1.55em}
\caption{Reference signal patterns of 2, 4 and 7 OFDM symbol units with coherent detection (a-d), frequency domain differential modulation (e-g), and time domain differential modulation (h-j). According to \cite{3GPP211}, the patterns (a) and (b) are used in both low and high mobility scenarios, (c) is used in low mobility scenarios, and (d) is used in high mobility scenarios.}
%DMRS patterns for mini-slot-assisted URLLC with coherent transmission (a-d), frequency domain differential OFDM (e-g) and time domain differential OFDM (h-j)
%DMRS patterns for frequency domain differential OFDM with 2-symbol unit, 4-symbol unit and 7-symbol unit
%{System model for OFDM with differential encoding in frequency domain.}%
\label{Model_minislot}
\end{figure*}

To simultaneously mitigate the costly channel estimation overhead and avoid the significant performance degradation in mini-slot-assisted SPT, one promising solution is to apply non-coherent detection, which does not require CSI\cite{Xu2019Sixty}.
%Since the ability to strike a balance between system complexity and performance, differential modulation has emerged as an appealing non-coherent detection scheme\cite{Wang2012Dispensing}.
Differential modulation (DM) has therefore emerged as an appealing non-coherent detection scheme over the years for various wireless communications systems due to its simple transceiver structure and low complexity (e.g. \cite{Xu2019Sixty} and \cite{Wang2012Dispensing}).
%due to the ability to strike a balance between system complexity and performance\cite{Wang2012Dispensing}.
%the ability to balance tradeoff between system complexity and performance.
In DM, the transmitter maps the information into the phase difference between consecutive symbols, while the receiver recovers the desired data by directly comparing the phases between two successively received symbols, thereby eliminating the need for CSI (e.g. \cite{Baeza2019Non,Avendi2014Performance,Himsoon2008Differential}).
%Thus, differential modulation eliminates the need for CSI, reducing the reference signal overhead in short packet URLLC.
Furthermore, depending on whether DM is performed on adjacent subcarriers of the same OFDM symbol or on the same subcarrier of adjacent OFDM symbols,
differential OFDM systems can be classified into frequency domain differential OFDM\cite{Kun2004A} and time domain differential OFDM\cite{Osaki2019Differentially}.
%Furthermore, depending on whether the differential modulated symbols are placed on adjacent subcarriers of the same OFDM symbol or on the same subcarrier of adjacent OFDM symbols, frequency domain differential OFDM and time domain differential OFDM can be distinguished.
%frequency domain differential OFDM\cite{Kun2004A} and time domain differential OFDM\cite{Osaki2019Differentially} can be distinguished.
%Fig.1(e-j) depict the reference signal configurations for the mini-slot with frequency domain differential OFDM and time domain differential OFDM.
%The reference signal configurations for the mini-slot employing frequency domain differential OFDM and time domain differential OFDM are depicted in Fig.1(e-j).
%Even though non-coherent detection schemes are well-studied in the general area of wireless communications, the potential benefits of employing differential modulation in massive MIMO with FBL for achieving URLLC have rarely been reported in the literature so far.
These existing works on DM, however, are all aimed at conventional long packet/slot-based transmission. An interesting question is therefore how DM could be applied to SPT and how it would perform. For the former, based on the 3GPP mini-slot structures, DM can be implemented in the frequency domain, given the relatively small channel variation over two adjacent sub-carriers (especially for mmWave bands) - hence termed frequency domain differential OFDM (FDDi-OFDM) in this paper, as shown in Figs.1(e-g); or realized in the time domain, given the possibly small channel variation over two adjacent OFDM symbols - hence termed time domain differential OFDM (TDDi-OFDM) in this paper, as shown in Figs.1(h-j).

On the other hand, it is also worth noting that, in low mobility scenarios with 4 and 7 OFDM symbol units (as illustrated in Fig.1), the DM scheme (especially the TDDi-OFDM scheme) still consumes a sizable (although less) reference signal overhead compared to the pilot-based coherent counterpart.
This implies that, given the well known 3dB issue with DM, the pilot-based coherent detection scheme may still remain competitive in a slow fading environment when using a relatively medium or large sized mini-slot.
%As a result, it is beneficial for obtaining satisfactory performance and reducing the reference signal overhead in mini-slot-assisted short packet URLLC to adaptively switch between differential and coherent detection scheme based on application scenarios, channel statistics, information payloads, and service requirements.
Consequently, it may well be necessary to adaptively switch between differential and coherent modes based on application scenarios, channel statistics, information payloads, mini-slot deployment options and service requirements, minimizing the reference signal overhead and the impact on the reliability performance.
%reducing the reference signal overhead and obtaining satisfactory performance in mini-slot-assisted short packet URLLC.
%affecting the minimal impact on reliable data transmission.
%affecting the minimal impact on the duration of data transmission period.
%Although
Despite both non-coherent and coherent detection having in the past been investigated in wireless communications, the potential advantages of utilizing adaptively differential and coherent modulation modes for mini-slot-assisted short packet URLLC have not been explored in the literature, which constitutes the main motivation of this paper.

\begin{figure*}[!htp]
\centering
\includegraphics [width=0.9\textwidth]{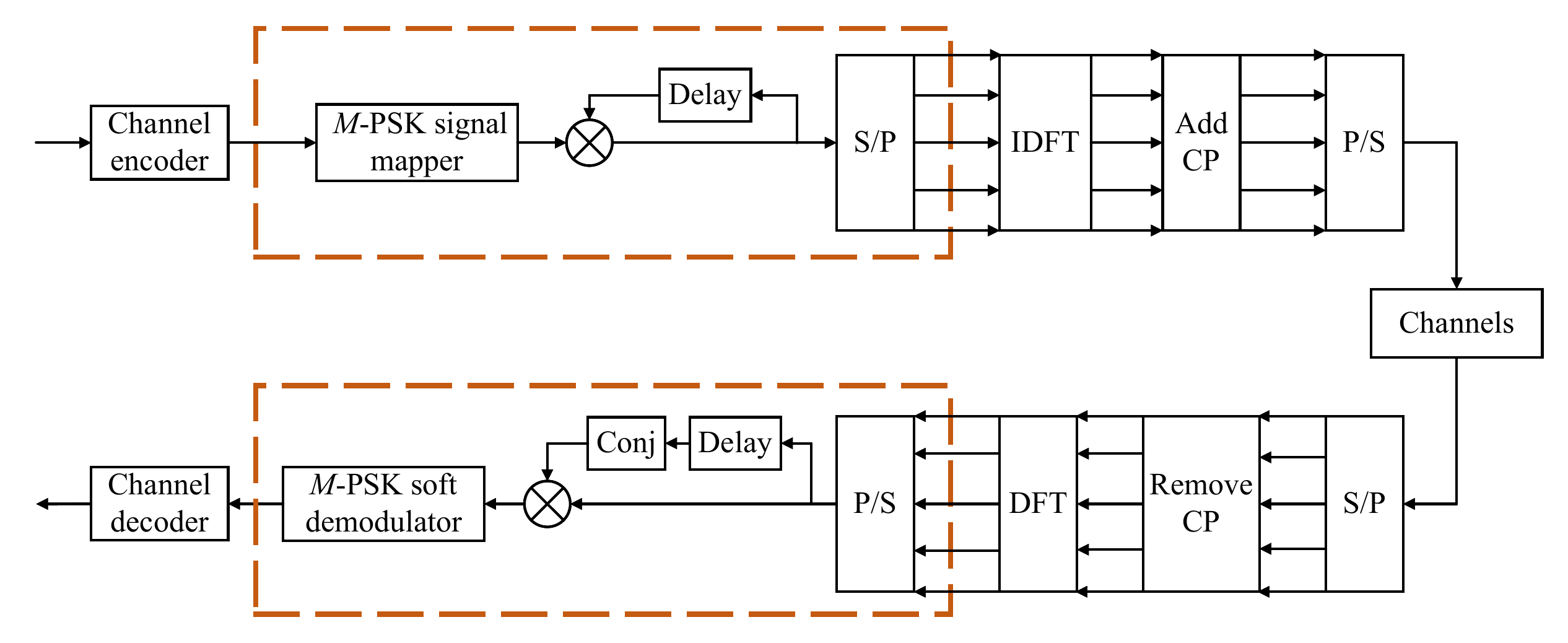}
%\vspace{-1.55em}
\caption{Block diagram of FDDi-OFDM.}
%{System model for OFDM with differential encoding in frequency domain.}%
\label{DPSKwithMC}
\end{figure*}

\subsection{Related Works and Our Contributions}
1) Pilot and coherent detection based SPT:
%Coherent detection-based SPT:
%Considerable research has investigated the performance of coherent detected SPT under the assumption of perfect CSI
Substantial research has delved into the performance of coherent detection for short packet transmission (SPT) with perfect CSI over various types of channels, including ergodic fading channels\cite{Polyanskiy2011Scalar}, block fading channels\cite{Collins2019Coherent} and quasi-static fading channels\cite{Yang2014Quasi}.
Building on these foundations, the impact of pilot lengths and mismatched CSI on the maximum achievable coded rate and packet error rate performance of SPT was studied for multiple antenna systems in \cite{Ferrante2018Pilot}, for massive multiple-input multiple-output (MIMO) systems in \cite{Kislal2023Efficient} and \cite{Johan2021URLLC}, and for cell-free massive MIMO systems in \cite{Lancho2023Cell}.
%in \cite{Johan2017Finite} for single antenna systems,
Additionally,
%several existing works have explored the optimization of channel estimation overhead for coherent detected SPT.
%the optimization of channel estimation overheads has been explored in some existing works for coherent detected URLLC.
there have been some attempts to optimize the channel estimation overhead for coherent SPT detection and URLLC.
%\cite{Mousaei2017Optimizing,Cao2022Independent,Zhou2023Joint,Lee2018Packet,Johan2019Short}.
For example,
in \cite{Mousaei2017Optimizing},
%the optimization of pilot length was investigated, considering the packet size and the error probability.
the pilot length optimization of short packets considering the packet size and the error probability was investigated.
%the pilot overhead optimization of short packets is investigated in \cite{Mousaei2017Optimizing} for a point-to-point communications system while taking into account the packet size and the error probability.
%the pilot length optimization of short packets taking into account the packet size and the error probability was investigated in \cite{Mousaei2017Optimizing}.
%the pilot overhead optimization of short packets was studied in
%In \cite{Ren2020Joint}, joint pilot and payload power allocation for short packet URLLC users was explored, and a low complexity optimization algorithm was proposed to maximize the achievable rate of the uplink massive MIMO system.
%\cite{Ren2020Joint} explored joint pilot and payload power allocation for short packet URLLC users, proposing a low complexity optimization algorithm to maximize the achievable rate of the uplink massive MIMO system.
In \cite{Cao2022Independent}, joint optimization of pilot length and block length was explored for multi-device URLLC, developing dynamic and static optimization algorithms to maximize effective throughput with independent pilot or shared pilot frame structures.
%where both dynamic and static optimization algorithms were developed to maximize the effective throughput with independent-pilot or shared-pilot frame structures.
In \cite{Zhou2023Joint}, a partial superimposed pilot frame structure was introduced for SPT, proposing a joint pilot length, pilot power and payload power allocation algorithm to maximize the weighted sum rate of the presented scheme.
Moreover, \cite{Lee2018Packet} investigated a data-assisted channel estimation approach to enhance the channel estimation quality with reduced training overheads in SPT.
However, the non-asymptotic information theoretic results presented in \cite{Johan2019Short} reveal that, for block-fading channels, non-coherent SPT is more energy efficient than its pilot-assisted counterpart with maximum likelihood (ML) channel estimation, even with pilot length and power optimized.
Indeed, due to the inherent tradeoff between performance and signaling overhead under coherent SPT, optimizing the pilot length alone remains insufficient to minimize the reference signal overhead, particularly when the channel conditions change rapidly.
%, especially in rapidly fading channels
%high mobility environments with

2) Non-coherent detection-based SPT:
%Non-pilot based transmission and non-coherent detection
Given the high training cost in coherent detection schemes, research on non-coherent SPT detection has emerged as a topic of interest in URLLC.
Specifically,
in \cite{Lancho2020On} and \cite{Lancho2020Saddlepoint}, the maximum achievable coded rate of non-coherent SPT detection over single antenna Rayleigh block-fading channels was derived using normal approximation and saddlepoint methods, respectively.
The results of non-coherent SPT detection under single antenna systems were subsequently extended to MIMO systems in \cite{Qi2020A}.
%\cite{Lancho2020On} and \cite{Lancho2020Saddlepoint} employed normal approximation and saddlepoint methods to derive the maximum achievable coding rate of non-coherent SPT over single antenna Rayleigh block-fading channels, respectively.
Moreover,
several studies have explored pilot-free designs for SPT in URLLC.
For instance, in \cite{Wu2020Pilot} and \cite{Ji2019Pilot},
a sparse vector coding approach without pilots was proposed for SPT
%relying on identifying nonzero elements of the received signals for decoding.
where the decoding process relies on identifying nonzero elements of the received signals.
%the information bits are detected by finding out the nonzero positions of a sparse vector.
%Two major disadvantages of this scheme are its low spectral efficiency and poor performance in high-mobility scenarios, particularly when the length of the sparse vector is relatively large.
However, this approach suffers from two main drawbacks: low spectral efficiency (especially when the sparse vector length is larger), and considerably diminished performance in high mobility scenarios.
%However, this scheme suffers from low spectral efficiency and considerably diminished performance in high-mobility scenarios
In \cite{Liu2019Fast}, an iterative semi-blind receiver structure was proposed to enable mini-slot-assisted URLLC in short frame full duplex systems with carrier frequency offsets.
In addition,
%\cite{Li2021Constellation} studied a non-coherent ML receiver for massive single-input multiple-output URLLC systems, designing the optimal constellation structures to improve symbol detection reliability.
in \cite{Walk2019MOCZ} and \cite{Siddiqui2024Spectrally},
%\cite{Walk2020MOCZ},
a novel blind SPT strategy was introduced where the data is modulated onto the zero structure of the $z$-transform of the transmitted discrete-time baseband signal
%the complex zeros of the transmitted discrete-time baseband signal's $z$-transform
and the information is recovered by using either an ML detector or a simple direct zero-testing detector at the receiver.
%the receiver detects the information by applying a ML detector or a simple direct zero-testing detector.
%However, the computational complexity of the ML detector may be unacceptable high for URLLC, while the direct zero-testing detector with low complexity suffers significant performance degradation, especially for frequency-selective fading channels.
Unfortunately, the high computational complexity of the ML detector might render it unsuitable for URLLC. Conversely, while the direct zero-testing detector offers lower complexity, it experiences significant performance degradation, particularly in frequency-selective fading channels.
%Nevertheless, how to optimally balance tradeoff among overheads, system performance and computational complexity of mini-slot-assisted URLLC across a wide range of application scenarios remains a open problem.

%the direct zero-testing detector, although lower in complexity,

%the DiZeT method ignores the correlation and simplifies (12) to independent zero decisions, which results in non-negligible performance degradation, especially for
%frequency-selective fading channels.

3) DM-based SPT:
%URLLC with differential modulation:
%There are several studies in the literature aiming to implement the mini-slot-assisted URLLC with low training overheads.
%Recently, in \cite{Ullah2023Implementation}, the low overhead implementation of a mini-slot-assisted short packet physical layer network coding system over software-defined radio platforms was investigated where a small number of pilots is still required in an OFDM symbol to address issues including channel estimation, time offsets and carrier frequency offsets (CFO).
%In addition, an iterative semi-blind receiver structure was proposed in \cite{Liu2019Fast} to enable mini-slot-assisted URLLC in short frame full duplex systems with CFO.
Surprisingly, DM has been largely overlooked in the literature related to URLLC except for \cite{Choi2021Generalized} and \cite{Jiang2019Packet}.
In \cite{Choi2021Generalized}, a combination of DM and index modulation was proposed for OFDM-based pilot-free SPT,
while in \cite{Jiang2019Packet}, differential preamble detection was applied to minimize the synchronization overhead of SPT. %and perform accurate detection.
Despite these efforts,
%that differential modulation is considered in the aforementioned studies to achieve low overhead URLLC,
the theoretical analysis and comparative insight of DM based SPT have never been fully established.
Our recent work \cite{Zheng2023Differential} on single carrier systems has demonstrated, from an information theoretic perspective,
that DM is a promising solution to achieve efficient SPT with low training overheads, providing helpful design insights for URLLC.
%based SPT can outperform pilot-assisted SPT in terms of data rate, power fairness and latency, providing helpful design insights for DM based SPT.
%The aforementioned studies either focus on the coherent detection or the non-coherent detection, in which the issue of pilot overhead in SPT is considered seriously, but the solutions are still imperfect in
%striking a balance among overheads, system performance and computational complexity.
Regrettably, although the aforementioned studies have considered the issue of pilot overhead, the solutions are still imperfect in striking a balance among signaling overhead, system performance and computational complexity either by
%through
%employing either coherent detection or non-coherent detection\cite{Wu2020Pilot,Ji2019Pilot,Liu2019Fast,Walk2019MOCZ,Siddiqui2024Spectrally,Choi2021Generalized,Jiang2019Packet,Zheng2023Differential}.
optimizing the pilot length\cite{Mousaei2017Optimizing,Cao2022Independent,Zhou2023Joint,Lee2018Packet} or by employing  non-coherent detection\cite{Wu2020Pilot,Ji2019Pilot,Liu2019Fast,Walk2019MOCZ,Siddiqui2024Spectrally,Choi2021Generalized,Jiang2019Packet,Zheng2023Differential}.
Therefore, this paper aims to make a first step towards bridging this gap and tries to explore an efficient transmission paradigm based on adaptive non-coherent detection and coherent detection to achieve URLLC with an optimum reference signal overhead.

\section{System Model and Assumptions}

%\begin{figure*}[!htp]
%\centering
%\includegraphics [width=1\textwidth]{Fig3}
%%\vspace{-1.55em}
%\caption{Block diagram of frequency domain differential OFDM.}
%%{System model for OFDM with differential encoding in frequency domain.}%
%\label{DPSKwithMC}
%\end{figure*}

\begin{figure*}[!ht]
\setlength{\abovecaptionskip}{5pt}%2
\setlength{\belowcaptionskip}{-5pt}%-5
\centering
\subfloat[Time domain differential modulation and demodulation]{
\includegraphics[width=0.4\textwidth]{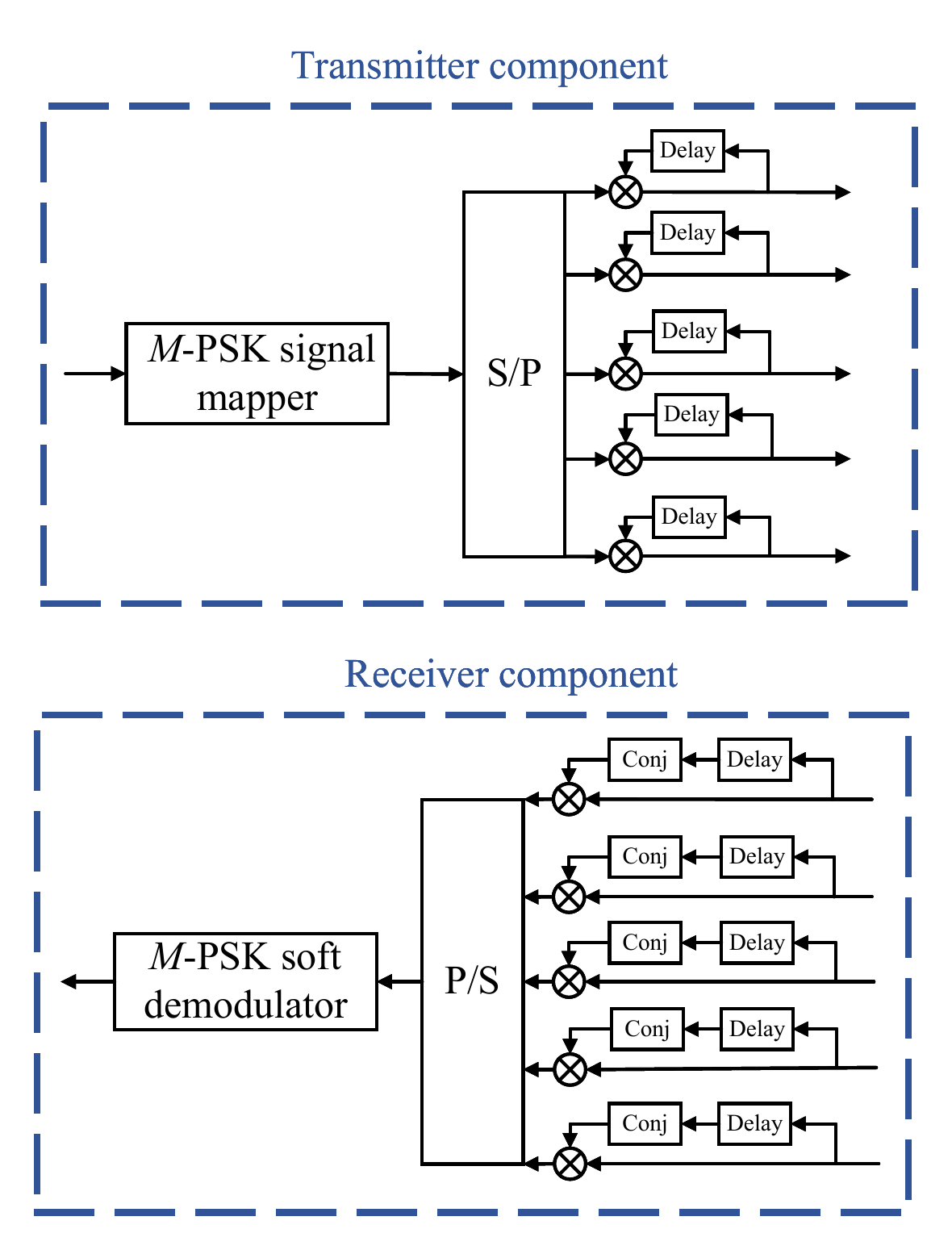}
}%\vspace{-2ex}
\subfloat[pilots-based PSK/QAM modulation and demodulation]{
\includegraphics[width=0.4\textwidth]{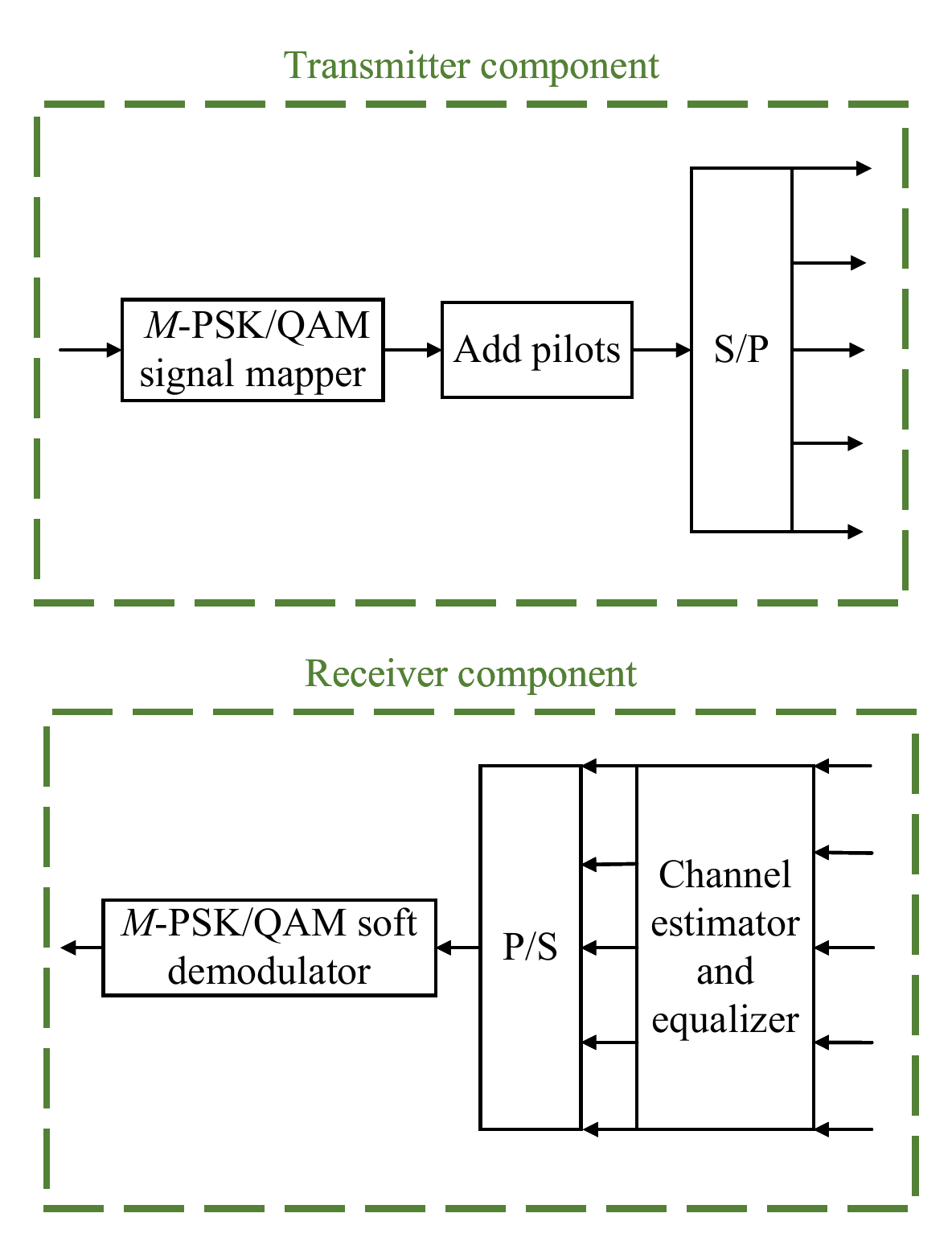}
}
\caption{Transceiver components of (a) TDDi-OFDM and (b) PA-OFDM.}
%%OFDM with differential encoding in time domain and (b) OFDM with pilot symbol-assisted modulation.}
%\vspace{-1.25em}
%\label{pic_k}
\end{figure*}
\setlength{\textfloatsep}{26pt}

%\begin{figure*}[!ht]
%\setlength{\abovecaptionskip}{5pt}%2
%\setlength{\belowcaptionskip}{-5pt}%-5
%\centering
%\subfloat[Time domain differential modulation and demodulation]{
%\includegraphics[height=0.25\linewidth]{Fig_2b_OFDM}
%\includegraphics[height=0.25\linewidth]{Fig_2c_OFDM}
%}
%\subfloat[Pilot symbol-assisted modulation and demodulation]{
%\includegraphics[height=0.25\linewidth]{Fig_2d_OFDM}
%\includegraphics[height=0.25\linewidth]{Fig_2e_OFDM}
%}
%\caption{Transceiver components of (a) time domain differential OFDM and (b) pilot symbol-assisted OFDM.}
%%OFDM with differential encoding in time domain and (b) OFDM with pilot symbol-assisted modulation.}
%%\vspace{-1.25em}
%%\label{pic_k}
%\end{figure*}

In this paper, we consider a mini-slot-assisted single-input single-output URLLC system with $K$ subcarriers and $T$ consecutive OFDM symbols.
%Without loss of generality, a sequence of emergency information $\mathbf{b} = \left\{ {{b_l}} \right\}_{l = 1}^{L}$ with a length of $L$ is assumed at the transmitter.
%We assume the presence of a sequence of emergency information $\mathbf{b} = \left{ {{b_l}} \right}{l = 1}^{L}$, where $L$ denotes the length of the sequence.
%We assume a sequence of emergency information $\mathbf{g} = \left\{ {{g_b}} \right\}_{b = 1}^{B}$ at the transmitter, where $B$ represents the length of the sequence.
We assume a bit sequence of emergency information $\mathbf{g} = \left\{ {g_1},{g_2},...,{g_B} \right\}$ at the transmitter and $B$ is the sequence length.
%$\mathbf{g} = \left\{ {{g_b}} \right\}_{b = 1}^{B}$ of $B$ length at the transmitter.
%Without loss of generality, we assume a sequence of emergency information bits of $B$ length at the transmitter.
%$L$-bit emergency information
%Without loss of generality, we assume the presence of a sequence of $B$-bit emergency information at the transmitter.
%Without loss of generality, a sequence of emergency information bits of $B$ length is assumed at the transmitter.
%we assume a $B$-length emergency information bit sequence at the transmitter
In order to enhance the system reliability, a channel encoder is employed, transforming $B$ information bits into a sequence of $J$ coded bits.
%we utilize a channel encoder that encodes $L$ information bits into a group of $J$ coded bits.
Furthermore, we consider an adaptive differential and pilot symbol-assisted OFDM transmission scheme to accommodate varying channel conditions and information payloads,
%minimizing the channel estimation overhead and the impact on the system performance in short packet URLLC
minimizing the channel estimation overheads in short packet URLLC.
Fig. 2 illustrates the block diagram of FDDi-OFDM.
In addition, the transceiver components within the two dashed boxes of Fig. 2 can also be replaced with TDDi-OFDM \cite{Ding2006Performance} and conventional pilot symbol-assisted OFDM (PA-OFDM), as shown in Fig. 3. The details are as follows.
%the transceiver components within the two dashed boxes of Fig. 1 is distinct from that of time domain differential and pilot symbol-assisted OFDM
%the transceiver components of time domain differential and pilot symbol-assisted OFDM is distinct from that within the two dashed boxes of Fig. 1, as shown in Fig. 2.
%as illustrated in Fig. 2.

%a group of coded bits of length $J$.
%Additionally, $J$ coded bits are mapped into a complex symbol sequence $\mathbf{c} = \left\{ {{c_ 1},{c_2 },...,{c_J }} \right\}$ of length $N_d = J/{\log _2}M$
%In addition, the adaptive differential/pilot symbol-assisted OFDM transmission scheme that adapt to the varying channel statistics and information payloads is considered in our system model to ease the costly channel estimation overheads in mini-slot-assisted URLLC.

%We use $\mathcal{X}^\text{PSK} = \left\{ {{e^{j{\varphi _m}}}} \right\}_{m = 0}^{M-1}$ to be the constellation set of $M$-PSK symbols.
\subsection{Differentially modulated OFDM}
For differentially modulated OFDM, we first map the $J$ coded bits into an $M$-ary phase-shift keying ($M$-PSK) symbol block $\mathbf{c} = \left\{ {{c_1, c_2,...,c_N}} \right\}$ of length $N= J/{\log _2}M$, where $c_n \in \mathcal{X}^\text{PSK}$ and $N$ is assumed to be an integer.
Here, $\mathcal{X}^\text{PSK} = \left\{ {{e^{j{\varphi _m}}}} \right\}_{m = 0}^{M-1}$ is the constellation alphabet of $M$-PSK symbols and ${\varphi _m} = {{2\pi m} / M}$ is the information-carrying phase.
%$J$ coded bits are first mapped into a block of $M$-PSK symbols $\mathbf{c} = \left\{ {{c_q}} \right\}_{q = 1}^{Q}$ of length $Q = J/{\log _2}M$, where $c_q  \in \mathcal{X}^\text{PSK}$ and $Q$ is assumed to be an integer multiple of ${\log _2}M$.
%$\mathbf{c} = \left\{ {{c_ 1},{c_2 },...,{c_{N_s} }} \right\}$
Moreover, depending on whether the differentially modulated symbols are placed on adjacent subcarriers within the same OFDM symbol duration or on the same subcarrier but across adjacent OFDM symbols, we can make a distinction between frequency domain and time domain differential OFDM.
%frequency domain or time domain differential OFDM can be constructed
Specifically, we rearrange $N$-length PSK symbol block into a matrix ${\left( {{v_{k,t}}} \right)}$ of size ${\left( {K- 1} \right) \times T}$ for FDDi-OFDM.
%\footnote{As illustrated in Fig. 1 and 2(a), the implementation of frequency domain or time domain differential OFDM.have a different sequence of arrangement, as illustrated in Fig. 1(b).}.
By contrast, in TDDi-OFDM, the $N$-length PSK symbol block is rearranged into a matrix ${\left( {{v_{k,t}}} \right)}$ with dimensions ${K \times \left( {T- 1} \right)}$.
%Depending on whether the differential modulated symbols are placed on adjacent subcarriers of the same OFDM symbol or on the same subcarrier of sequential OFDM symbols, time domain and frequency domain differential OFDM can be distinguished.
Then, the signal assigned to the $k$-th subcarrier of the $t$-th OFDM symbol, denoted as ${d_{k,t}}$, is given by\cite[Eq.(15)]{Ding2006Performance}
%assign
\begin{equation}
\begin{aligned}
%{d_{t,k}} = \left\{ {\begin{array}{*{20}{c}}
%{{v_{t,k}}{d_{t-1,k }},~\text{frequency-domain differential OFDM}}\\
%{{v_{t,k}}{d_{t,k-1 }},~~~~~~~\text{time-domain differential OFDM}}
%\end{array}} \right.
{d_{k,t}} = \left\{ {\begin{array}{*{20}{c}}
{{v_{k,t}}{d_{k-1,t }},~~~~~~~~~~~~~\text{FDDi-OFDM}}\\
{{v_{k,t}}{d_{k,t-1 }},~~~~~~~~~~~~~\text{TDDi-OFDM}}
\end{array}} \right.
\label{equ_OFDM_DPSK}
\end{aligned}
\end{equation}
%Specifically, we use ${d_{n,k}}$ to denote the transmitted signal in the $n$-th subcarrier of the $k$-th OFDM symbol and rearrange $Q$-length PSK symbol block to a matrix ${\left( {{v_{n,k}}} \right)}$ of the ${\left( {N - 1} \right) \times K}$ size for frequency domain differential OFDM.
%On the other hand, for time domain differential OFDM, $Q$-length PSK symbol block is rearranged to a matrix ${\left( {{v_{n,k}}} \right)}$ with the size of ${N \times \left( {K - 1} \right)}$.
%Then, the transmitted signal ${d_{n,k}}$ is given by
%\begin{equation}
%\begin{aligned}
%{d_{i,k}} = \left\{ {\begin{array}{*{20}{c}}
%{{v_{i,k}}{d_{i,k - 1}},~~~n=0,...,N-1 ~\text{and}~k=1,...,K-1}\\
%{{v_{i,k}}{d_{i-1,k }},~~~n=1,...,N-1 ~\text{and}~k=0,...,K-1}
%\end{array}} \right.
%\label{equ_OFDM_DPSK}
%\end{aligned}
%\end{equation}
where ${d_{0,t}}=1$ and ${d_{k,0}}=1$ are the initial reference signal of the $t$-th OFDM symbol in FDDi-OFDM and of the $k$-th subcarrier in TDDi-OFDM, respectively.
%Equivalently,
%${d_{0,k}}=1$ is the initial reference signal of the $k$-th OFDM symbol for frequency domain differential OFDM and ${d_{n,0}}=1$ is the initial reference signal of the $n$-th subcarrier for time domain differential OFDM

%Specifically, we use ${d_{n,k}}$ to denote the transmitted signal in the $n$-th subcarrier of the $k$-th OFDM symbol and rearrange $Q$-length PSK symbol block to a matrix ${\left( {{v_{n,k}}} \right)}$ of the ${\left( {N - 1} \right) \times K}$ size.
%%a (${\left( {N - 1} \right) \times K}$)-size matrix ${\left( {{v_{i,k}}} \right)_{\left( {N - 1} \right) \times K}}$
%Then, the transmitted signal in frequency domain differential OFDM is given by
%\begin{equation}
%\begin{aligned}
%{d_{n,k}} = {v_{n,k}}{d_{n - 1,k}},
%\label{equ_fre_domain_DPSK}
%\end{aligned}
%%\vspace{-0.5em}
%\end{equation}
%for $n=1,...,N-1$ and $k=0,...,K-1$, where ${d_{0,k}}=1$ is the initial reference signal of the $k$-th OFDM symbol.
%For time domain differential OFDM, $Q$-length PSK symbol block is rearranged to a matrix ${\left( {{v_{n,k}}} \right)}$ with the size of ${N \times \left( {K - 1} \right)}$ and the transmitted signal can be obtained as follows:
%\begin{equation}
%\begin{aligned}
%{d_{n,k}} = {v_{n,k}}{d_{n,k-1}},
%\label{equ_time_domain_DPSK}
%\end{aligned}
%%\vspace{-0.5em}
%\end{equation}
%for $n=0,...,N-1$ and $k=1,...,K-1$, where ${d_{n,0}}=1$ is the initial reference signal of the $n$-th subcarrier.

%On the other hand, for pilot symbols-assisted OFDM,
%To obtain the transmitted signal for OFDM systems, $N$-point inverse discrete Fourier transform (IDFT) is performed for

In OFDM systems, the transmitted sequence is obtained by performing an $K$-point inverse discrete Fourier transform (IDFT) on $\left\{ {{d_{k,t}}} \right\}_{k = 0}^{K - 1}$.
Let $L$ denote the maximum number of channel paths. A cyclic prefix (CP) of length $L$ is then added in the transmitted sequence.
%To mitigate ISI caused by multipath propagation, a cyclic prefix (CP) of length $T_\text{CP}$ is appended to the transmitted sequence.
%By performing $N$-point inverse discrete Fourier transform (IDFT) for $\left\{ {{d_{n,k}}} \right\}_{n = 0}^{N - 1}$, the transmitted signal is obtained
%We denote $N_\text{CP}$ denote the length of the cyclic prefix (CP).
%Let $L$ denote the number of channel paths.
%Let $L$ denote the number of channel taps.
%We further assume that a cyclic prefix (CP) of length $N_\text{CP}$ is appended in the transmitted sequence, reducing inter symbol interference (ISI) incurred by multipath delays.
%, as in
%we employ a cyclic prefix (CP) of length $N_\text{CP}\ge L$ to avoid inter symbol interference (ISI) caused by multipath delays
%We further assume that a cyclic prefix (CP) of length $N_\text{CP}>L$ is added to avoid inter symbol interference (ISI) caused by multipath delays.%interference among OFDM symbols.
%Consequently, the transmitted sequence of the $k$-th OFDM symbol with the addition of CP can be expressed as
Consequently, the transmitted sequence of the $t$-th OFDM symbol, including the CP, can be expressed as
\begin{equation}
\begin{aligned}
{s_{t}\left( i \right)} = \frac{1}{{\sqrt K }}\sum\limits_{k = 0}^{K - 1} {{d_{k,t}}{e^{j{{2\pi ki} \mathord{\left/
 {\vphantom {{2\pi ki} K}} \right.
 \kern-\nulldelimiterspace} K}}}} ,
\label{equ_OFDMsymbol}
\end{aligned}
%\vspace{-0.5em}
\end{equation}
for $-L\le i\le K-1$.
%, where {s_{k}\left( i \right)} is the .
%Let $L$ denote the number of channel paths and we assume perfect synchronization at the receiver.
%Let $L$ denote the number of multipath components and assume the wide-sense stationary uncorrelated scattering (WSSUS) channel in our system model
%We use $L$ and ${h_k}\left( {i,l} \right)$ respectively represent the number of multipath components and the channel impulse response of the $l$-th path during the $i$-th sample of the $k$-th OFDM symbol.
%In addition, we assume that the channel follows the wide-sense stationary uncorrelated scattering (WSSUS) model.
Moreover, the time-varying wide sense stationary uncorrelated scattering (WSSUS) Rayleigh fading channels with arbitrary power delay profiles are assumed, which can be modeled by tapped delay line models.
%For WSSUS Rayleigh fading channels, the channel gain of different paths is spatially uncorrelated and independent.
%In particular, we use $L$ to denote the number of channel tap and ${h_k}\left( {i,l} \right) \sim \mathcal{CN}\left( {0,\sigma _l^2} \right)$ to represent the tap gain of the $l$-th path during the $i$-th sample of the $k$-th OFDM symbol, where $\sigma _l^2$ is the normalized fading power of the $l$-th path with the constraint $\sum\limits_{l = 1}^L {\sigma _l^2}  = 1$. Then, we have
% with coefficients
In particular, we use ${h_t}\left( {l} \right)$ to denote the tap gain of the $l$-th path during the $t$-th OFDM symbol.
For WSSUS Rayleigh fading channels, ${h_t}\left( {l} \right)$ of different channel paths is spatially uncorrelated and independent.
%Let $\sigma _l^2$ and $G$ represent the normalized fading power of the $l$-th path and the bandwidth, respectively. %such that
%We denote the normalized fading power of the $l$-th path and the bandwidth as $\sigma _l^2$ and $G$, respectively.
We denote the normalized fading power of the $l$-th path as $\sigma _l^2$.
Then, we have
%$\sum\limits_{l = 1}^L {\sigma _l^2}  = 1$
$\sum\nolimits_{l = 1}^L {\sigma _l^2 = 1} $
and ${h_t}\left( {l} \right) \sim \mathcal{CN}\left( {0,\sigma _l^2} \right)$.
%As in , the time-varying nature of the $l$-th path is further characterized by the following correlation function:
%Furthermore, we characterize the time-varying nature of the $l$-th path based on the following correlation function:
Furthermore, Jakes' model is utilized to characterize the time-varying nature of the channels and the autocorrelation function of the $l$-th path is given by
\begin{equation}
\begin{aligned}
\mathbb{E}\left\{ {{h_{t+\Delta t}}\left( {l} \right){h^*_t}\left( {l} \right)} \right\}=\sigma _l^2{{\cal J}_0}\left( {{{2\pi {f_d}{T_{s}}\Delta t} }} \right),
%${\mathcal{J}_0}\left( {\frac{{2\pi {f_d}{T_{{\rm{sym}}}}\Delta i}}{N}} \right)$,
%= \sigma _l^2{\mathcal{J}_0}\left( {2\pi {f_d}{T_s}\Delta i} \right)
\label{channel_correlation}
\end{aligned}
\end{equation}
where $\mathcal{J}_{0}\left( x  \right)$ is the zero-th order Bessel function of the first kind, $f_d$ is the Doppler frequency and ${T_s}$ is the sampling period.
%${T_s} = {{1}}/{G}$ is the symbol duration.
%${T_s} = \frac{{{T_{{\rm{sym}}}}}}{T}$ and ${T_{{\rm{sym}}}} = \frac{{T + L}}{G}$ is the OFDM symbol duration.

Based on ${h_t}\left( {l} \right)$, the channel frequency response at the $k$-th subcarrier of the $t$-th OFDM symbol can be written as
\begin{equation}
\begin{aligned}
{H_{ {k,t} }} = \sum\limits_{l = 0}^{L - 1} {{h_t}\left( {l} \right)} {e^{-j{{2\pi kl} \mathord{\left/ {\vphantom {{2\pi kl} K}} \right. \kern-\nulldelimiterspace}K}}}.
\label{CFR}
\end{aligned}
\end{equation}
%In this work, the channel is assumed to be composed of
%Dh + 1 independent multipath components each of which has
%a gain hm  and delay m  Ts
%As in, the channel is assumed to follow the wide-sense stationary uncorrelated scattering (WSSUS) model.
%In particular, we use $L$ and ${h_k}\left( {i,l} \right)$ to respectively denote the number of multipath components and the channel tap gain of the $l$-th path during the $i$-th sample of the $k$-th OFDM symbol.
%Then, the channel frequency response is given by
%\begin{equation}
%\begin{aligned}
%{H_k}\left( {i,n} \right) = \sum\limits_{l = 0}^{L - 1} {{h_k}\left( {i,l} \right)} {e^{-j{{2\pi nl} \mathord{\left/ {\vphantom {{2\pi nl} N}} \right. \kern-\nulldelimiterspace} N}}}.
%\label{CFR}
%\end{aligned}
%\end{equation}
%In addition,
%%For the WSSUS channel model, the time-domain correlation function is given by
%we characterize the time-varying nature of the channel based on the following time domain correlation function:
%For the WSSUS channel model,
Using (\ref{channel_correlation}) and (\ref{CFR}), the time-frequency correlation function of ${H_{ {k,t} }}$ for different symbols and subcarriers can be further expressed as
%the product of a time domain correlation function and a frequency domain correlation function:
\begin{equation}
\begin{aligned}
{\rho }\left( {\Delta k,\Delta t} \right)&=\mathbb{E}\left\{ {{H_{k+\Delta k,t +\Delta t} },{H^*_{ {k,t} }}} \right\}\\
%= {\rho _t}\left( {\Delta i} \right){\rho _f}\left( {\Delta n} \right),
%= \sigma _l^2{\mathcal{J}_0}\left( {2\pi {f_d}{T_s}\Delta i} \right)
&= {{\cal J}_0}\left( {{{2\pi {f_d}{T_{s}}\Delta t} }} \right) \sum\limits_{l = 0}^{L-1} {\sigma _l^2{e^{ - j2\pi l\Delta k /K}}}.
%=\underbrace {{J_0}\left( {2\pi {f_d}{T_s}\Delta k} \right)}_{{\rho _t}\left( {\Delta k} \right)}\underbrace {\sum\limits_{l = 0}^{L - 1} {\sigma _l^2{e^{ - j2\pi l\Delta t/T}}} }_{{\rho _f}\left( {\Delta t} \right)}
\label{corr_fre_time}
\end{aligned}
\end{equation}
It is worth noting that (\ref{corr_fre_time}) reduces to
\begin{equation}
\begin{aligned}
{\rho _f}\left( {\Delta k} \right)={\rho }\left( {\Delta k,0} \right)= \sum\limits_{l = 0}^{L-1} {\sigma _l^2{e^{ - j2\pi l\Delta k/K}}}
\label{corr_fre}
\end{aligned}
\end{equation}
for $\Delta t=0$ and
\begin{equation}
\begin{aligned}
{\rho _t}\left( {\Delta t} \right)={\rho }\left( {0,\Delta t} \right)={{\cal J}_0}\left( {{{2\pi {f_d}{T_{s}}\Delta t} }} \right)
\label{corr_time}
\end{aligned}
\end{equation}
for $\Delta k=0$.
As a result, we can separate ${\rho }\left( {\Delta k,\Delta t} \right)$ as the product of the frequency domain correlation function ${\rho _f}\left( {\Delta k} \right)$ (depending on the multi-path delay profile) and the time domain correlation function ${\rho _t}\left( {\Delta t} \right)$ (depending on the mobility).
Especially, the frequency domain correlation for two consecutive subcarriers of the same OFDM symbol is given by
\begin{equation}
\begin{aligned}
{{\rho _f} = {\rho _f}\left( 1 \right) = \sum\limits_{l = 0}^{L - 1} {\sigma _l^2{e^{ - j2\pi l/K}}} }
\label{corr_fre_two}
\end{aligned}
\end{equation}
and the time domain correlation for two adjacent OFDM symbols on the same subcarrier is
\begin{equation}
\begin{aligned}
{{\rho _t} = {\rho _t}\left( 1 \right) = {J_0}\left( {2\pi {f_d}{T_s}} \right)}
\label{corr_time_two}
\end{aligned}
\end{equation}

%the time-frequency correlation function can be represented as the product of the time domain correlation function and the frequency domain correlation function
%
%the time-frequency correlation function can be further expressed as the product of a time domain correlation function and a frequency domain correlation function:
%\begin{equation}
%\begin{aligned}
%\mathbb{E}\left\{ {{H_k}\left( {i,n} \right),{H^*_k}\left( {i+\Delta i,n+\Delta n} \right)} \right\}= {\rho _t}\left( {\Delta i} \right){\rho _f}\left( {\Delta n} \right),
%%= \sigma _l^2{\mathcal{J}_0}\left( {2\pi {f_d}{T_s}\Delta i} \right)
%\label{channel_WSSUS}
%\end{aligned}
%\end{equation}
%where ${\rho _f}\left( {\Delta n} \right)$ is the frequency domain correlation function of ${H_k}\left( {i,n} \right)$ given by
%\begin{equation}
%\begin{aligned}
%{\rho _f}\left( {\Delta n} \right) &=\mathbb{E}\left\{ {{H_k}\left( {i,n} \right),{H^*_k}\left( {i,n+\Delta n} \right)} \right\}\\
%&= \sum\limits_{l = 1}^L {\sigma _l^2{e^{ - j2\pi l\Delta n}}}
%\label{FD_correlation}
%\end{aligned}
%\end{equation}
%and $\sigma _l^2$ is the power of the $l$-th path.

With the assumption of perfect symbol synchronization, the corresponding received signal of ${s_{t}\left( i \right)}$ after removing the CP is
%can be written as
\begin{equation}
\begin{aligned}
{r_t}\left( i \right) = \sum\limits_{l = 0}^{L - 1} {{h_t}\left( {l} \right){s_t}\left( {i - l} \right)}  + {w_t}\left( i \right),
\label{TD_rece_signal_with_CP}
\end{aligned}
\end{equation}
for $0\le i \le K-1$, where ${w_t}\left( i \right)$ represents the complex additive white Gaussian noise (AWGN) with zero mean and variance $\sigma _w^2$.
Then, computing a $K$-point discrete Fourier transform (DFT) of $\left\{ {r_t}\left( i \right) \right\}_{i = 0}^{K- 1}$ yields the following expression for the received signal at the $k$-th subcarrier of the $t$-th OFDM symbol:
\begin{equation}
\begin{aligned}
{z_{k,t}} = \frac{1}{{\sqrt K }}\sum\limits_{i = 0}^{K - 1} {{r_t}\left( i \right)} {e^{{{ - j2\pi ki} \mathord{\left/
 {\vphantom {{ - j2\pi ki} K}} \right.
 \kern-\nulldelimiterspace} K}}}.
\label{FD_rece_signal}
\end{aligned}
\end{equation}
%For differential modulated OFDM, the information is further recovered from the phase difference of two adjacent signals on the the same sub-carrier or of two sub-carrier signals of the same OFDM symbol
%to complete differential demodulation
%Furthermore, for differential modulated OFDM, the information is retrieved from the phase difference of two adjacent signals on the the same sub-carrier or of two sub-carrier signals of the same OFDM symbol
%to complete differential demodulation
Furthermore, by substituting (\ref{equ_OFDMsymbol}) and (\ref{TD_rece_signal_with_CP}) into (\ref{FD_rece_signal}), the received symbol after DFT can be rewritten as
\begin{equation}
\begin{aligned}
&{z_{k,t}}\\
&= \underbrace {\left[ { {\sum\limits_{l = 0}^{L - 1} {{h_t}\left( {l} \right){e^{{{ - j2\pi kl} \mathord{\left/
 {\vphantom {{ - j2\pi kl} K}} \right.
 \kern-\nulldelimiterspace} K}}}} } } \right]}_{{\rm{Channel~frequency~response}}}{d_{t,k}}  + \underbrace {\frac{1}{{\sqrt K }}\sum\limits_{i = 0}^{K - 1} {{w_t}\left( i \right){e^{{{ - j2\pi ki} \mathord{\left/
 {\vphantom {{ - j2\pi ki} K}} \right.
 \kern-\nulldelimiterspace} K}}}} }_{{\rm{Additive~noise}}}\\
&= {H_{k,t}}{d_{k,t}}  + {W_{k,t}}.
\label{rece_signal_DFT_without_ISI}
\end{aligned}
\end{equation}
%we can easily decompose the received symbol after DFT into the desired signal ${d_{t,k}}$, multiplicative distortion ${H_{t,k}}$, ICI term $I^{{\rm{ICI}}}$ and additional noise term ${W_{t,k}}$, as expressed by
%\begin{equation}
%\begin{aligned}
%{z_{t,k}}& = \underbrace {\frac{1}{T}\left[ {  \sum\limits_{i = 0}^{T - 1} {\sum\limits_{l = 0}^{L - 1} {{h_k}\left( {i,l} \right){e^{{{ - j2\pi tl} \mathord{\left/
% {\vphantom {{ - j2\pi tl} T}} \right.
% \kern-\nulldelimiterspace} T}}}} } } \right]}_{{\rm{Multiplicative~distortion }}}{d_{t,k}}+ \\
%&\underbrace {\frac{1}{T}\sum\limits_{t' \ne t} {{d_{t',k}}\left[ {\sum\limits_{i = 0}^{T - 1} {\sum\limits_{l = 0}^{L - 1} {{h_k}\left( {i,l} \right){e^{{{ - j2\pi t'l} \mathord{\left/
% {\vphantom {{ - j2\pi t'l} T}} \right.
% \kern-\nulldelimiterspace} T}}}{e^{{{j2\pi i\left( {t' - t} \right)} \mathord{\left/
% {\vphantom {{j2\pi i\left( {t' - t} \right)} T}} \right.
% \kern-\nulldelimiterspace} T}}}} }  } \right]} }_{{\rm{ICI }}}\\
%&  + \underbrace {\frac{1}{{\sqrt T }}\sum\limits_{i = 0}^{T - 1} {{w_k}\left( i \right){e^{{{ - j2\pi ti} \mathord{\left/
% {\vphantom {{ - j2\pi ti} T}} \right.
% \kern-\nulldelimiterspace} T}}}} }_{{\rm{Additive~noise}}}\\
%&= {H_{t,k}}{d_{t,k}} + I^{{\rm{ICI}}} + {W_{t,k}}.
%\label{rece_signal_DFT_without_ISI}
%\end{aligned}
%\end{equation}
For the variance of ${W_{k,t}}$, we have
\begin{equation}
\begin{aligned}
\mathbb{E}\left\{ {{W_{k,t}}W_{k,t}^ * } \right\} = \frac{1}{K}\sum\limits_{i = 0}^{K - 1} {\mathbb{E}\left\{ {{w_t}\left( i \right)w_t^ * \left( i \right)} \right\}} = \sigma _w^2.
\label{variance_noise}
\end{aligned}
\end{equation}
%Let $\sigma _{H}^2$ and $\sigma _{{\rm{ICI}}}^2$ denote the variance of ${H_{t,k}}$ and $I^{{\rm{ICI}}}$, respectively.
Therefore, the SNR for the transmitted signals in mini-slot-assisted OFDM is
\begin{equation}
\begin{aligned}
\gamma  = \frac{1}{{ \sigma _w^2}}.
%{{\sigma _H^2}}/\left( {{\sigma _{{\rm{ICI}}}^2 + \sigma _w^2}}\right).
\label{SINR_coh}
\end{aligned}
\end{equation}

%Based on ${z_{t,k}}$,
For differentially modulated OFDM,
the information is detected at the receiver by comparing two received signals, which can be two adjacent signals on the same subcarrier, or two subcarrier signals during the same OFDM duration.
%Furthermore, the information in differential modulated OFDM is detected by comparing two received signals, which are two subcarrier signals of the same OFDM symbol for frequency domain differential OFDM or two adjacent signals on the same subcarrier for time domain differential OFDM.
Hence, the detected signal is given by
% can be written as
%mathematically
\begin{equation}
\begin{aligned}
%{a_{t,k}} = \left\{ {\begin{array}{*{20}{c}}
%{{z_{t,k}}{z_{t-1,k }^*},~\text{frequency-domain differential OFDM}}\\
%{{z_{t,k}}{z_{t,k-1 }^*},~~~~~~~\text{time-domain differential OFDM}}
{a_{k,t}} = \left\{ {\begin{array}{*{20}{c}}
{{z_{k,t}}{z_{k-1,t }^*},~~~~~~~~~~~~~\text{FDDi-OFDM}}\\
{{z_{k,t}}{z_{k,t-1 }^*},~~~~~~~~~~~~~\text{TDDi-OFDM}}
\end{array}} \right.
\label{dete_OFDM_DPSK}
\end{aligned}
\end{equation}
Subsequently, the parallel detected signal is rearranged into a serial sequence and passed through the soft decision detector and the channel decoder. Finally, an estimate for the $B$-bit information, $\widehat{ \mathbf{g}} = \left\{ {{\widehat g}_1},{{\widehat g}_2},...,{{\widehat g}_B} \right\}$, is obtained at the receiver.

\subsection{PA-OFDM}
On the other hand, if coherent detection is used, the pilot symbols have to be inserted in mini-slot-assisted SPT to estimate CSI and detect information.
%perform equalization.
The channel estimation performance, however, may heavily depend on the pilot patterns and the estimation schemes employed.
For PA-OFDM, we employ the pilot patterns defined in 3GPP (i.e., \cite{3GPP912} and \cite{3GPP211}) for mini-slot-assisted URLLC.
Specifically, let ${\mathcal{T}}$ be the set of the pilot-carrying OFDM symbols. We assume that the pilot symbols are placed at equally spaced subcarriers of each OFDM symbol in $ {\mathcal{T}}$, as shown in Figs. 1(a-d).
%Furthermore, we denote the total number of the pilot symbols in mini-slot as ${\lambda _p}$.
%Then, the interval of pilot subcarriers is ${\delta _p}=T\left|{\mathcal{K}_p}\right|/{\lambda _p}$.
%the number of pilots in a pilot-carrying OFDM symbol is ${\lambda _p}/\left| {\mathcal{K}_p} \right|$ and
%we use ${\delta _p}$ to denote the total number of the pilot symbols in mini-slot-assisted OFDM
Furthermore, we denote the interval between pilot subcarriers as ${\delta _{sub}}$ and consider two widely used digital modulation schemes, namely PSK and QAM, in the system model for PA-OFDM.
%represent the OFDM symbol interval between two adjacent pilot-carrying OFDM symbols in ${\mathcal{K}}$ as ${\delta _{sym}}$.
%Especially, if there is only one pilot-carrying OFDM symbol in mini-slot (i.e.,$\left|{\mathcal{K}}\right|=1$), we have ${\delta _{sym}}=K$.
%In this paper we consider MQAM and MPSK modulation schemes.
%Thus, the total number of the pilot symbols in mini-slot-assisted OFDM is ${ \lambda_p}=T\left|{\mathcal{K}_p}\right|/{\delta _p}$.
Then, the number of pilots for each pilot-carrying OFDM symbol is ${\lambda_p } = K/{\delta_{sub}}$ and the total number of the pilot symbols in the entire mini-slot-assisted OFDM is ${ \lambda_{\rm{total}}}={\lambda_p }\left|{\mathcal{T}}\right|$.
%In addition, two widely used digital modulation schemes, namely PSK and QAM, are considered in our system model for pilot symbol-assisted OFDM.

At the receiver, we adopt a combination of LMMSE channel estimation and linear interpolation to obtain the estimated CSI of mini-slot-assisted OFDM.
For the convenience of understanding, we divide resource elements in mini-slot-assisted OFDM into the following three distinct types.
\begin{itemize}
\item Pilots: Channel estimates at pilot subcarriers are computed using the LMMSE estimator.
%The channel estimates at pilot subcarriers are based on the LMMSE estimator.
    Let $ \mathbf{H}_{t_p}= \left[ {{H_{0 ,t_p}}~{H_{\delta_{sub} ,t_p}}~...~{H_{\left( { \lambda_p- 1} \right) \delta_{sub} ,t_p}}} \right]^T$ denote the channel responses at pilot subcarriers of a pilot-carrying OFDM symbol and $ {\widehat {\mathbf {H}}_{{\rm{LMMSE}}}}= \left[ {{\widehat{H}_{0 ,t_p}}~{\widehat{H}_{\delta_{sub} ,t_p}}~...~{\widehat{H}_{\left( { \lambda_p- 1} \right) \delta_{sub} ,t_p}}} \right]^T$ denote the corresponding LMMSE estimates, where the index $t_p \in {\mathcal{T}}$.
% Let $ \mathbf{H}_{k_p}= \left[ {{H_{0 ,k_p}}~{H_{\delta ,k_p}}~...~{H_{\left( { \lambda_p- 1} \right) \delta ,k_p}}} \right]^T$ and $ {\widehat {\mathbf {H}}_{{\rm{LMMSE}}}}= \left[ {{\widehat{H}_{0 ,k_p}}~{\widehat{H}_{\delta ,k_p}}~...~{\widehat{H}_{\left( { \lambda_p- 1} \right) \delta ,k_p}}} \right]^T$ respectively denote the channel responses and the corresponding LMMSE estimates at pilot subcarriers of a pilot-carrying OFDM symbol, where the index $k_p \in {\mathcal{K}}$.
    %Then, ${\widehat {\mathbf {H}}_{{\rm{LMMSE}}}}$ can be expressed by
    Then, the expression for ${\widehat {\mathbf {H}}_{{\rm{LMMSE}}}}$ is given by \cite[Eq.(11)]{Liu2014Channel}
    %\cite[Eq.(5)]{Edfors1998OFDM}
\begin{equation}
\begin{aligned}
%{\widehat {\mathbf {H}}_{{\rm{LMMSE}}}} = {{\mathbf {R}}_{{\mathbf {H}_k}{\mathbf {H}_k}}}{\left( {{{\mathbf {R}}_{{\mathbf {H}_k}{\mathbf {H}_k}}} + \frac{\alpha }{{\gamma}}\mathbf {I}} \right)^{ - 1}}{\widehat {\mathbf {H}}_{{\rm{LS}}}}
{\widehat {\mathbf {H}}_{{\rm{LMMSE}}}} = {{\mathbf {R}}_{{\mathbf {H}_{t_p}}}}{\left( {{{\mathbf {R}}_{{\mathbf {H}_{t_p}}}} + \frac{1 }{{\gamma}}\mathbf {I}} \right)^{ - 1}}{\widehat {\mathbf {H}}_{{\rm{LS}}}}
\label{LMMSE_esti}
\end{aligned}
\end{equation}
where ${{\mathbf {R}}_{{\mathbf {H}_{t_p}}}}= \mathbb{E}\left\{ {{\mathbf {H}_{t_p}}{\mathbf {H}_{t_p}^H}} \right\}$ is the channel autocorrelation matrix at the pilot subcarriers, $\mathbf{I}$ is the identity matrix and ${\widehat {\mathbf {H}}_{{\rm{LS}}}}= \left[ {\frac{{{z_{0,t_p}}}}{{{d_{0,t_p}}}}~\frac{{{z_{\delta_{sub} ,t_p}}}}{{{d_{\delta_{sub} ,t_p}}}}~...~\frac{{{z_{\left( { \lambda_p- 1} \right) \delta_{sub} ,t_p}}}}{{{d_{\left( { \lambda_p- 1} \right) \delta_{sub} ,t_p}}}}} \right]$ is the least-squares (LS) estimate of $\mathbf{H}_{t_p}$.
%, $\alpha=\mathbb{E}\left\{ {{{\left| {{d_{t,k}}} \right|}^2}} \right\}\mathbb{E}\left\{ {{{\left| {1/{d_{t,k}}} \right|}^2}} \right\}$

\item Data symbols within pilot-carrying OFDM symbols: The linear interpolation technique is utilized to acquire CSI of data subcarriers lying between two pilot subcarriers. With the linear interpolation technique, the channel estimate of $\left( \lambda \delta_{sub} + {k_d} \right) $-th data subcarriers is given by \cite[Eq.(3)]{Kim2005Performance}
\begin{equation}
\begin{aligned}
{\widehat H_{\lambda \delta_{sub}  + {k_d},{t_p}}} = \frac{{\delta_{sub}  - {k_d} }}{\delta_{sub} }{\widehat H_{\lambda \delta_{sub} ,{t_p}}} + \frac{{k_d} }{\delta_{sub} }{\widehat H_{\left( {\lambda  + 1} \right)\delta_{sub} ,{t_p}}}
\label{linear_esti}
\end{aligned}
\end{equation}
for $\lambda=0,...,\lambda_p-2$ and ${k_d}=1,...,\delta_{sub}-1$.
Note that the channel estimates for the data subcarriers that lie after the last pilot subcarrier cannot be obtained directly from (\ref{linear_esti}).
%For this case, we can use the last and second last pilot subcarriers and compute the channel estimates of the data subcarriers at the edge as\cite[Eq.(4)]{Kim2005Performance}
In such cases, we utilize the last and second-last pilot subcarriers to compute channel estimates for the data subcarriers at the edge, as expressed by\cite[Eq.(4)]{Kim2005Performance}:
\begin{equation}%\nonumber
\begin{aligned}
&{{\widehat H}_{\left( {{\lambda _p} - 1} \right)\delta_{sub}  + {k_d},{t_p}}} =\\
&- \frac{{{k_d}}}{\delta_{sub} }{{\widehat H}_{\left( {{\lambda _p} - 2} \right)\delta_{sub} ,{t_p}}} + \frac{{\delta_{sub}  + {k_d}}}{\delta_{sub} }{{\widehat H}_{\left( {{\lambda _p} - 1} \right)\delta_{sub} ,{t_p}}}
\label{edge_esti}
\end{aligned}
\end{equation}
for ${k_d}=1,...,\delta_{sub}$.
%there are a few of the data subcarriers at the edge
%they are very few in number, they can been ignored in this analysis.

\item Data symbols that do not belong to pilot-carrying OFDM symbols: The channel estimates for data symbols that do not belong to pilot-carrying OFDM symbols are assumed to be a reuse of the channel estimates from pilot-carrying OFDM symbols. Specifically, data symbols that do not belong to pilot-carrying OFDM symbols can be further divided into two types: data symbols at the region A and B, as illustrated in Fig.1(d). In Region A, the estimated CSI of the pilot sub-carriers is directly used to detect data symbols while in Region B, the channel estimates of data symbols are obtained by directly utilizing the estimated CSI of data sub-carriers using the linear interpolation.

%    the estimated CSI of the pilot sub-carriers is directly using to detect data symbols at the region A. On the other hand, the channel estimates of data symbols at the region B are obtained by directly utilizing the estimated CSI of data sub-carriers using the linear interpolation subcarrier. In other words, the channel estimates for data symbols that do not belong to pilot-carrying OFDM symbols are essentially a reuse of the channel estimates from pilot-carrying OFDM symbols.

%    the channel estimates of data symbols at the region A are obtained by directly using the estimated CSI at the pilot subcarrier. On the other hand, the estimated CSI of data sub-carriers using the linear interpolation
%    the receiver directly employs the estimated CSI from the pilot sub-carriers to detect data symbols at the region A. On the other hand, the detection of data symbols at the region B relies on
%    For data symbols at the region A, the estimated CSI is obtained by reusing

%The channel estimation and detection for data symbols that do not belong to pilot-carrying OFDM symbols relies on
%The estimated CSI of a pilot-carrying OFDM symbol is utilized to detect several subsequent symbols until another pilot-carrying OFDM symbol occurs.
\end{itemize}
Based on the estimated CSI, we can detect the transmitted information for PA-OFDM.

\section{Performance analysis of mini-slot-assisted short packet OFDM}%{Preliminaries of Codebook Design}

\newcounter{TempEqCnt7}                        % 创建临时变量TempEqCnt
\setcounter{TempEqCnt7}{\value{equation}} % 将当前公式序号 赋给TempEqCnt
\setcounter{equation}{27}
\begin{figure*}[hb]
\hrulefill
%\textcolor{blue}{
\begin{equation}
\begin{aligned}
\sigma _{ e}^2=& \frac{{{\lambda _p}}}{{K{\delta _{sym}}}}{\Phi _{{\rm{LMMSE}}}} + \left( {\frac{{K - {\lambda _p}}}{{K{\delta _{sym}}}} - \frac{{{\delta _{sub}} - 1}}{{K{\delta _{sym}}}}} \right){\Phi _{{\rm{linear}}}} + \frac{{{\delta _{sub}} - 1}}{{K{\delta _{sym}}}}{\Phi _{{\rm{edge}}}} + \frac{{{\lambda _p}\left( {{\delta _{sym}} - 1} \right)}}{{K{\delta _{sym}}}}{\Phi _{{\rm{A}}}} + \frac{{\left( {K - {\lambda _p}} \right)\left( {{\delta _{sym}} - 1} \right)}}{{K{\delta _{sym}}}}{\Phi _{{\rm{B}}}}
\label{var_chan_error}
\end{aligned}
\end{equation}
\end{figure*}
\setcounter{equation}{\value{TempEqCnt7}}

\subsection{Preliminaries on Non-Asymptotic Information Theory}
%In this subsection, we briefly summarize the relevant literature on the non-asymptotic short packet information theory.
%the non-asymptotic short packet information theoretic results in the relevant references
%the background literature on non-asymptotic as well as first-order and second-order asymptotic channel coding rates for several point-to-point (P2P) and multi-user problems in information theory.
%In the classic information theory, Shannon’s second theorem (i.e., the channel coding theorem) points that any reliable communications can be achieved with an infinite block length as long as the transmission rate remains below the channel capacity. As a result, the
%In this subsection, we give a brief review for the non-asymptotic short packet information theory in the relevant references.
In information theory, a well-established result is that any reliable communications can be achieved with an infinite blocklength as long as the transmission rate remains below the channel capacity.
Consequently, the classical Shannon's capacity formula serves as a performance metric for designing communication systems with an asymptotically large blocklength.
%capture the behaviour of the maximum coding rate in the limit as the blocklength tends to infinity
However, such a canonical principle is unsuitable for the performance analysis and system design of SPT in URLLC due to the inherently constrained blocklength.
%However, the performance analysis and system design for SPT in URLLC cannot adhere to such a canonical principle due to the inherently constrained blocklengths.
%The performance analysis and system design for SPT in URLLC, however, cannnot rely on such a classical asymptotic information theoretic result due to the strictly limited blocklength.
%To enable the analysis of the interplay among the rate, reliability and blocklength in SPT, Polyanskiy et al. in \cite{Polyanskiy2010Channel} developed several non-asymptotic information-theoretic bounds and derived an approximation formula of the maximum achievable coding rate for a point-to-point (P2P) AWGN channel with given BLER and blocklength.
%To facilitate the performance analysis of SPT,
To address this limitation in SPT,
%To facilitate the performance analysis of SPT,
Polyanskiy et al. in \cite{Polyanskiy2010Channel} developed several non-asymptotic upper and lower bounds on the maximum coded rate and the smallest error probability. %facilitating the analysis of the interplay among the rate, reliability and blocklength in SPT.
%Polyanskiy et al. in \cite{Polyanskiy2010Channel} established several non-asymptotic information-theoretic bounds by obtaining upper and lower bounds on the maximum coding rate or the smallest error probability of short packets.
%Here, we briefly review a pair of upper and lower bounds on the smallest error probability introduced in \cite{Polyanskiy2010Channel}, which are commonly referred to as the IS and DT bounds.
Here, we briefly review a specific pair of upper and lower bounds on the smallest error probability introduced in \cite{Polyanskiy2010Channel}, commonly referred to as the information spectrum (IS) and dependence testing (DT) bounds, which play a crucial role in our subsequent analysis.
%are later utilized in our analysis.
%Here, we briefly review a pair of upper and lower bounds on the smallest error probability introduced in \cite{Polyanskiy2010Channel}, which are commonly referred to as the IS and DT bounds and later utilized in our analysis.
%Before stating these bounds, we need to define a few notations.
%Let $\varepsilon  = \mathbb{P}\left[ {{\bf{g}} \ne {\bf{\hat g}}} \right]$ denote the probability that the receiver makes an erroneous guess about the information bit sequence $\mathbf{b}$ (i.e., BLER).
%In addition, we use
%Let $X \in \mathcal{X}$ and $Y\in \mathcal{Y}$ respectively denote the input and output symbol of the channel, where $\mathcal{X}$ is the input alphabet and $\mathcal{Y}$ is the output alphabet.
%Then, the channel can be modeled with a conditional transition probability ${{\mathbb{P}}_{Y^{N}\left| X^{N} \right.}}\left( {y^{N}\left| x^{N} \right.} \right) :\mathcal{X}^{N}  \to \mathcal{Y}^{N}$ for blocklength $N$.
%In addition, we define
%\begin{equation}
%\begin{aligned}
%i\left( {X^{N};Y^{N}} \right) = \log_2 \frac{{{\mathbb{P}}_{Y^{N}\left| X^{N} \right.}}\left( {y^{N}\left| x^{N} \right.} \right)}{{{\mathbb{P}}_{Y^{N}}}\left( y^{N} \right)}
%\label{Th1_inf_dens}
%\end{aligned}
%%\vspace{-0.1em}
%\end{equation}
%as the information density. Then,
For the convenience of our derivation next, the IS and DT bounds, as derived in \cite{Polyanskiy2010Channel}, are included below.
\newtheorem{theorem}{Theorem}
\begin{theorem}
(IS lower bound \cite[Th.11]{Polyanskiy2010Channel} ): For a general P2P channel consists of an input alphabet $\mathcal{X}$, an output alphabet $\mathcal{Y}$ and a conditional channel transition probability ${{\mathbb{P}}_{Y^{\overline N}\left| X^{\overline N} \right.}}\left( {y^{\overline N}\left| x^{\overline N} \right.} \right) :\mathcal{X}^{\overline N}  \to \mathcal{Y}^{\overline N}$ with blocklength $\overline N $,
%an $N$ blocklength ${{\mathbb{P}}_{Y^{N}\left| X^{N} \right.}}\left( {y^{N}\left| x^{N} \right.} \right)$
%${{\mathbb{P}}_{Y^{N}\left| X^{N} \right.}}\left( {y^{N}\left| x^{N} \right.} \right)$,
the minimum achievable BLER $\varepsilon$ with information bits of length $B$ is lower-bounded as
%\vspace{-0.1em}
\begin{equation}
\begin{aligned}
%\varepsilon  \geqslant \sum\limits_{{x^N} \in {\mathcal{X}^N}} {{{\mathbb{P}}_{{X^N}}}\left( {{x^N}} \right) {{\mathbb{P}}_{Y^{N}\left| X^{N} \right.}}\left[ { i\left( {X^{N};Y^{N}} \right) > {\log _2}{\gamma _N}} \right] - \frac{{{\gamma _N}}}{{{2^L}}}}
\varepsilon  \geqslant \mathop {\sup }\limits_{\beta  > 0} \left\{ {\mathop {\inf }\limits_{{{\mathbb{P}}_X}} {\mathbb{P}}\left[ {i\left( {X^{\overline N};Y^{\overline N}} \right) \leqslant {{\log }_2}\beta } \right] - \frac{\beta }{{{2^B}}}} \right\}
\label{Th0_lower}
\end{aligned}
%\vspace{-0.1em}
\end{equation}
where
%\vspace{-0.1em}
$\beta>0$ is an arbitrary positive constant and
\begin{equation}
\begin{aligned}
i\left( {X^{\overline N};Y^{\overline N}} \right) = \log_2 \frac{{{\mathbb{P}}_{Y^{\overline N}\left| X^{\overline N} \right.}}\left( {y^{\overline N}\left| x^{\overline N} \right.} \right)}{{{\mathbb{P}}_{Y^{\overline N}}}\left( y^{\overline N} \right)}
\label{infor_density}
\end{aligned}
\end{equation}
is termed the information density.
%where
%\begin{equation}
%\begin{aligned}
%i\left( {X^{N};Y^{N}} \right) = \log_2 \frac{{{\mathbb{P}}_{Y^{N}\left| X^{N} \right.}}\left( {y^{N}\left| x^{N} \right.} \right)}{{{\mathbb{P}}_{Y^{N}}}\left( y^{N} \right)}
%\label{Th1_inf_dens}
%\end{aligned}
%\end{equation}
%is the information density
\end{theorem}

%\vspace{-1.55em}
\newtheorem{theorem0}{Theorem}
\begin{theorem}
(DT upper bound \cite[Th.17]{Polyanskiy2010Channel}): For a general P2P channel ${{\mathbb{P}}_{Y^{\overline N}\left| X^{\overline N} \right.}}\left( {y^{\overline N}\left| x^{\overline N} \right.} \right)$, the minimum achievable BLER $\varepsilon$ with information bits of length $B$ is upper-bounded as
%\vspace{-0.1em}
\begin{equation}
\begin{aligned}
\varepsilon  \leqslant \mathbb{E}\left[ {\exp \left\{ { - {{\left[ {i\left( {{X^{\overline N}};{Y^{\overline N}}} \right) - {{\log }_2}\left( {\left({{{2^{B}} - 1}}\right)/{2}} \right)} \right]}^ + }} \right\}} \right],
\label{Th1_DT}
\end{aligned}
%\vspace{-0.1em}
\end{equation}
where ${\left[ A \right]^ + }$ = $A$ if $A \geqslant 0$ and 0 otherwise.
%${\left[ A \right]^ + } = \left\{ {\begin{array}{*{20}{c}}
%  {A,~~~A \geqslant 0} \\
%  {0,~~~A < 0}
%\end{array}} \right.$.
\end{theorem}

Obviously, the IS lower bound signifies an impossible result: for a given blocklength, encoding and decoding schemes that lie below the IS bound are unattainable. Conversely,
%the DT bound indicates the presence of a feasible encoding and decoding scheme, despite the method of constructing such an encoding and decoding scheme does not specified in the upper bound.
the upper bound indicates there exist some encoding and decoding schemes whose performance is at least as good as the DT bound, despite the method of constructing such these schemes being unknown.
%does not specified in the upper bound.
%the upper bound does not prescribe specific methods for constructing such an encoding and decoding scheme.
%to construct such an encoding and decoding scheme.
%In information theory, the IS and DT bounds are generally referred to as converse and achievability bounds on the error probability, respectively \cite{Cover2006Elements}.
%This section reviews the achievability bounds that will be used in Chapters 5 and 6.
Note that the non-asymptotic bounds presented in \cite{Polyanskiy2010Channel} are applicable for SPT under any P2P channels, enabling us to utilize the IS and DT bounds to evaluate the error probability of mini-slot-assisted short packet URLLC.
%However, although tight non-asymptotic bounds in Theorems 1 and 2 can help with precise analysis and design of SPT systems, their numerical computation is usually cumbersome and does not explicitly reflect the interplay among the rate, reliability and blocklength in SPT.
Moreover, to describe the asymptotic behavior of the non-asymptotic upper and lower bounds, \cite{Polyanskiy2010Channel} derived an elegant approximation formula for the maximum coded rate and the minimum error probability under P2P Gaussian channels.
%on the maximum coding rate and the minimum error probability,
%by employing a central-limit-theorem analysis.
%as a function of
%Moreover, to capture the behavior of the non-asymptotic upper and lower bounds, \cite{Polyanskiy2010Channel} derived an approximation formula of the maximum achievable coding rate as a function of blocklength, error probability and SNR
%It is therefore essential to provide a precise approximation to capture the behavior of the non-asymptotic upper and lower bounds on the minimum error probability.
%To this end, \cite{Polyanskiy2010Channel} derived an elegant approximation formula for the maximum achievable coding rate under P2P AWGN channels with given BLER and blocklength
In particular,
%To explicitly reflect the interplay among the rate, reliability and blocklength in SPT, \cite{Polyanskiy2010Channel}
%To capture the
%the effect of system parameters (e.g., modulation order, SNR, blocklength and coding rate) on the performance of SPT from an engineering point of view.
%However, as observed in Theorem 1 and 2, obtaining BLER results based on the IS and DT bounds does not explicitly reflect the impact of system parameters, such as modulation order, hierarchical codebook, SNR, blocklength, and coding rate, on the BLER performance.
%from an engineering perspective.
%function
%the maximum achievable rate (bits/symbol, bits/s/Hz or bits/channel use) under P2P Gaussian channels with given BLER $\varepsilon$ and blocklength $N$ can be tightly approximated as\cite[Eq.(296)]{Polyanskiy2010Channel}
the maximum achievable rate (bits/symbol, bits/s/Hz or bits/channel use) for P2P Gaussian channels can be tightly approximated as the following function of blocklength $\overline N$, error probability $\varepsilon$ and SNR $\gamma$ \cite[Eq.(296)]{Polyanskiy2010Channel}:
%This conventional information theoretic result, however, is unsuitable to characterize the performance of SPT
%transmission and performance optimization
\begin{equation}
\begin{aligned}
{ R} \approx C\left( \gamma \right) - \sqrt {\frac{V\left( \gamma \right) }{\overline N}} {Q^{ - 1}}\left( \varepsilon \right)+\frac{{{{\log }_2}{\overline N}}}{{2{\overline N}}}.
\label{rate_fbl}
\end{aligned}
\end{equation}
Here $C\left( \gamma \right) = {\log _2}\left( {1 + \gamma } \right)$
%, where $\rho$ denotes the SNR,
is the complex AWGN channel capacity,
${V\left( \gamma \right) }=\gamma \left( {2 + \gamma } \right)/{\left( {1 + \gamma } \right)^2}$ is known as
%referred to as
channel dispersion and ${Q^{ - 1}}\left( \cdot \right)$ is the inverse of the Gaussian Q-function.
Furthermore, for a fixed rate $R = {B}/{\overline N}$ and blocklength, the minimum achievable BLER under P2P Gaussian channels can be equivalently expressed as\cite[Eq.(23)]{Durisi2016Toward}
%the minimum achievable BLER $\widehat {\varepsilon  }$ for P2P Gaussian channels with given blocklength $N$ and information bit length $L$ can be equivalently expressed as
\begin{equation}
\begin{aligned}
\varepsilon  \approx Q\left( {\sqrt{\frac{\overline N}{V\left( \gamma \right)}}\left( {C\left( \gamma \right) - R}+\frac{{{{\log }_2}{\overline N}}}{{2{\overline N}}}\right)} \right).
%\varepsilon  \approx Q\left( {\frac{{N\left( {C - R} \right) + \frac{1}{2}{{\log }_2}N}}{{\sqrt {NV} }}} \right).
\label{bler_fbl}
\end{aligned}
%\vspace{-0.1em}
\end{equation}
%where $R = \frac{L}{N}$ (bits/symbol).
%The expression (\ref{rate_fbl}) and (\ref{bler_fbl}), which are usually termed normal approximation, rely on a central-limit-theorem analysis and describe the asymptotic behavior of the non-asymptotic upper and lower bounds on the maximum coding rate and the minimum error probability.

The expression (\ref{rate_fbl}) and (\ref{bler_fbl}), which are usually termed normal approximation, rely on a central-limit-theorem analysis and characterize the interplay among the rate, reliability and blocklength in SPT under AWGN channels.
%and the non-asymptotic upper and lower bounds on the maximum coding rate and the minimum error probability
%perform asymptotic expansions of the maximum coding rate or the error probability
%describe the asymptotic behavior of the non-asymptotic upper and lower bounds on the maximum coding rate and the minimum error probability.
%Building on this seminal contribution, subsequent studies have generalized the achievable rate and the minimum achievable BLER formula under AWGN to various channels, including scalar coherent fading channels \cite{Polyanskiy2011Scalar, Collins2019Coherent}, quasi-static fading channels \cite{Yang2014Quasi}, non-coherent block fading channels \cite{Johan2019Short,Lancho2020On} and multi-access channels \cite{MolavianJazi2015A} by employing the non-asymptotic information-theoretic bounds proposed in \cite{Polyanskiy2010Channel}.
%It is worth noting that most existing SPT performance results, including \cite{Polyanskiy2011Scalar, Collins2019Coherent,Yang2014Quasi,Ferrante2018Pilot,Kislal2023Efficient, Johan2021URLLC,Lancho2023Cell,Johan2019Short,Lancho2020On,Lancho2020Saddlepoint,Qi2020A,Polyanskiy2010Channel} and \cite{MolavianJazi2015A}, are based on the assumption of continuous Gaussian signals as channel inputs, except \cite{A2014Jazi}, which presented the maximum achievable rate for Gaussian channels with QAM or PSK inputs.
%focused on transmission design and performance optimization based on the achievable rate formula in SPT.
Building on this seminal contribution, subsequent studies (e.g. \cite{Polyanskiy2011Scalar, Collins2019Coherent,Yang2014Quasi,Ferrante2018Pilot,Kislal2023Efficient, Johan2021URLLC,Lancho2023Cell,Johan2019Short,Lancho2020On,Lancho2020Saddlepoint,Qi2020A} and \cite{MolavianJazi2015A,Xiong2023Status,Zheng2021Open}) have generalized the achievable rate and the minimum achievable BLER formula under AWGN to various channels and communication models by employing the non-asymptotic information-theoretic bounds and normal approximation results proposed in \cite{Polyanskiy2010Channel}.
However, it is worth noting that most existing SPT performance results are based on the assumption of continuous Gaussian signals as channel inputs, except \cite{A2014Jazi} which presented the maximum achievable rate for AWGN channels with QAM or PSK inputs.
Nevertheless, extending these SPT performance results to mini-slot-assisted OFDM systems with discrete inputs is still challenging due to imperfect CSI, frequency selective and temporally correlated channels.
\newcounter{TempEqCnt5}                        % 创建临时变量TempEqCnt
\setcounter{TempEqCnt5}{\value{equation}} % 将当前公式序号 赋给TempEqCnt
\setcounter{equation}{29}
\begin{figure*}[hb]
\hrulefill
%\textcolor{blue}{
\begin{equation}
\begin{aligned}
&{\Phi _{{\rm{linear}}}} = \frac{1}{{{\delta _{sub}} - 1}}\sum\limits_{{k_d} = 1}^{{\delta _{sub}} - 1} \frac{1}{{\lambda_p-1}}\sum\limits_{\lambda = 0}^{\lambda_p - 2}  {{\mathbb{E}}\left\{ {{{\left\| {{{\widehat H}_{\lambda {\delta _{sub}} + {k_d},{t_p}}} - {H_{\lambda {\delta _{sub}} + {k_d},{t_p}}}} \right\|}^2}} \right\}}\\
&\mathop  = \limits^{\left( a \right)}\frac{1}{{{\delta _{sub}} - 1}}\sum\limits_{{k_d} = 1}^{{\delta _{sub}} - 1} \frac{1}{{\lambda_p-1}}\sum\limits_{\lambda = 0}^{\lambda_p - 2} {{\mathbb{E}}\left\{ {H_{\lambda {\delta _{sub}} + {k_d},{t_p}}^2} \right\} + {\mathbb{E}}\left\{ {{{\left\| {\frac{{{\delta _{sub}} - {k_d}}}{{{\delta _{sub}}}}{H_{\lambda {\delta _{sub}},{t_p}}}} \right\|}^2}} \right\}}  + {\mathbb{E}}\left\{ {{{\left\| {\frac{{{k_d}}}{{{\delta _{sub}}}}{H_{\left( {\lambda  + 1} \right){\delta _{sub}},{t_p}}}} \right\|}^2}} \right\}\\
& + 2\Re \left\{{\mathbb{E}}\left\{ {\frac{{\left( {{\delta _{sub}} - {k_d}} \right){k_d}}}{{\delta _{sub}^2}}H_{\lambda {\delta _{sub}},{t_p}}^ * {H_{\left( {\lambda  + 1} \right){\delta _{sub}},{t_p}}}} \right\} \right\}+ {\mathbb{E}}\left\{ {{{\left\| { \frac{{\delta_{sub}  - {k_d} }}{\delta_{sub} } { {{e}_{\lambda {\delta _{sub}},{t_p}}}}  + \frac{{k_d} }{\delta_{sub} } { {{e}_{\left( {\lambda  + 1} \right){\delta _{sub}},{t_p}}}}  } \right\|}^2}} \right\}\\
&- 2\Re \left\{{\mathbb{E}}\left\{ {H_{\lambda {\delta _{sub}} + {k_d},{t_p}}^ * \left( {\frac{{{\delta _{sub}} - {k_d}}}{{{\delta _{sub}}}}{H_{\lambda {\delta _{sub}},{t_p}}} + \frac{{{k_d}}}{{{\delta _{sub}}}}{H_{\left( {\lambda  + 1} \right){\delta _{sub}},{t_p}}}} \right)} \right\}\right\}\\
&\mathop  = \limits^{\left( b \right)}\mathcal{L}-\frac{2}{{{\delta _{sub}} - 1}}\sum\limits_{{k_d} = 1}^{{\delta _{sub}} - 1} {\left[ {\frac{{{\delta _{sub}} - {k_d}}}{{{\delta _{sub}}}}\Re \left( {{\rho }\left( {{k_d},0} \right)} \right) + \frac{{{k_d}}}{{{\delta _{sub}}}}\Re \left( {{\rho}\left( {{k_d} - {\delta _{sub}},0} \right)} \right)} \right]}
%&\mathop  = \limits^{\left( b \right)}\frac{{5{\delta _{sub}} - 1}}{{3{\delta _{sub}}}}\sigma _H^2 + \frac{{{\delta _{sub}} + 1}}{{3{\delta _{sub}}}}\Re \left( {{\rho _H}\left( {{\delta _{sub}},0} \right)} \right)-\frac{2}{{{\delta _{sub}} - 1}}\sum\limits_{{t_d} = 1}^{{\delta _{sub}} - 1} {\left[ {\frac{{{\delta _{sub}} - {t_d}}}{{{\delta _{sub}}}}\Re \left( {{\rho _H}\left( {{t_d},0} \right)} \right) + \frac{{{t_d}}}{{{\delta _{sub}}}}\Re \left( {{\rho _H}\left( {{t_d} - {\delta _{sub}},0} \right)} \right)} \right]} \\
%&+ \frac{{2{\delta _{sub}} - 1}}{{6{\delta _{sub}}\left( {{\lambda _p} - 1} \right)}}\sum\limits_{\lambda  = 0}^{{\lambda _p} - 2} {\left( {\frac{{{\tau _\lambda }}}{{\gamma {\tau _\lambda } + 1}} + \frac{{{\tau _{\lambda  + 1}}}}{{\gamma {\tau _{\lambda  + 1}} + 1}}} \right)}
%\frac{{2{\delta _{sub}} - 1}}{{3{\delta _{sub}}}}{\Phi _{{\rm{LMMSE}}}}
%&= \underbrace {\frac{1}{{{\delta _{sub}} - 1}}\sum\limits_{{t_d} = 1}^{{\delta _{sub}} - 1} {E\left\{ {{{\left\| {\frac{{{\delta _{sub}} - {t_d}}}{{{\delta _{sub}}}}{H_{\lambda {\delta _{sub}},{k_p}}} + \frac{{{t_d}}}{{{\delta _{sub}}}}{H_{\left( {\lambda  + 1} \right){\delta _{sub}},{k_p}}} - {H_{\lambda {\delta _{sub}} + {t_d},{k_p}}}} \right\|}^2}} \right\}} }_{{L_1}}
\label{MSE_linear}
\end{aligned}
\end{equation}
\end{figure*}
\setcounter{equation}{\value{TempEqCnt5}}

\subsection{Differentially Modulated Short Packet OFDM}
%In this subsection, we derive the performance of differentially modulated short packet OFDM.
Based on the aforementioned non-asymptotic information-theoretic bounds,
%and statistical properties,
we can derive the performance of differentially modulated short packet OFDM.
%Based on the aforementioned non-asymptotic information-theoretic bounds and statistical properties, we can derive the performance of differentially modulated short packet OFDM.
Let $N_{\text{dif}}$ denote the number of available data symbols in differential OFDM.
Thus, we have ${N_{\text{dif}}}=\left( {K - 1} \right)T$ for FDDi-OFDM and ${N_{\text{dif}}}=K\left( {T - 1} \right)$ for TDDi-OFDM.
Additionally, we use $I_{\text{dif}}\left( \gamma  \right)$ and $V_{\text{dif}}\left( \gamma  \right)$ to denote the expectation and variance of the information density for differential OFDM with a single channel use, respectively.
%which are respectively derived in (\ref{infor_density_mean}) and (\ref{infor_density_variance})
Specifically, $I_{\text{dif}}\left( \gamma  \right)$ and $V_{\text{dif}}\left( \gamma  \right)$ for FDDi-OFDM are given in (\ref{infor_density_mean}) and (\ref{infor_density_variance}) in Appendix A.
Similarly, $I_{\text{dif}}\left( \gamma  \right)$ and $V_{\text{dif}}\left( \gamma  \right)$ for TDDi-OFDM can be obtained using the same procedures with the variable $\rho_f$ in (\ref{infor_density_mean}) and (\ref{infor_density_variance}) replaced by $\rho_t$.
Here, ${\rho _t}$ is the time domain correlation defined in (\ref{corr_time_two}).
%denote the expectation and variance of the information density for differential OFDM with a single channel use as $I_{\text{dif}}$ and $V_{\text{dif}}$, respectively.
Then, the performance of differentially modulated short packet OFDM can be described in the following theorem:
%based on the aforementioned non-asymptotic information-theoretic bounds and statistical properties, we obtain the following theorem for the performance of differentially modulated short packet OFDM:

\newtheorem{theorem1}{Theorem}
\begin{theorem}
(Differentially modulated OFDM):
%For given subcarriers $T$ and OFDM symbols$K$
%For mini-slot-assisted SPT, the maximum achievable rate of differentially modulated OFDM can be determined as
For both frequency and time domain differential OFDM, the maximum achievable rate can be determined as
%Supposing that the memory of the modulation scheme and the channel is ignored, the maximum achievable rate of differential modulation with hard decisions over a point-to-point Gaussian channel can be determined by employing the information spectrum bound and the DT bound as
%\vspace{-0.5em}
\begin{equation}
\begin{aligned}
R  \approx  I_{\text{dif}}\left( \gamma  \right)- {{\sqrt {\frac{ {V_{\text{dif}}}\left( \gamma  \right)  }{N_{\text{dif}}}}   }}Q^{-1}\left( \varepsilon \right)+\frac{{{{\log }_2}{N_{\text{dif}}}}}{{2{N_{\text{dif}}}}}.
%\leqslant
%\varepsilon  \approx Q\left( {\frac{{N\left( {C - R} \right) + \frac{1}{2}{{\log }_2}N}}{{\sqrt {NV} }}} \right).
\label{rate_FD_OFDM}
\end{aligned}
%\vspace{-0.5em}
\end{equation}
%\begin{equation}
%\begin{aligned}
%R  \approx  I_{\text{diff}}- {{\sqrt {\frac{ {V_{\text{diff}}}   }{N_{{\text{diff}}}}}   }}Q^{-1}\left( \varepsilon \right)+\frac{{{{\log }_2}{N_{{\text{diff}}}}}}{{2{N_{{\text{diff}}}}}}.
%%\leqslant
%%\varepsilon  \approx Q\left( {\frac{{N\left( {C - R} \right) + \frac{1}{2}{{\log }_2}N}}{{\sqrt {NV} }}} \right).
%\label{rate_TD_OFDM}
%\end{aligned}
%%\vspace{-0.5em}
%\end{equation}
The minimum achievable BLER of differentially modulated OFDM for given rate $R$ can be approximately expressed as
%\vspace{-0.5em}
\begin{equation}
\begin{aligned}
%\varepsilon  \approx  Q\left( {\sqrt{{N}/{V_{\text{coh}} \left( \rho \right)}}\left( {I_{\text{coh}} \left( \rho \right) - R}\right)} \right).
 \varepsilon  \approx  Q\left( {\sqrt{\frac{N_{\text{dif}}}{ V_{\text{dif}} \left( \gamma  \right)}}\left( { I_{\text{dif}}\left( \gamma  \right) - R}+\frac{{{{\log }_2}N_{\text{dif}}}}{{2N_{\text{dif}}}}\right)} \right).
%\leqslant
%\varepsilon  \approx Q\left( {\frac{{N\left( {C - R} \right) + \frac{1}{2}{{\log }_2}N}}{{\sqrt {NV} }}} \right).
\label{bler_FD_OFDM}
\end{aligned}
%\vspace{-0.5em}
\end{equation}
\end{theorem}
%}

\emph{Proof:}
See Appendix A.
$\hfill\blacksquare$

We note that the results in Theorem 3 are similar to the normal approximation given by (\ref{rate_fbl}) and (\ref{bler_fbl}) for the optimal continuous input AWGN channel, with capacity and dispersion replaced by $I_{\text{dif}}\left( \gamma  \right)$ and $V_{\text{dif}}\left( \gamma  \right)$.
Therefore, $I_{\text{dif}}\left( \gamma  \right)$ and $V_{\text{dif}}\left( \gamma  \right)$ can be respectively regarded as the channel capacity and dispersion for differentially modulated OFDM.

\begin{figure*}[!ht]
\setlength{\abovecaptionskip}{5pt}%2
\setlength{\belowcaptionskip}{-5pt}%-5
\centering
\subfloat{
\includegraphics[width=0.32\textwidth]{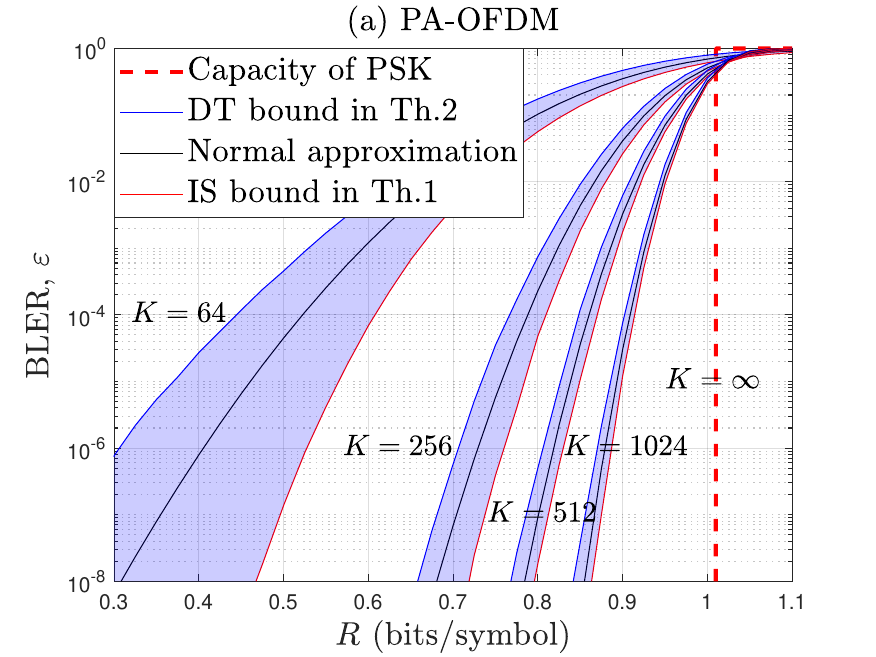}
}%\vspace{-2ex}
\subfloat{
\includegraphics[width=0.32\textwidth]{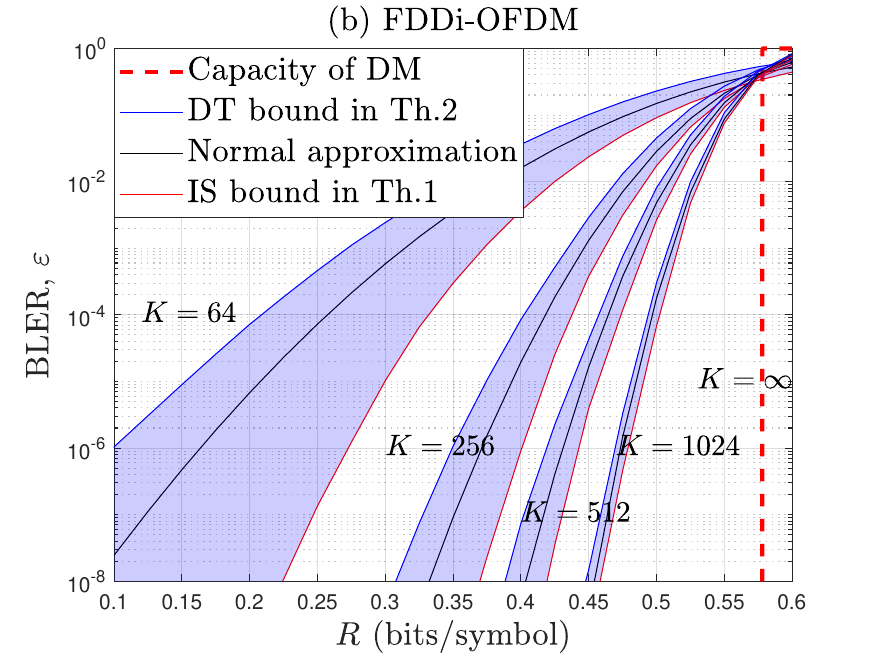}
}
\subfloat{
\includegraphics[width=0.32\textwidth]{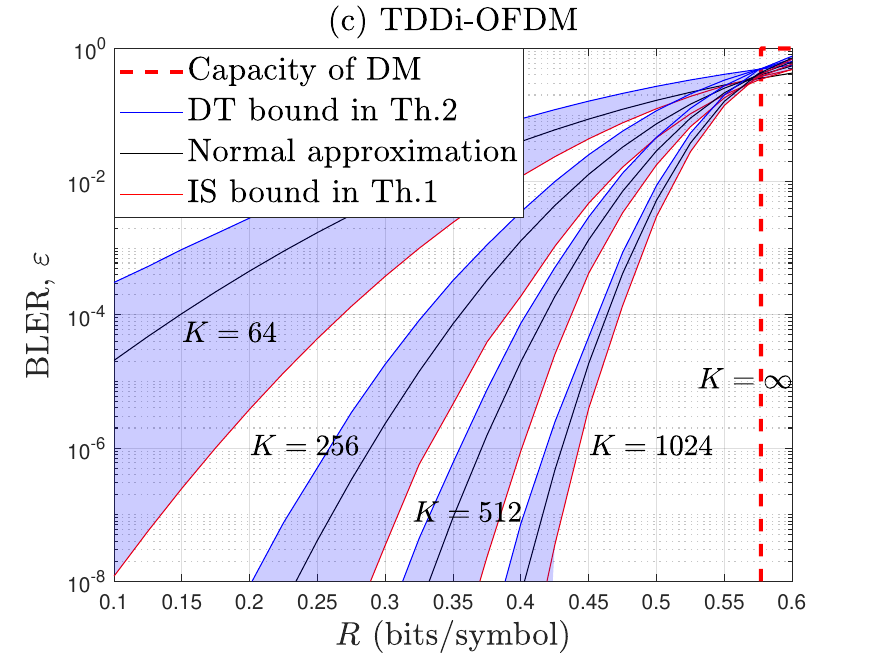}
}
\caption{The BLER performance of PA-OFDM, FDDi-OFDM and TDDi-OFDM for different $K$ under  a 2 OFDM symbol unit with modulation order $M = 4$, $f_dT_s =0.01$, $L=5$ and SNR $\gamma=2$dB.}
%\label{pic_k}
\end{figure*}
\setlength{\textfloatsep}{26pt}

\begin{figure*}[!ht]
\setlength{\abovecaptionskip}{5pt}%2
\setlength{\belowcaptionskip}{-5pt}%-5
\centering
\subfloat{
\includegraphics[width=0.5\textwidth]{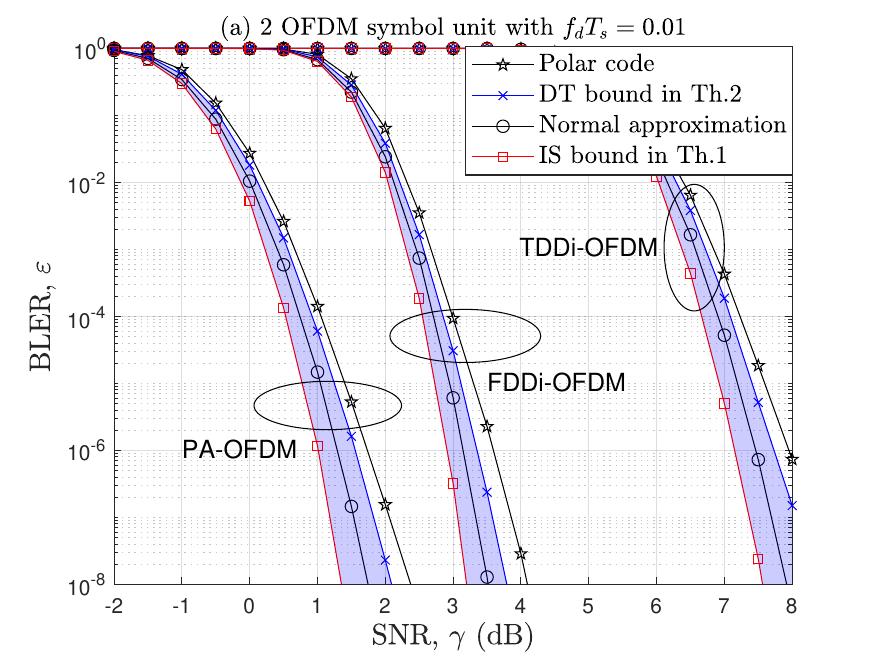}
}%\vspace{-2ex}
\subfloat{
\includegraphics[width=0.5\textwidth]{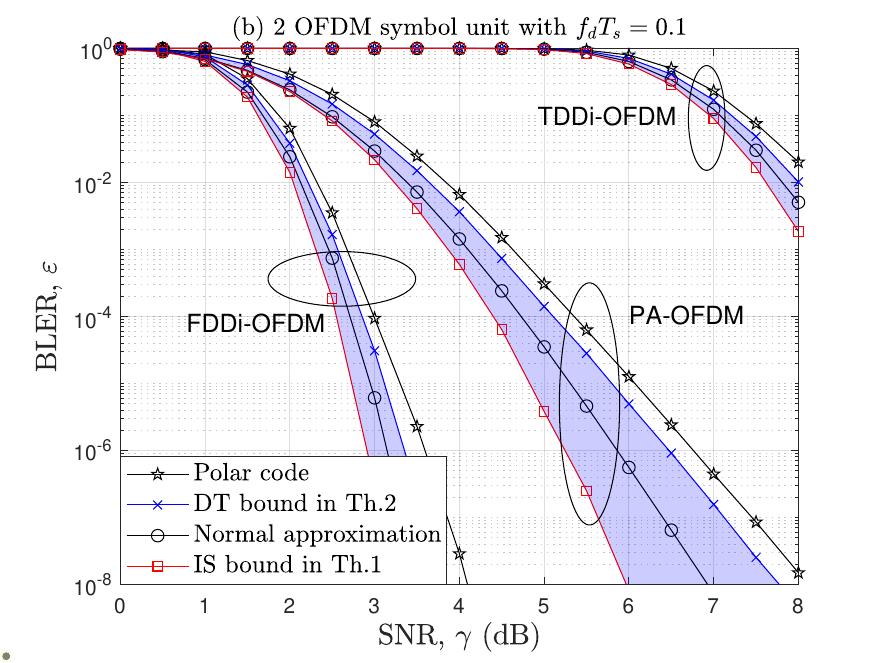}
}
\caption{BLER performance comparisons of PA-OFDM, FDDi-OFDM and TDDi-OFDM under a 2 OFDM symbol unit: (a) $f_d T_s=0.01$, and (b)  $f_d T_s=0.1$.}
%\vspace{-1.25em}
%\label{pic_k}
\end{figure*}
\setlength{\textfloatsep}{26pt}

%Here, $I_{\text{dif}}$ and $V_{\text{dif}}$ are the expectation and variance of the information density for differential OFDM with a single channel use.
%For frequency differential OFDM, $I_{\text{dif}}$ and $V_{\text{dif}}$
\subsection{Pilot Symbol-assisted Short Packet OFDM}
In this subsection, the performance of pilot symbol-based short packet OFDM is analysed.
%we derive closed form expressions for the performance of pilot symbol-based short packet OFDM with imperfect channel estimation.
Specifically, the average channel estimation MSE is first evaluated for the pilot patterns and estimation schemes considered in Section II-B, which helps in quantifying the SNR penalty due to imperfect channel estimation in PA-OFDM.
%To quantify the impact of channel estimation errors on the system performance, the channel estimation MSE is first evaluated for the pilot patterns and estimation schemes considered in Section II-B, which helps in characterizing the SINR penalty in pilot symbol-based short packet OFDM.
%helps in characterizing the SINR penalty due to imperfect channel estimation in pilot symbol-based short packet OFDM.
%, which is a factor that contributes significantly to the system performance.
Then, based on the average channel estimation MSE and the effective SNR, we can derive closed-form expressions (i.e., normal approximation) for the performance of pilot symbol-based short packet OFDM with imperfect channel estimation.

%Let ${\widehat H_{t,k}}={H_{t,k}}-{\widetilde H_{t,k}}$ denote the channel estimate of ${H_{t,k}}$, where ${\widetilde H_{t,k}}$ is the channel estimation error.
%According to the channel estimation scheme adopted in Section II-B, ${\widetilde H_{t,k}}$ is affected by three error types: the channel estimation error at the pilot subcarrier, the interpolation error and the error incurred by the time-varying nature of the channel.
%To characterize the performance degradation due to these three error types, we use the average channel estimation MSE to evaluate the variance of ${\widetilde H_{t,k}}$, as in \cite{Rao2018Adaptive} and \cite{Simko2013Adaptive}.
%In particular, we denote the variance of ${\widetilde H_{t,k}}$ as ${\widetilde H_{t,k}}$.

Let ${\widehat H_{k,t}}={H_{k,t}}-{ e_{k,t}}$ denote the channel estimate of ${H_{k,t}}$, where ${e_{k,t}}$ is the corresponding channel estimation error.
%and ${\widetilde H_{t,k}}={H_{t,k}}-{\widehat H_{t,k}}$ denote the corresponding channel estimation error.
%Let ${\widehat H_{t,k}}$ and ${\widetilde H_{t,k}}$ denote the corresponding channel estimate and error of ${H_{t,k}}$, respectively.
%capacity of the OFDM system.
Then, for PA-OFDM, the expression in (\ref{rece_signal_DFT_without_ISI}) can be rewritten as
\begin{equation}
\begin{aligned}
{z_{k,t}}= {\widehat H_{k,t}}{d_{k,t}} +{e_{k,t}} {d_{k,t}} + {W_{k,t}}.
\label{rece_signal_PAOFDM}
\end{aligned}
\end{equation}
%In addition, we use $\sigma _{\widehat H_{t,k}}^2$ and $\sigma _{ \widetilde H_{t,k}}^2$ to represent the variance of ${\widehat H_{t,k}}$ and ${\widetilde H_{t,k}}$, respectively.
For simplicity, we use $\sigma _{ e}^2$ to represent the variance of ${e_{k,t}}$ and assume that ${e_{k,t}}$ and ${ H_{k,t}}$ are statistically independent.
From (\ref{rece_signal_PAOFDM}), the effective SNR for PA-OFDM becomes
\begin{equation}
\begin{aligned}
\widehat \gamma
%= \frac{\sigma _{\widehat H_{t,k}}^2}{{\sigma _{ \widetilde H_{t,k}}^2+\sigma _{{\rm{ICI}}}^2 + \sigma _w^2}}
= \frac{1-\sigma _{ e}^2}{{\sigma _{ e}^2 + \sigma _w^2}}.
%{{\sigma _H^2}}/\left( {{\sigma _{{\rm{ICI}}}^2 + \sigma _w^2}}\right).
\label{SINR_PAOFDM}
\end{aligned}
\end{equation}
%According to the channel estimation scheme adopted in Section II-B, $\sigma _{ \widetilde H_{t,k}}^2$ is mainly determined by three error types: the channel estimation error at the pilot subcarrier, the interpolation error and the error incurred by the time-varying nature of the channel.
According to the channel estimation scheme in Section II-B and the pilot pattern in Fig.1, $\sigma _{ e}^2$ is mainly determined by the channel estimation error on pilot subcarriers, data subcarriers using the linear interpolation, data subcarriers at the edge and data symbols at the regions A and B.
%three error types: the channel estimation error at the pilot subcarrier, the interpolation error and the error incurred by the time-varying nature of the channel.
%should include three distinct error types
Thus, to characterize the SNR penalty due to the channel estimation error at these resource elements, we use the average channel estimation MSE to evaluate $\sigma _{ e}^2$, as in \cite{Rao2018Adaptive} and \cite{Simko2013Adaptive}.
%To evaluate $\sigma _{ \widetilde H_{t,k}}^2$, we use the average channel estimation MSE
%Similarly to \cite{Rao2018Adaptive} and \cite{Simko2013Adaptive}, we use the average channel estimation MSE to evaluate $\sigma _{ \widetilde H_{t,k}}^2$.
%calculate $\sigma _{ \widetilde H_{t,k}}^2$ by the average channel estimation MSE.
%According to the channel estimation scheme adopted in Section II-B,
In particular, the average channel estimation MSE can be expressed as a weighted mean of the MSE of the different resource element types.
We represent the interval between two adjacent pilot-carrying OFDM symbols in ${\mathcal{T}}$ as ${\delta _{sym}}$.
Especially, if there is only one pilot-carrying OFDM symbol in mini-slot (i.e., $\left|{\mathcal{T}}\right|=1$), the symbol interval is ${\delta _{sym}}=T$.
Then, for the resource elements at the OFDM symbol interval ${\delta _{sym}}$, the average channel estimation MSE is given by (\ref{var_chan_error}),
%\begin{equation}
%\begin{aligned}
%&\sigma _{ \widetilde H_{t,k}}^2= \frac{{{\lambda _p}}}{{T{\delta _{sym}}}}{\Phi _{{\rm{LMMSE}}}} + \left( {\frac{{T - {\lambda _p}}}{{T{\delta _{sym}}}} - \frac{{{\delta _{sub}} - 1}}{{T{\delta _{sym}}}}} \right){\Phi _{{\rm{linear}}}} + \\
%&\frac{{{\delta _{sub}} - 1}}{{T{\delta _{sym}}}}{\Phi _{{\rm{edge}}}} + \frac{{{\lambda _p}\left( {{\delta _{sym}} - 1} \right)}}{{T{\delta _{sym}}}}{\Phi _{{\rm{A}}}} + \frac{{\left( {T - {\lambda _p}} \right)\left( {{\delta _{sym}} - 1} \right)}}{{T{\delta _{sym}}}}{\Phi _{{\rm{B}}}}
%%\sum\limits_{\delta  = 1}^{{\delta _{sym}} - 1} {\frac{1}{{{\delta _{sym}}}}{\Phi _{{\rm{time}}}}\left( \delta  \right)}
%%= \frac{{{\lambda _p}}}{{T{\delta _{sym}}}}{\Phi _{{\rm{LMMSE}}}} + \left( {\frac{{T - {\lambda _p}}}{{T{\delta _{sym}}}} - \frac{{{\delta _{sub}} - 1}}{{T{\delta _{sym}}}}} \right){\Phi _{{\rm{linear}}}} + \frac{{{\delta _{sub}} - 1}}{{T{\delta _{sym}}}}{\Phi _{{\rm{edge}}}} + \sum\limits_{\delta  = 1}^{{\delta _{sym}} - 1} {\frac{1}{{{\delta _{sym}}}}{\Phi _{{\rm{time}}}}\left( \delta  \right)}
%\label{var_chan_error}
%\end{aligned}
%\end{equation}
where ${\Phi _{{\rm{LMMSE}}}}$, ${\Phi _{{\rm{linear}}}}$, ${\Phi _{{\rm{edge}}}}$ ${{\Phi _{{\rm{A}}}}}$ and ${{\Phi _{{\rm{B}}}}}$ are respectively the channel estimation MSE for pilot subcarriers, data subcarriers using the linear interpolation, data subcarriers at the edge and data symbols at regions A and B.
%that are $\delta$ OFDM symbols away from the pilot-carrying OFDM symbol.
Since the number of the data subcarriers at the edge is very small in mini-slot-assisted OFDM, we ignore the difference between ${\Phi _{{\rm{linear}}}}$ and ${\Phi _{{\rm{edge}}}}$ for simplicity.
Therefore, (\ref{var_chan_error}) can be simplified as
\setcounter{equation}{28}
\begin{equation}
\begin{aligned}
&\sigma _{ e}^2= \frac{{{\lambda _p{\Phi _{{\rm{LMMSE}}}}}}}{{K{\delta _{sym}}}} +  {\frac{\left({K - {\lambda _p}}\right){\Phi _{{\rm{linear}}}}}{{K{\delta _{sym}}}} } + \frac{{{\lambda _p}\left( {{\delta _{sym}} - 1} \right)}}{{K{\delta _{sym}}}}{\Phi _{{\rm{A}}}} \\
&+ \frac{{\left( {K- {\lambda _p}} \right)\left( {{\delta _{sym}} - 1} \right)}}{{K{\delta _{sym}}}}{\Phi _{{\rm{B}}}}.
%= \frac{{{\lambda _p}}}{{T{\delta _{sym}}}}{\Phi _{{\rm{LMMSE}}}} + \left( {\frac{{T - {\lambda _p}}}{{T{\delta _{sym}}}} - \frac{{{\delta _{sub}} - 1}}{{T{\delta _{sym}}}}} \right){\Phi _{{\rm{linear}}}} + \frac{{{\delta _{sub}} - 1}}{{T{\delta _{sym}}}}{\Phi _{{\rm{edge}}}} + \sum\limits_{\delta  = 1}^{{\delta _{sym}} - 1} {\frac{1}{{{\delta _{sym}}}}{\Phi _{{\rm{time}}}}\left( \delta  \right)}
\label{var_chan_error_simp}
\end{aligned}
\end{equation}

For the LMMSE channel estimator in (\ref{LMMSE_esti}), the MSE ${\Phi _{{\rm{LMMSE}}}}$ is given by\cite{Edfors1998OFDM}
\setcounter{equation}{28}
\begin{equation}
\begin{aligned}
&{\Phi _{{\rm{LMMSE}}}}=\frac{1}{{\lambda_p}}\sum\limits_{\lambda = 0}^{\lambda_p - 1}  {{\mathbb{E}}\left\{ {{{\left\| {{{\widehat H}_{\lambda {\delta _{sub}} ,{t_p}}} - {H_{\lambda {\delta _{sub}} ,{t_p}}}} \right\|}^2}} \right\}}\\
&=\frac{1}{{{\lambda _p}}}{\rm{tr}}\left( {{{\bf{R}}_{{{\bf{H}}_{{t_p}}}}}\left( {{\bf{I}} - {{\left( {{{\bf{R}}_{{{\bf{H}}_{{t_p}}}}} + \frac{{\bf{I}}}{\gamma }} \right)}^{ - 1}}{{\bf{R}}_{{{\bf{H}}_{{t_p}}}}}} \right)} \right)\\
&=\frac{1}{{{\lambda _p} }}\sum\limits_{\lambda=0} ^{{\lambda _p-1}} {\frac{{{\psi _\lambda }}}{{{\gamma \psi _\lambda } + 1}}} ,
\label{MSE_LMMSE}
\end{aligned}
\end{equation}
where $\left\{ {{\psi  _\lambda }} \right\}_{\lambda  = 0}^{{\lambda _p} - 1}$ are the eigenvalues of ${{\bf{R}}_{{{\bf{H}}_{{t_p}}}}}$.
In addition, the MSE ${\Phi _{{\rm{linear}}}}$ can be derived as in (\ref{MSE_linear}), where $(a)$ follows from (\ref{linear_esti}) and the fact that
${\mathbb{E}}\left\{ {{{\left\| {a{H_{\lambda {\delta _{sub}},{t_p}}} + b{H_{\left( {\lambda  + 1} \right){\delta _{sub}},{t_p}}}} \right\|}^2}} \right\} = {\mathbb{E}}\left\{ {{{\left\| {a{H_{\lambda {\delta _{sub}},{t_p}}}} \right\|}^2}} \right\} + 2\Re\left\{ {\mathbb{E}}\left\{ {abH_{\lambda {\delta _{sub}},{t_p}}^ * {H_{\left( {\lambda  + 1} \right){\delta _{sub}},{t_p}}}} \right\}\right\}+ {\mathbb{E}}\left\{ {{{\left\| {b{H_{\left( {\lambda  + 1} \right){\delta _{sub}},{t_p}}}} \right\|}^2}} \right\}$ for any constants $a$ and $b$,
%and the multiplicative distortion ${H_{t,k}}$ and the estimation error ${\widetilde H_{t,k}}$ are assumed to be statistically independent
while $(b)$ is due to the result of (\ref{corr_fre_time}) and ${\mathcal L}=\frac{{5{\delta _{sub}} - 1}}{{3{\delta _{sub}}}} + \frac{{{\delta _{sub}} + 1}}{{3{\delta _{sub}}}}\Re \left( {{\rho }\left( {{\delta _{sub}},0} \right)} \right)+ \frac{{2{\delta _{sub}} - 1}}{{3{\delta _{sub}}}}{\Phi _{{\rm{LMMSE}}}} $.
%%It is worth noting that if $\sigma _H^2=1$ and the LS channel estimator is used, the result of (\ref{MSE_linear}) can be simplified to \cite[Eq.(7)]{Kim2005Performance} or \cite[Eq.(14)]{Rao2018Adaptive}.
%of the linear interpolation scheme in
%\begin{equation}
%\begin{aligned}
%%{\Phi _{{\rm{linear}}}} = \frac{1}{{{\delta _{sub}} - 1}}\sum\limits_{{t_d} = 1}^{{\delta _{sub}} - 1} {{\mathbb{E}}\left\{ {{{\widehat H}_{\lambda {\delta _{sub}} + {t_d},{k_p}}} - {H_{\lambda {\delta _{sub}} + {t_d},{k_p}}}} \right\}}
%&{\Phi _{{\rm{linear}}}} = \frac{1}{{{\delta _{sub}} - 1}}\sum\limits_{{t_d} = 1}^{{\delta _{sub}} - 1} {{\mathbb{E}}\left\{ {{{\left\| {{{\widehat H}_{\lambda {\delta _{sub}} + {t_d},{k_p}}} - {H_{\lambda {\delta _{sub}} + {t_d},{k_p}}}} \right\|}^2}} \right\}}\\
%&= \frac{1}{{{\delta _{sub}} - 1}}\sum\limits_{{t_d} = 1}^{{\delta _{sub}} - 1} {{\mathbb{E}}\left\{ {{{\left\| { \frac{{\delta_{sub}  - {t_d} }}{\delta_{sub} }{\widehat H_{\lambda \delta_{sub} ,{k_p}}} + \frac{{t_d} }{\delta_{sub} }{\widehat H_{\left( {\lambda  + 1} \right)\delta_{sub} ,{k_p}}} - {H_{\lambda {\delta _{sub}} + {t_d},{k_p}}}} \right\|}^2}} \right\}}
%\label{MSE_linear}
%\end{aligned}
%\end{equation}
Similarly to the derivation in (\ref{MSE_linear}), ${{\Phi _{{\rm{A}}}}}$ and ${{\Phi _{{\rm{B}}}}}$ can be respectively expressed as
\setcounter{equation}{30}
\begin{equation}
\begin{aligned}
&{{\Phi _{{\rm{A}}}}}=\frac{1}{{\lambda_p}\left( {{\delta _{sym}} - 1} \right)}\\
&\times \sum\limits_{{\delta} = 1}^{{\delta _{sym}} - 1} \sum\limits_{\lambda = 0}^{\lambda_p - 1}  {{\mathbb{E}}\left\{ {{{\left\| {{{\widehat H}_{\lambda {\delta _{sub}} ,{t_p}}} - {H_{\lambda {\delta _{sub}},{t_p}+ {\delta}}}} \right\|}^2}} \right\}}\\
&=2 +{\Phi _{{\rm{LMMSE}}}}-\frac{2}{ {{\delta _{sym}} - 1} }\sum\limits_{{\delta} = 1}^{{\delta _{sym}} - 1}{\rho }\left( {0,\delta} \right),
\label{MSE_A}
\end{aligned}
\end{equation}
and
\begin{equation}
\begin{aligned}
&{{\Phi _{{\rm{B}}}}}=\frac{1}{\left( {{\lambda_p} - 1} \right)\left( {{\delta _{sym}} - 1} \right)\left( {{\delta _{sub}} - 1} \right)}\times\\
& \sum\limits_{{\delta} = 1}^{{\delta _{sym}} - 1} \sum\limits_{{k_d} = 1}^{{\delta _{sub}} - 1} \sum\limits_{\lambda = 0}^{\lambda_p - 1}  {{\mathbb{E}}\left\{ {{{\left\| {{{\widehat H}_{\lambda {\delta _{sub}} + {k_d} ,{t_p}}} - {H_{\lambda {\delta _{sub}} + {k_d},{t_p}+ {\delta}}}} \right\|}^2}} \right\}}\\
&=\mathcal{L} - \frac{2}{{\left( {{\delta _{sym}} - 1} \right)\left( {{\delta _{sub}} - 1} \right)}}\sum\limits_{\delta  = 1}^{{\delta _{sym}} - 1} {\sum\limits_{{k_d} = 1}^{{\delta _{sub}} - 1} {\left[ {\frac{{{\delta _{sub}} - {k_d}}}{{{\delta _{sub}}}} } \right.} }\\
&\times \left. {\Re \left( {{\rho}\left( {{k_d},\delta } \right)} \right) + \frac{{{k_d}}}{{{\delta _{sub}}}}\Re \left( {{\rho }\left( {{k_d} - {\delta _{sub}},\delta } \right)} \right)} \right].
\label{MSE_B}
\end{aligned}
\end{equation}
Finally, combining (\ref{SINR_PAOFDM}) and (\ref{var_chan_error_simp})-(\ref{MSE_B}) leads to the effective SNR $\widehat \gamma$ for PA-OFDM.

Note that the single carrier SPT performance of ergodic fading channels with independent and identically distributed (i.i.d) PSK/QAM channel inputs and perfect CSI has been analysed in \cite{Zheng2023Differential}.
Similarly, the performance results derived in \cite{Zheng2023Differential} for coherent fading channels with i.i.d PSK/QAM channel inputs are basically analogous to the normal approximation (\ref{rate_fbl}) and (\ref{bler_fbl}), with ${I \left( \gamma \right)}$ and ${V \left( \gamma \right)}$ replaced by the channel capacity and dispersion of i.i.d PSK/QAM inputs.
%${I_{\text{coh}} \left( \rho \right)}$ and dispersion ${V_{\text{coh}} \left( \rho \right)}$.
To be specific, the coherent channel capacity with i.i.d PSK/QAM inputs is given by\cite[Eq.(15)]{Zheng2023Differential}
\begin{equation}
\begin{aligned}
&I_{\text{coh}}\left(  \gamma \right) ={\log _2}M -\\
%={\log _2}M - \frac{1}{M}\sum\nolimits_{j = 1}^M {\mathbb{E}\left[ {{\log _2}\frac{{\sum\nolimits_{i = 1}^M {{{\mathbb{P}}_{Y\left| X \right.}}\left( {y\left| x_i \right.} \right)} }}{{{{\mathbb{P}}_{Y\left| X \right.}}\left( {y\left| x_j \right.} \right)}}} \right]}
&\frac{1}{M}\sum\limits_{j = 1}^M {\mathbb{E}}\left[ {\log _2}\sum\limits_{i = 1}^M {\exp \left( {{w^2} - {{\left( {w + \sqrt  {\gamma}  h\left( {{x_j} - {x_i}} \right)} \right)}^2}} \right)}  \right]
\label{capa_coh}
\end{aligned}
\end{equation}
and the coherent channel dispersion with i.i.d PSK/QAM inputs can be written as\cite[Eq.(16)]{Zheng2023Differential}
\begin{equation}
\begin{aligned}
&V_{\text{coh}}\left(  \gamma \right) =- {\left( { {{\log }_2}M- I_{\text{coh}}\left(  \gamma \right) } \right)^2}+\frac{1}{M}\times\\
%={\log _2}M - \frac{1}{M}\sum\nolimits_{j = 1}^M {\mathbb{E}\left[ {{\log _2}\frac{{\sum\nolimits_{i = 1}^M {{{\mathbb{P}}_{Y\left| X \right.}}\left( {y\left| x_i \right.} \right)} }}{{{{\mathbb{P}}_{Y\left| X \right.}}\left( {y\left| x_j \right.} \right)}}} \right]}
&\sum\limits_{j = 1}^M {\mathbb{E}}\left[ \left\{ { {\log _2}\sum\limits_{i = 1}^M {\exp \left( {{w^2} - {{\left( {w + \sqrt {\gamma}  h\left( {{x_j} - {x_i}} \right)} \right)}^2}} \right)} } \right\}^2  \right]
\label{disp_coh}
\end{aligned}
\end{equation}
where ${w} \sim \mathcal{CN}\left( {0,1} \right)$ is the unit-power noise component, the variables $x_i$ and $x_j \in \mathcal{X}^\text{PSK}$ for PSK inputs and the variables $x_i$ and $x_j \in \mathcal{X}^\text{QAM}$ for QAM inputs. Here, $\mathcal{X}^\text{QAM} =\sqrt {\frac{1}{{{\xi  }}}}  \times \left\{ { \pm \left( {2m - 1} \right) \pm \left( {2m - 1} \right)j} \right\}$ is the input alphabet of an $M$-ary QAM constellation, where $m \in \left\{ {1,...,\frac{{\sqrt M }}{2}} \right\}$ and ${\xi } = \frac{{2\left( {M - 1} \right)}}{3}$ is a constant to normalize the average transmit power to unity.
Then, based on the results in \cite{Zheng2023Differential} and the effective SNR for PA-OFDM, we have the following corollary:
\newtheorem{corollary}{Corollary}
\begin{corollary}
\newtheorem*{Proof4}{proof}
(Pilot symbol-assisted short packet OFDM):
For the pilot patterns and the channel estimation schemes considered in Section II-B, the maximum achievable rate with PSK/QAM inputs can be written as
\begin{equation}
\begin{aligned}
&{ R} \approx  I_{\text{coh}}\left( \widehat \gamma \right)- {{\sqrt {\frac{ {V_{\text{coh}}\left( \widehat \gamma \right)}  }{N_{\text{coh}}}}   }}Q^{-1}\left( \varepsilon \right)+\frac{{{{\log }_2}N_{\text{coh}}}}{{2N_{\text{coh}}}},
%\leqslant
%\varepsilon  \approx Q\left( {\frac{{N\left( {C - R} \right) + \frac{1}{2}{{\log }_2}N}}{{\sqrt {NV} }}} \right).
\label{rate_fbl_pilot}
\end{aligned}
\end{equation}
where $N_{\text{coh}}=KT-{ \lambda_{\rm{total}}}$ is the number of available data symbols in PA-OFDM.
The minimum achievable BLER can be approximately determined as
\begin{equation}
\begin{aligned}
\varepsilon  \approx  Q\left( {\sqrt{\frac{N_{\text{coh}}}{ V_{\text{coh}} \left( \widehat \gamma \right)}}\left( { I_{\text{coh}} \left( \widehat \gamma \right) - R}+\frac{{{{\log }_2}N_{\text{coh}}}}{{2N_{\text{coh}}}}\right)} \right).
%\leqslant
%\varepsilon  \approx Q\left( {\frac{{N\left( {C - R} \right) + \frac{1}{2}{{\log }_2}N}}{{\sqrt {NV} }}} \right).
\label{bler_fbl_pilot}
\end{aligned}
\end{equation}
\end{corollary}

\begin{figure*}[!ht]
\setlength{\abovecaptionskip}{5pt}%2
\setlength{\belowcaptionskip}{-5pt}%-5
\centering
%\vspace{-0.15in}
\begin{minipage}{1\linewidth}	% linewidth就是栏宽
\centering
\subfloat{
\includegraphics[width=0.45\textwidth]{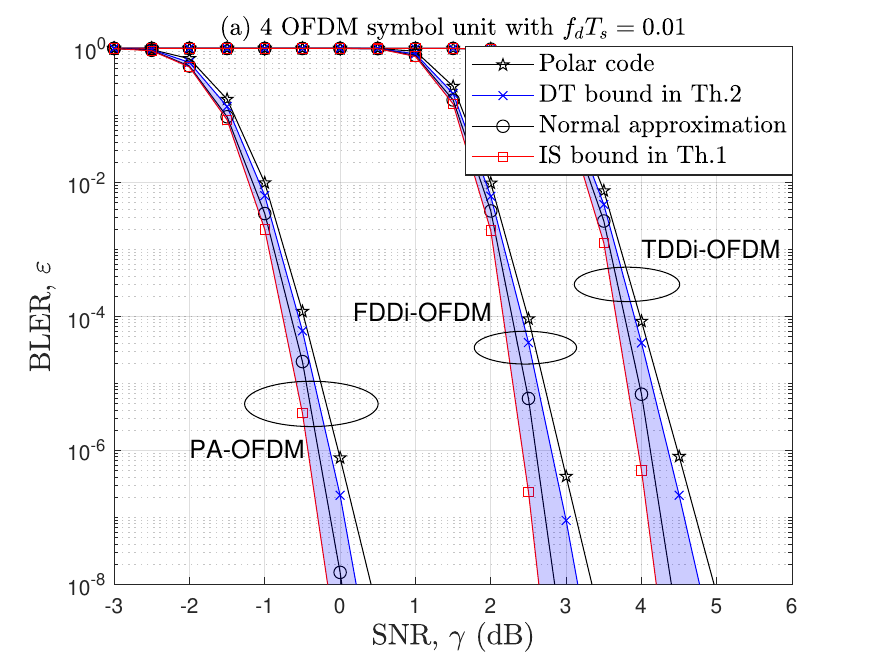}
}
\subfloat{
\includegraphics[width=0.45\textwidth]{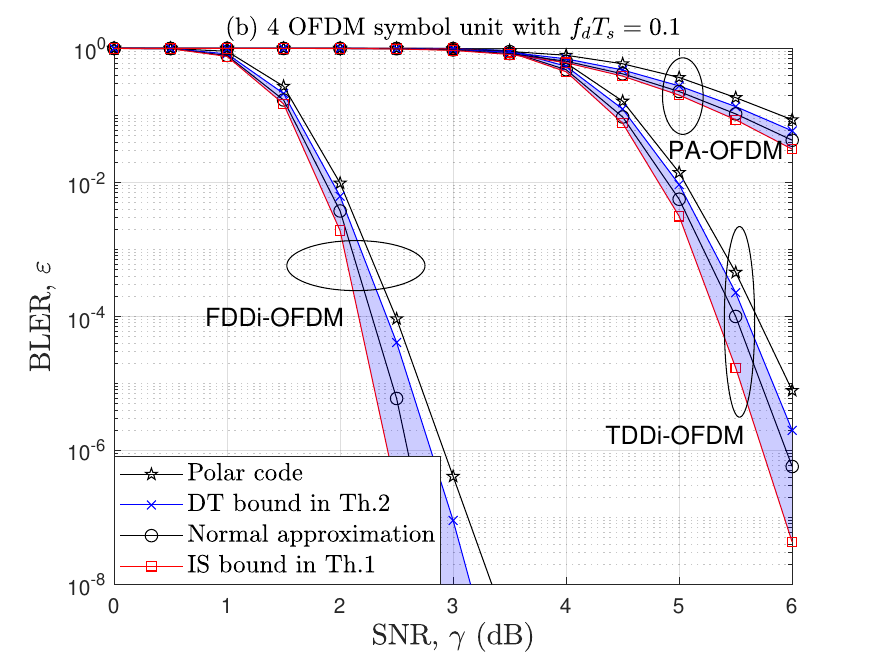}
}
\end{minipage}
\vskip 0cm % 用于调整两个minipage之间的垂直间距
\begin{minipage}{1\linewidth }
\centering
\subfloat{
\includegraphics[width=0.45\textwidth]{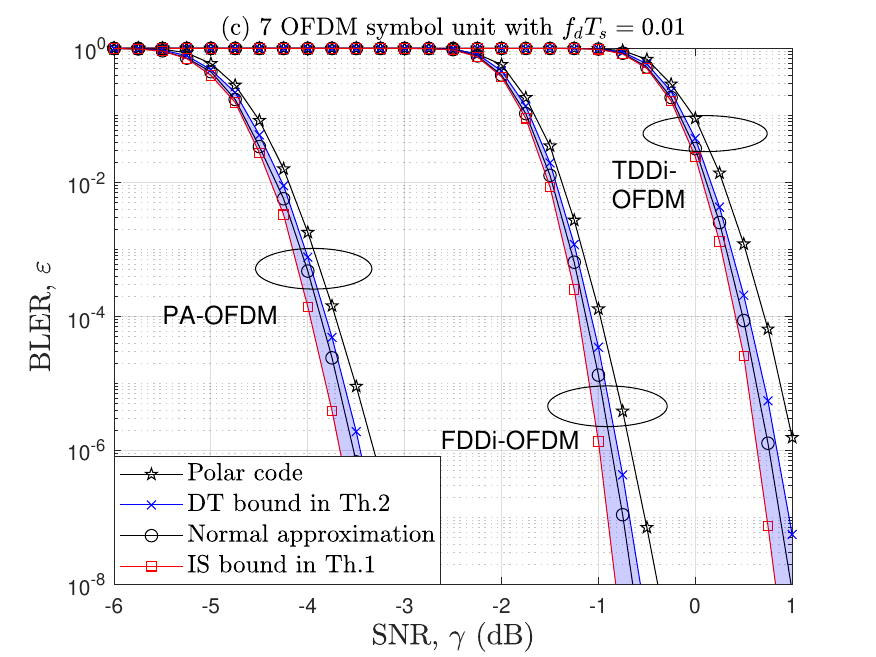}
}
\subfloat{
\includegraphics[width=0.45\textwidth]{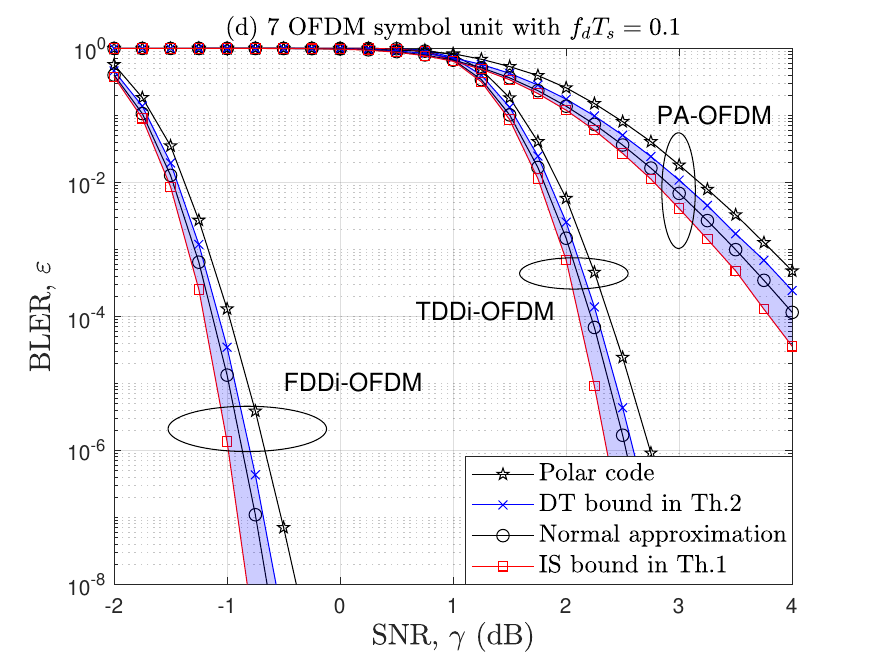}
}
\end{minipage}
%\vspace{0pt}	% 调整大标题和图片之间的距离，单位有cm in pt
\caption{BLER performance comparisons of PA-OFDM, FDDi-OFDM and TDDi-OFDM under a 4 or 7 OFDM symbol unit.}
%\vspace{0cm}		% 调整正文部分和标题（图片之间的距离）
%\vspace{-1.25em}
%\label{fig:1234}
%\vspace{-1.5em}
\end{figure*}

\section{Simulation Results}
In this section, we compare the analytical BLER performance (i.e., normal approximation) of PA-OFDM, FDDi-OFDM and TDDi-OFDM.
Furthermore, we employ Monte Carlo simulations to obtain the simulated BLER results and verify the correctness of our analysis.
To be specific, for each SNR value, we utilize $10^{9}$ packet samples to compute the IS bound in Theorem 1 and the DT bound in Theorem 2.
On the other hand, since the non-asymptotic bounds developed in \cite{Polyanskiy2010Channel} are based on optimal encoder-decoder pairs, there may be a gap between the normal approximation results and the performance achievable by actual coding schemes.
To show how close actual encoder-decoder pairs perform to the normal approximation results derived in this paper, we employ cyclic redundancy check (CRC)-aided polar encoding and successive cancellation list (SCL) decoding schemes, as in \cite{Yuan2021Polar}.
Specifically, a 6-bit CRC code with the generator polynomial ${x^6} + {x^5} + 1$ is appended to $B$ information bits. The CRC-appended information bits are then polar-encoded and rate-matched to output $J$ information bits, forming a ($J$, $B$) polar encoder. For decoding, we use an SCL algorithm with a list size of 32 to generate the estimated information bits. In the polar coding-based Monte Carlo simulations, the simulated BLER for each SNR value is calculated after 100 packet errors have been observed.
In addition, we adopt an exponentially decaying power-delay profile for the multi-path channel.

Fig.4 illustrates the BLER performance of PA-OFDM, FDDi-OFDM and TDDi-OFDM for different numbers of sub-carriers $K$ (also means different values of the blocklength) under a 2 OFDM symbol unit.
Especially, the pilot pattern in Fig.1(a) is adopted for PA-OFDM (i.e., the set of the pilot-carrying OFDM symbols $ {\mathcal{T}} =\left\{ 1 \right\}$ and the interval between pilot sub-carriers ${\delta _{sub}}=2$).
In the figure, the SNR value is set to be $\gamma =2$dB and the dashed vertical red line denotes the corresponding channel capacity result.
As shown in Fig.4, the normal approximation, IS bound and DT bound-based BLER results  gradually approach the dashed vertical red line with the increase of $K$ (also blocklength).
Moreover, it is evident that the normal approximation results for PA-OFDM, FDDi-OFDM and TDDi-OFDM lie within a specific region (highlighted in blue) bounded by their respective IS and DT bounds. Therefore, the normal approximation results proposed in this paper provide satisfactory accuracy and low computational complexity in characterizing the minimum achievable BLER performance of PA-OFDM, FDDi-OFDM and TDDi-OFDM.

%\begin{figure*}[!ht]
%\setlength{\abovecaptionskip}{5pt}%2
%\setlength{\belowcaptionskip}{-5pt}%-5
%\centering
%\subfloat{
%\includegraphics[width=0.32\textwidth]{2OFDM_fdTs}
%}%\vspace{-2ex}
%\subfloat{
%\includegraphics[width=0.32\textwidth]{4OFDM_fdTs}
%}
%\subfloat{
%\includegraphics[width=0.32\textwidth]{7OFDM_fdTs}
%}
%\caption{BLER comparisons of PA-OFDM and FD-OFDM for different normalized Doppler values: (a) 2 OFDM symbol unit with $\gamma =3$ dB, (b) 4 OFDM symbol unit with $\gamma =2.5$ dB and (c) 7 OFDM symbol unit with $\gamma =-1$ dB.}
%%\label{pic_k}
%\end{figure*}
%\setlength{\textfloatsep}{26pt}

\begin{figure*}[!ht]
\setlength{\abovecaptionskip}{5pt}%2
\setlength{\belowcaptionskip}{-5pt}%-5
\centering
\subfloat{
\includegraphics[width=0.5\textwidth]{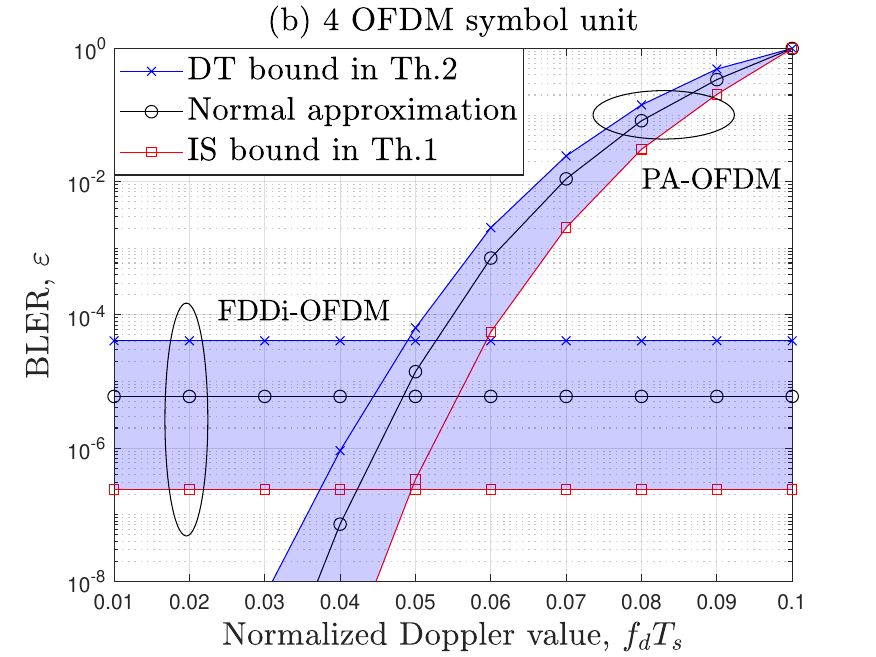}
}%\vspace{-2ex}
\subfloat{
\includegraphics[width=0.5\textwidth]{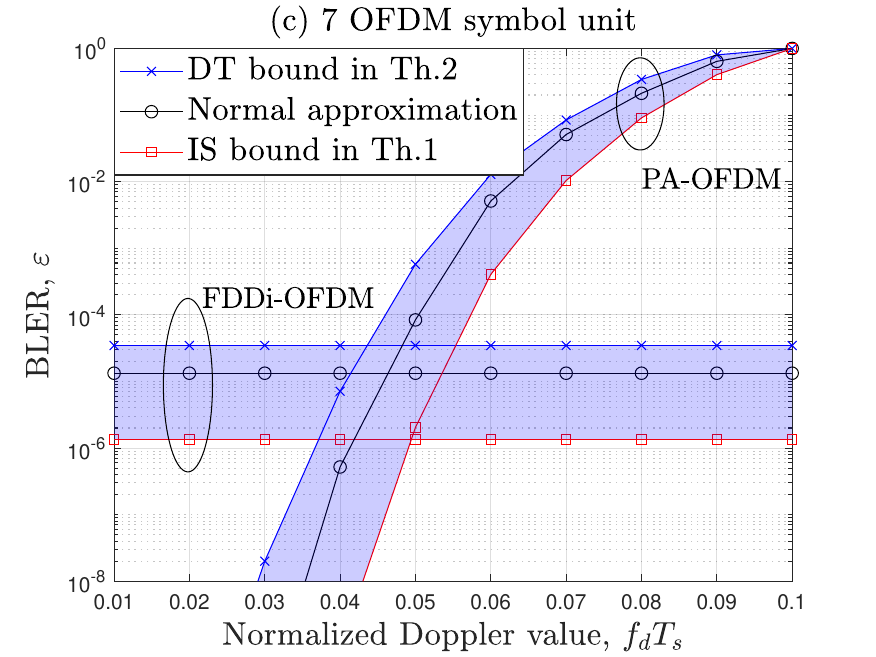}
}
\caption{BLER comparisons of PA-OFDM and FDDi-OFDM for different normalized Doppler values: (a) 4 OFDM symbol unit with $\gamma =2.5$ dB and (b) 7 OFDM symbol unit with $\gamma =-1$ dB.}
%\vspace{-1.25em}
%\label{pic_k}
\end{figure*}
\setlength{\textfloatsep}{26pt}

Figs.5-7 provide a BLER performance comparison of pilot-based, frequency domain differential and time domain differential OFDM under different normalized Doppler values, mini-slot deployment options and modulation orders.
Notably, PA-OFDM suffers from a capacity and energy loss compared to differentially modulated OFDM due to the use of pilot symbols.
%It is worth noting that since the pilot symbols are utilized, the pilot symbol-assisted OFDM suffers from a capacity and energy loss compared to differentially modulated OFDM.
Therefore, to ensure a fair comparison, it is essential to make the net data rates and power (or energy per block) of PA-OFDM, FDDi-OFDM, and TDDi-OFDM consistent.
%we need to make the net data rates and power of PA-OFDM, FD-OFDM and TD-OFDM consistent.
In general, there are two design approaches for the pilot-assisted scheme and DM to achieve the same data rate and power. One approach is to adopt different modulation orders for the pilot-assisted scheme and DM, as utilized in \cite{Zheng2023Differential}. Another easier approach is to use different coding rate for the pilot-assisted scheme and DM. For example, if BPSK and a code of rate $\frac{1}{5}$ are used in the pilot-assisted scheme to transmit 100 information bits and the number of pilots for the channel estimation is assumed to be 100 bits, the transmitting blocklength is thus 600 bits. For DM, a channel code of rate $\frac{1}{6}$ can be utilized to obtain the same transmit blocklenth, achieving the same data rate as the pilot-assisted scheme. In our simulation result, the second approaches is employed to design PA-OFDM, FDDi-OFDM and TDDi-OFDM.

In Fig.5, the BLER performance of PA-OFDM, FDDi-OFDM and TDDi-OFDM with a 2 OFDM symbol unit is  illustrated under normalized Doppler value $f_d T_s =0.01$, $0.1$. Specifically, the order of modulation is $M=4$ and
the numbers of the paths, sub-carriers and information bits are $L=5$, $K=256$ and $B=256$, respectively.
Additionally, for PA-OFDM, the set of the pilot-carrying OFDM symbols is $ {\mathcal{T}} =\left\{ 1 \right\}$ (i.e., pilots are placed at the first OFDM symbol of the mini-slot) and the interval between pilot sub-carriers is ${\delta _{sub}}=2$ (corresponds to the pilot pattern in Fig.1(a)). Consequently, the available number of data symbol is $N_{\text{coh}}=384$ for PA-OFDM, $N_{\text{dif}}=510$ for FDDi-OFDM and $N_{\text{dif}}=256$ for TDDi-OFDM. Moreover, the coding rate is $\frac{{256}}{{384 \times 2}}$ for PA-OFDM, $\frac{{256}}{{510 \times 2}}$ for FDDi-OFDM and $\frac{{256}}{{256 \times 2}}$ for TDDi-OFDM.
From Fig.4(a), PA-OFDM shows a significant advantage over frequency domain and time domain differential OFDM in terms of the normal approximation, IS bound, DT bound and polar coding-based BLER results.
Moreover, we note that PA-OFDM achieves an approximately 1-3 dB performance advantage compared with FDDi-OFDM for a low normalized Doppler value.
%However, the BLER performance of FD-OFDM outperforms PA-OFDM for a high normalized Doppler value in Fig.4(b).
However, the BLER performance of FDDi-OFDM surpasses that of PA-OFDM for a high normalized Doppler value, as shown in Fig. 5(b).
This advantage arises from two primary factors: (1) The detection of FDDi-OFDM mainly relies on the received symbols of two adjacent sub-carriers, making it robust against time-selective fading channels. (2) Unlike PA-OFDM, FDDi-OFDM does not incur the costly pilot overhead in SPT. Consequently, the saved bandwidth, power, and signal overhead can be utilized to generate a stronger channel code (i.e., a lower coding rate), improving the system reliability performance.

Fig.6 shows that the BLER performance of PA-OFDM, FDDi-OFDM and TDDi-OFDM with a 4 or 7 OFDM symbol unit.
The system parameters for Fig.5 are: $L=5$, $K=256$, ${\delta _{sub}}=2$ and the number of the information bits $B=512$.
In addition, 8-ary PSK and 16-ary QAM are adopted in PA-OFDM for 4 OFDM symbol units and 7 OFDM symbol units, respectively.
%In addition, PA-OFDM adopts 8-ary PSK for a 4 OFDM symbol unit and 16-ary QAM for a 7 OFDM symbol unit.
Similarly, the modulation order of FDDi-OFDM and TDDi-OFDM is set to be $M=8$ for a 4 OFDM symbol unit and $M=16$ for a 7 OFDM symbol unit.
%For Fig.5(a) and (b), the number of the information bits is $B=1204$, while for Fig.5(c) and (d), the number of the information bits is $B=2048$.
Especially, for 7 OFDM symbol units with $f_d T_s =0.1$, the set of the pilot-carrying OFDM symbols is $ {\mathcal{T}} =\left\{ 1, 5 \right\}$ (i.e., pilots are placed at the first and 5-th OFDM symbol of the mini-slot, which corresponds to the pilot pattern in Fig.1(d)). By contrast, the set of the pilot-carrying OFDM symbols for Fig.6(a-c) is $ {\mathcal{T}} =\left\{ 1\right\}$ (corresponds to the pilot pattern in Fig.1(b) and Fig.1(c)).
From Fig.6, the normal approximation, IS bound, DT bound and polar coding-based results show that the BLER performance of PA-OFDM outperforms that of FDDi-OFDM and TDDi-OFDM in a low Doppler environment for both 4 or 7 OFDM symbol units.
%Moreover, it can be seen that FD-OFDM offer a significant advantage over PA-OFDM for a low Doppler environment.
%In all simulations, TD-OFDM has yielded the worst results.
Instead, for a high Doppler environment (e.g., $f_d T_s =0.1$ in Fig.6(b) and Fig.6(d)), PA-OFDM has yielded the worst results.
This is because, compared to the 2 OFDM symbol unit in Fig.1(a), PA-OFDM of 4 or 7 symbol units in Fig.1(b-d) introduce more data symbol (i.e., decreases the pilot percentage) and thus far more sensitive to channel variation.

\begin{figure*}[!htp]
\centering
\includegraphics [width=0.8\textwidth]{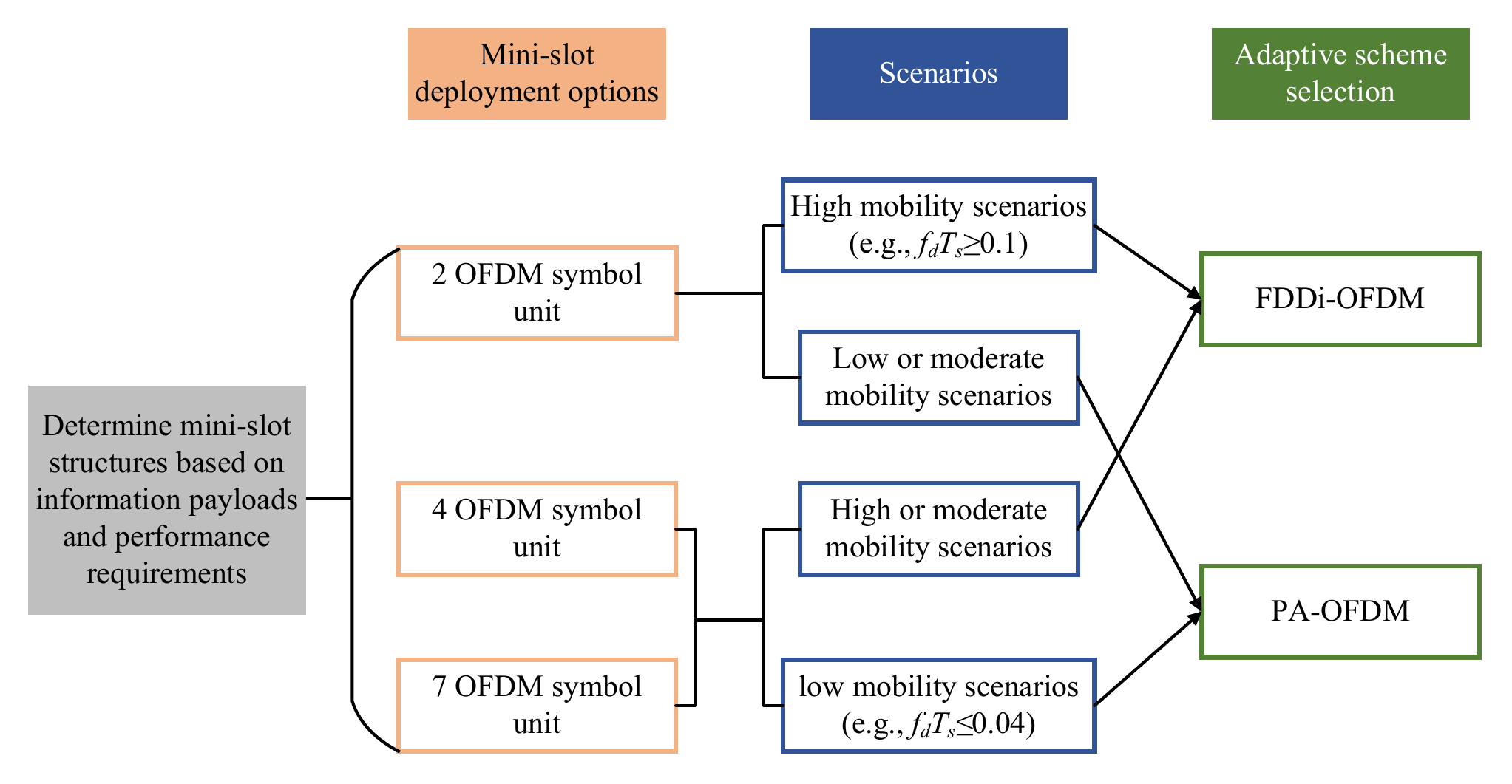}
%\vspace{-1.55em}
\caption{Illustration of adaptive scheme selection for mini-slot-assisted SPT.}
%{System model for OFDM with differential encoding in frequency domain.}%
%\label{DPSKwithMC}
\end{figure*}

%differential modulation needs no channel estimation and thus far more sensitive to channel variation.
%making the pilot patterns in Fig.()
Fig.7 illustrates a comparison of the BLER performance for PA-OFDM and FDDi-OFDM under different normalized Doppler values.
Specifically,
%we use the same simulation parameters of Fig.5
the simulation parameters of Fig.7 are set to be consistent with that of Fig.6(b) and Fig.6(d), except $f_d T_s$.
From Fig.7, the analytical and simulated BLER of FDDi-OFDM stays constant with the increase of $f_d T_s$. This is because, for FDDi-OFDM, data detection is only affected by received signals over two adjacent sub-carriers, which is independent of channel variation over two adjacent OFDM symbols.
It is also seen that the BLER of PA-OFDM is better than that of FDDi-OFDM, when $f_d T_s$ is small.
However, when $f_d T_s$ increases to a certain value, the BLER performance of PA-OFDM becomes worse than that of FDDi-OFDM.
As a result, we can adaptively activate or switch between PA-OFDM and FDDi-OFDM to achieve efficient URLLC based on information payloads, mini-slot deployment options, user mobility,  and service requirements, as illustrated in Fig.8.

\section{Conclusions}
In this paper, we have proposed an adaptive scheme that integrates both differential and coherent detection for mini-slot-assisted short packet URLLC, effectively balancing training overhead, reliability, and latency performance.
To evaluate the performance of the proposed schemes, we derived the BLER performance using non-asymptotic information-theoretic bounds.
The simulation results verified the accuracy of the analysis and showed the feasibility and effectiveness of adaptive differential and coherent detection.

The results of this paper are based on single BS association(i.e. single connection). One further issue is the impact of multi-BS association (i.e. macro diversity), which we are currently investigating.
%differential modulation combined with hierarchical codebook-based beam training scheme for URLLC with FBL systems, applicable to the scenario of mmWave massive MIMO.

\ifCLASSOPTIONcaptionsoff
  \newpage
\fi

\section*{Appendix A: proof of Theorem 3}
%For simplicity,
%the difference between
%similar steps can be used for the derivation of time domain differential OFDM
For convenience, we take FDDi-OFDM as an example and the performance analytical results of TDDi-OFDM can be obtained following similar steps.
The proof consists of the four steps below.

\textbf{Step 1. Equivalent representation of DM:}
For ease of utilizing the non-asymptotic information-theoretic bounds in Theorem 1 and 2 to evaluate the SPT performance, we first give an equivalent representation of DM.
Since DM uses the phase difference between two consecutive symbols to transmit information, we, without loss of generality, assume that the phases of the input data ${v_{k,t}}$ and symbol ${d_{k-1,t}}$ in (\ref{equ_OFDM_DPSK}) are $\Delta \varphi $ and $\varphi$, respectively. Thus,
the complex representation of the transmitted symbol ${d_{k,t}}$ in (\ref{equ_OFDM_DPSK}) is ${e^{ j\left( {\varphi  + \Delta \varphi } \right)}}$.
Based on (\ref{rece_signal_DFT_without_ISI}),
the corresponding received signal of ${d_{k-1,t}}$ and ${d_{k,t}}$ can be rewritten as
\begin{equation}
\begin{aligned}
\left\{ {\begin{array}{*{20}{c}}
%\!\!{{z_{k - 1,t}} = {e^{j\varphi }}{{U_{k - 1,t}}}}\\
%{{z_{k,t}} = {e^{j\left( {\varphi  + \Delta \varphi } \right)}}{{U_{k,t}}}}
\!\!{{z_{k - 1,t}} = {{H_{k - 1,t}}{e^{j{ \varphi } }} + { {W_{k - 1,t}}} }}\\
{{z_{k,t}} = {{H_{k,t}}}{e^{ j\left( {\varphi  + \Delta \varphi } \right)}} +{ {W_{k,t}}} }
\end{array}}. \right.
\label{Equ_repre_received}
\end{aligned}
\end{equation}
Furthermore, we decompose the complex random variables ${z_{k - 1,t}}$, ${z_{k,t}}$, ${ {W_{k - 1,t}}} $ and ${ {W_{k,t}}} $ of (\ref{Equ_repre_received}) into their real and imaginary parts:
\begin{equation}\nonumber
\begin{aligned}
\left\{ {\begin{array}{*{20}{c}}
%\!\!\!\!{{z_{k-1,t}} = {x_1} + j{y_1}}\\
%~{{z_{k,t}} = {x_2} + j{y_2}}\\
%~{{W_{k-1,t}} = {w_{r,1}} + j{w_{i,1}}}\\
%~~~~{{W_{k,t}} = {w_{r,2}} + j{w_{i,2}}}
\!\!\!\!\!\!{{z_{k-1,t}} = {x_1} + j{y_1}}\\
{{z_{k,t}} = {x_2} + j{y_2}}\\
{{W_{k-1,t}} = {w_{r,1}} + j{w_{i,1}}}\\
~~~{{W_{k,t}} = {w_{r,2}} + j{w_{i,2}}}
\end{array}}. \right.
\end{aligned}
\end{equation}

Let $r_1$, $\phi_1$ $r_2$ and $\phi_2$ be the amplitude and phase factors of ${H_{k - 1,t}}$ and ${H_{k,t}}$, respectively.
Then, we have
\begin{equation}
\begin{aligned}
\left\{ {\begin{array}{*{20}{c}}
{{x_1} = {r_1}\cos {\theta _1} + {w_{r,1}}}\\
{{y_1} = {r_1}\sin {\theta _1} + {w_{i,1}}}\\
{{x_2} = {r_2}\cos \left( {{\theta _2} + \Delta \varphi } \right) + {w_{r,2}}}\\
{{y_2} = {r_2}\sin \left( {{\theta _2} + \Delta \varphi } \right) + {w_{i,2}}}
\end{array}} \right.
\label{Equ_repre_DPSK}
\end{aligned}
\end{equation}
where ${\theta _1}=\phi_1+\varphi$ and ${\theta _2}=\phi_2+\varphi$.
Note that since $\phi_1$ and $\phi_2$ are uniformly distributed on $\left[ {0,2\pi } \right]$ for WSSUS Rayleigh fading channels, ${\theta _1}$ and ${\theta _2}$ are also uniformly distributed on $\left[ {0,2\pi } \right]$.
Let $\mathbf{z} = \left\{ {{x_1},{y_1},{x_2},{y_2}} \right\}$.
As expressed in (\ref{Equ_repre_DPSK}), DM can be equivalently described by the conditional transition probability ${\mathbb{P}_{\mathbf{z}\left| {\Delta \varphi } \right.}}$ with the channel input $\Delta \varphi$ and the channel output $\mathbf{z}$.
%channel output with the input $\Delta \varphi$ and the $\mathbf{z}$.
%As expressed in (\ref{Equ_repre_DPSK}), DM can be equivalently transformed as channels with the input $\Delta \varphi$ and the output $\mathbf{z}$.
This equivalent representation of DM in (\ref{Equ_repre_DPSK}) makes tractable the channel transition probability and information density in Theorems 1 and 2.

\textbf{Step 2. Conditional channel transition probability ${\mathbb{P}_{\mathbf{z}\left| {\Delta \varphi } \right.}}$ of (\ref{Equ_repre_DPSK}):}
From (\ref{Equ_repre_DPSK}), we obtain
\begin{equation}\nonumber
\begin{aligned}
\left\{ {\begin{array}{*{20}{c}}
{\mathbb{E}\left\{ {{x_1}{x_2}} \right\} = {\eta _f}\cos \Delta \varphi }\\
{\mathbb{E}\left\{ {{x_1}{y_2}} \right\} = {\eta _f}\sin \Delta \varphi }\\
~~{\mathbb{E}\left\{ {{y_1}{x_2}} \right\} =  - {\eta _f}\sin \Delta \varphi }\\
{\mathbb{E}\left\{ {{y_1}{y_2}} \right\} = {\eta _f}\cos \Delta \varphi }
\end{array}} \right.
\end{aligned}
\end{equation}
for FDDi-OFDM, where $\eta_f={\rho_f}/{2}$ and ${\rho _f}$ is the frequency domain correlation defined in (\ref{corr_fre_two}).
In addition, since ${{x_1},{y_1},{x_2},{y_2}}$ are a set of zero-mean Gaussian random variables with variances $\sigma^2=\left( 1+\gamma \right) /\left(2\gamma\right)$, the covariance matrix of $\mathbf{z}$ is given by
\begin{equation}
\begin{aligned}
\Sigma  = \left[ {\begin{array}{*{20}{c}}
{{\sigma ^2}}&0&{{\eta _f}\cos \Delta \varphi }&{{\eta _f}\sin \Delta \varphi }\\
0&{{\sigma ^2}}&{ - {\eta _f}\sin \Delta \varphi }&{{\eta _f}\cos \Delta \varphi }\\
{{\eta _f}\cos \Delta \varphi }&{ - {\eta _f}\sin \Delta \varphi }&{{\sigma ^2}}&0\\
{{\eta _f}\sin \Delta \varphi }&{{\eta _f}\cos \Delta \varphi }&0&{{\sigma ^2}}
\end{array}} \right]
\label{Equ_covariance}
\end{aligned}
\end{equation}
Based on (\ref{Equ_covariance}), we have $\left| \Sigma  \right| = {\left( {{\sigma ^4} - \eta _f^2} \right)^2}$ and
\begin{equation}\nonumber
\begin{aligned}
&\Sigma^{-1}  = \frac{1}{{{\sigma ^4} - \eta _f^2}}\times\\
&\left[ {\begin{array}{*{20}{c}}
{{\sigma ^2}}&0&{ - {\eta _f}\cos \Delta \varphi }&{ - {\eta _f}\sin \Delta \varphi }\\
0&{{\sigma ^2}}&{{\eta _f}\sin \Delta \varphi }&{ - {\eta _f}\cos \Delta \varphi }\\
{ - {\eta _f}\cos \Delta \varphi }&{{\eta _f}\sin \Delta \varphi }&{{\sigma ^2}}&0\\
{ - {\eta _f}\sin \Delta \varphi }&{ - {\eta _f}\cos \Delta \varphi }&0&{{\sigma ^2}}
\end{array}} \right]
%\label{Equ_covariance}
\end{aligned}
\end{equation}
As a result, the conditional channel transition probability ${\mathbb{P}_{\mathbf{z}\left| {\Delta \varphi } \right.}}$ of (\ref{Equ_repre_DPSK}) can be expressed as
\begin{equation}
\begin{aligned}
&{\mathbb{P}_{\mathbf{z}\left| {\Delta \varphi } \right.}}\left( {{x_1},{y_1},{x_2},{y_2}\left| {\Delta \varphi } \right.} \right)= \frac{{\exp \left( { - \frac{1}{2}\mathbf{z}{\Sigma ^{ - 1}}{\mathbf{z}}^{\rm T}} \right)}}{{\sqrt {16{\pi ^4}\left| \Sigma  \right|} }}\\
& = \frac{1}{{4{\pi ^2}\left( {{\sigma ^4} - \eta _f^2} \right)}}\exp \left[ {\frac{{{\eta _f}{\cal F}\left( {\Delta \varphi } \right)}}{{2\left( {{\sigma ^4} - \eta _f^2} \right)}}} \right] \times \\
&\exp \left[ { - \frac{{\left( {{\sigma ^2} + {\eta _f}} \right)\left( {x_1^2 + y_1^2 + x_2^2 + y_2^2} \right)}}{{2\left( {{\sigma ^4} - \eta _f^2} \right)}}} \right]\\
& = \frac{{{\gamma ^2}}}{{{\pi ^2}\left( {1 + 2\gamma  + {\gamma ^2} - {\gamma ^2}\rho _f^2} \right)}}\exp \left[ {\frac{{{\gamma ^2}{\rho _f}{\cal F}\left( {\Delta \varphi } \right)}}{{1 + 2\gamma  + {\gamma ^2} - {\gamma ^2}\rho _f^2}}} \right] \times \\
&\exp \left[ { - \frac{{\gamma \left( {1 + \gamma  + \gamma {\rho _f}} \right)\left( {x_1^2 + y_1^2 + x_2^2 + y_2^2} \right)}}{{1 + 2\gamma  + {\gamma ^2} - {\gamma ^2}\rho _f^2}}} \right]
\label{CCTP}
\end{aligned}
\end{equation}
where $\mathcal{F}\left( {\Delta \varphi } \right) = {\left( {{x_1} + {x_2}\cos \Delta \varphi  + {y_2}\sin \Delta \varphi } \right)^2} + {\left( {{y_1} - {x_2}\sin \Delta \varphi  + {y_2}\cos \Delta \varphi } \right)^2}$

\textbf{Step 3. Expectation and variance of the information density for (\ref{Equ_repre_DPSK}):}
Based on (\ref{infor_density}) and (\ref{CCTP}), we can calculate the expectation and variance of the information density for (\ref{Equ_repre_DPSK}). Both are key parameters to describe the asymptotic behavior of the non-asymptotic upper and lower bounds in Theorems 1 and 2, as shown next in Step 4.
According to the definition in (\ref{infor_density}), the information density for (\ref{Equ_repre_DPSK}) with a single channel use is
\begin{equation}
\begin{aligned}
i\left( {\Delta \varphi ;\mathbf{z} } \right) = {\log _2}\frac{{\mathbb{P}_{\mathbf{z}\left| {\Delta \varphi } \right.}}\left( {{x_1},{y_1},{x_2},{y_2}\left| {\Delta \varphi } \right.} \right)}{{\mathbb{P}_{\mathbf{z}}}\left( {{x_1},{y_1},{x_2},{y_2}} \right)}.
%{\log _2}M - {\log _2}\frac{{\sum\nolimits_{m = 0}^{M - 1} {\mathbb{P}\left( {\Delta \hat \varphi \left| {\Delta {\varphi _m}} \right.} \right)} }}{{\mathbb{P}\left( {\Delta \hat \varphi \left| {\Delta \varphi } \right.} \right)}}.
\label{infor_density_single_use_diff}
\end{aligned}
\end{equation}
Then, the expectation and variance of $i\left( {\Delta \varphi ;\mathbf{z} } \right)$ can be respectively determined as
\begin{equation}
\begin{aligned}
&{I_{{\rm{diff}}}}\left( \gamma  \right)= \mathbb{E}\left[ i\left( {\Delta \varphi ;\mathbf{z} } \right) \right]\\
&={\log _2}M - \mathbb{E}\left\{{\log _2}\frac{{\sum\nolimits_{m = 0}^{M-1} {{\mathbb{P}_{\mathbf{z}\left| {\Delta \varphi_m} \right.}}\left( {{x_1},{y_1},{x_2},{y_2}\left| {\Delta \varphi_m } \right.} \right)} }}{{\mathbb{P}_{\mathbf{z}\left| {\Delta \varphi_0 } \right.}}\left( {{x_1},{y_1},{x_2},{y_2}\left| {\Delta \varphi_0 } \right.} \right)}\right\}\\
%&={\log _2}M - \mathbb{E}\left\{{\log _2}\frac{{\sum\nolimits_{i = 0}^{M-1} {\exp \left[ {\frac{{{\rho _f}{\mathcal F}\left( {\Delta \varphi_i } \right)}}{{{{\left( {1 + \gamma } \right)}^2} - \rho _f^2}}} \right]} }}{\exp \left[ {\frac{{{\rho _f}{\mathcal F}\left( {\Delta \varphi_0 } \right)}}{{{{\left( {1 + \gamma } \right)}^2} - \rho _f^2}}} \right]}\right\}
&={\log _2}M - \mathbb{E}\left\{{\log _2}\frac{{\sum\nolimits_{m = 0}^{M-1} {\exp \left[ {\frac{{{\gamma ^2}{\rho _f}{\cal F}\left( {\Delta \varphi_m } \right)}}{{1 + 2\gamma  + {\gamma ^2} - {\gamma ^2}\rho _f^2}}} \right]} }}{\exp \left[ {\frac{{{\gamma ^2}{\rho _f}{\cal F}\left( {\Delta \varphi_0 } \right)}}{{1 + 2\gamma  + {\gamma ^2} - {\gamma ^2}\rho _f^2}}} \right]}\right\}
\label{infor_density_mean}
\end{aligned}
\end{equation}
and
\begin{equation}
\begin{aligned}
&{V_{{\rm{diff}}}}\left( \gamma  \right)= \mathbb{V}\text{ar}\left[  i\left( {\Delta \varphi ;\mathbf{z} } \right)\right] \\
%&=\mathbb{V}\text{ar}\left[ {{{\log }_2}M - {{\log }_2}\frac{{\sum\nolimits_{m = 0}^{M - 1} {\mathbb{P}\left( {\Delta \hat \varphi \left| {\Delta {\varphi _m}} \right.} \right)} }}{{\mathbb{P}\left( {\Delta \hat \varphi \left| {\Delta \varphi } \right.} \right)}}} \right]\\
%&\mathop  = \limits^{\left( c \right)}\mathbb{V}\text{ar}\left[ { {{\log }_2}\frac{{\sum\nolimits_{m = 0}^{M - 1} {\mathbb{P}\left( {\Delta \hat \varphi \left| {\Delta {\varphi _m}} \right.} \right)} }}{{\mathbb{P}\left( {\Delta \hat \varphi \left| {\Delta \varphi } \right.} \right)}}} \right]\\
&=\mathbb{E}\left\{ {\left[{\log _2}\frac{{\sum\nolimits_{m = 0}^{M-1} {{\mathbb{P}_{\mathbf{z}\left| {\Delta \varphi_m } \right.}}\left( {{x_1},{y_1},{x_2},{y_2}\left| {\Delta \varphi_m } \right.} \right)} }}{{\mathbb{P}_{\mathbf{z}\left| {\Delta \varphi_0 } \right.}}\left( {{x_1},{y_1},{x_2},{y_2}\left| {\Delta \varphi_0 } \right.} \right)}\right]^2}
\right\} \\
&-{\mathbb{E}^2}\left\{ {\log _2}\frac{{\sum\nolimits_{m= 0}^{M-1} {{\mathbb{P}_{\mathbf{z}\left| {\Delta \varphi_m } \right.}}\left( {{x_1},{y_1},{x_2},{y_2}\left| {\Delta \varphi_m } \right.} \right)} }}{{\mathbb{P}_{\mathbf{z}\left| {\Delta \varphi_0 } \right.}}\left( {{x_1},{y_1},{x_2},{y_2}\left| {\Delta \varphi_0 } \right.} \right)} \right\}\\
&=\mathbb{E}\left\{\left[{\log _2}\frac{{\sum\nolimits_{m = 0}^{M-1} {\exp \left[ {\frac{{{\gamma ^2}{\rho _f}{\cal F}\left( {\Delta \varphi_m } \right)}}{{1 + 2\gamma  + {\gamma ^2} - {\gamma ^2}\rho _f^2}}} \right]} }}{\exp \left[ {\frac{{{\gamma ^2}{\rho _f}{\cal F}\left( {\Delta \varphi_0 } \right)}}{{1 + 2\gamma  + {\gamma ^2} - {\gamma ^2}\rho _f^2}}} \right]}\right]^2\right\}\\
&-\mathbb{E}^2\left\{{\log _2}\frac{{\sum\nolimits_{m = 0}^{M-1} {\exp \left[ {\frac{{{\gamma ^2}{\rho _f}{\cal F}\left( {\Delta \varphi _m} \right)}}{{1 + 2\gamma  + {\gamma ^2} - {\gamma ^2}\rho _f^2}}} \right]} }}{\exp \left[ {\frac{{{\gamma ^2}{\rho _f}{\cal F}\left( {\Delta \varphi_0 } \right)}}{{1 + 2\gamma  + {\gamma ^2} - {\gamma ^2}\rho _f^2}}} \right]}\right\}
%&=\int_{ - \pi }^\pi  {{\mathbb{P}_\theta }\left( {\Delta \hat \varphi } \right){{\left[ {{{\log }_2}\frac{{\sum\limits_{m = 0}^{M - 1} {\mathbb{P}\left( {\Delta \hat \varphi \left| {\Delta {\varphi _m}} \right.} \right)} }}{{{\mathbb{P}_\theta }\left( {\Delta \hat \varphi } \right)}}} \right]}^2}d\left( {\Delta \hat \varphi } \right)} \\
%&-{\left[ {\int_{ - \pi }^\pi  {{\mathbb{P}_\theta }\left( {\Delta \hat \varphi } \right){{\log }_2}\frac{{\sum\limits_{m = 0}^{M - 1} {\mathbb{P}\left( {\Delta \hat \varphi \left| {\Delta {\varphi _m}} \right.} \right)} }}{{{\mathbb{P}_\theta }\left( {\Delta \hat \varphi } \right)}}d\left( {\Delta \hat \varphi } \right)} } \right]^2}\\
%&=\sum\limits_{i = 0}^{M - 1} {\int_{\left( {2i\pi /M} \right) - \pi }^{\left( {2\left( {i + 1} \right)\pi /M} \right) - \pi } {{\mathbb{P}_\theta }\left( {\Delta \hat \varphi } \right)} }\\
%& \times {{\left[ {{{\log }_2}\frac{{\sum\limits_{m = 0}^{M - 1} {\mathbb{P}\left( {\Delta \hat \varphi \left| {\Delta {\varphi _m}} \right.} \right)} }}{{{\mathbb{P}_\theta }\left( {\Delta \hat \varphi } \right)}}} \right]}^2}d\left( {\Delta \hat \varphi } \right)- {\left( { {{\log }_2}M- {I_{{\rm{diff}}}}} \right)^2}
\label{infor_density_variance}
\end{aligned}
\end{equation}
where ${\Delta \varphi _m} = {{2\pi m} / M}$.
%, $(a)$ is due to the symmetry, $(b)$ follows from $\mathbb{P}\left( {\Delta \hat \varphi } \right) = \frac{1}{M}\sum\limits_{m = 0}^{M - 1} {\mathbb{P}\left( {\Delta \hat \varphi \left| {\Delta {\varphi _m}} \right.} \right)} $, $(c)$ is obtained by using $\mathbb{V}\text{ar}\left( {\tau - X} \right) = \mathbb{V}\text{ar}\left( X \right)$ for any constant $\tau $ and random variable $X$, and $(d)$ is due to $\mathbb{V}\text{ar}(X) = {\mathbb{E}}({X^2}) - {{\mathbb{E}}^2}(X)$.
%$\int_{ - \pi }^\pi  {{P_\theta }\left( {\Delta \hat \varphi } \right){{\log }_2}\frac{{\sum\limits_{m = 0}^{M - 1} {P\left( {\Delta \hat \varphi \left| {\Delta {\varphi _m}} \right.} \right)} }}{{{P_\theta }\left( {\Delta \hat \varphi } \right)}}d\left( {\Delta \hat \varphi } \right)}$

\textbf{Step 4. Normal approximation for the IS and DT Bounds in Theorems 1 and 2:}
Based on ${I_{{\rm{diff}}}}\left( \gamma  \right)$ in (\ref{infor_density_mean}) and ${V_{{\rm{diff}}}}\left( \gamma  \right)$ in (\ref{infor_density_variance}), the normal approximation results can be finally obtained by using the derivation techniques in \cite[Appendix A]{Zheng2023Differential} and \cite{Polyanskiy2010Channel}.
$\hfill\blacksquare$

%% *** References section*** %%
\bibliographystyle{IEEEtran}
\bibliography{IEEEabrv,D2Dbib}

% Generated by IEEEtran.bst, version: 1.13 (2008/09/30)
\begin{thebibliography}{10}
\providecommand{\url}[1]{#1}
\csname url@samestyle\endcsname
\providecommand{\newblock}{\relax}
\providecommand{\bibinfo}[2]{#2}
\providecommand{\BIBentrySTDinterwordspacing}{\spaceskip=0pt\relax}
\providecommand{\BIBentryALTinterwordstretchfactor}{4}
\providecommand{\BIBentryALTinterwordspacing}{\spaceskip=\fontdimen2\font plus
\BIBentryALTinterwordstretchfactor\fontdimen3\font minus
  \fontdimen4\font\relax}
\providecommand{\BIBforeignlanguage}[2]{{%
\expandafter\ifx\csname l@#1\endcsname\relax
\typeout{** WARNING: IEEEtran.bst: No hyphenation pattern has been}%
\typeout{** loaded for the language `#1'. Using the pattern for}%
\typeout{** the default language instead.}%
\else
\language=\csname l@#1\endcsname
\fi
#2}}
\providecommand{\BIBdecl}{\relax}
\BIBdecl

\bibitem{doppler2009devicecm}
K.~Doppler, M.~Rinne, C.~Wijting, C.~Ribeiro, and K.~Hugl, ``Device-to-device
  communication as an underlay to {LTE-advanced} networks,'' \emph{{IEEE}
  Commun. Mag.}, vol.~47, no.~12, pp. 42--49, Dec. 2009.

\bibitem{Quawcm}
M.~Corson, R.~Laroia, J.~Li, V.~Park, T.~Richardson, and G.~Tsirtsis, ``Toward
  proximity-aware internetworking,'' \emph{IEEE Wireless Commun. Mag.},
  vol.~17, no.~6, pp. 26--33, Dec. 2010.

\bibitem{Vcm}
M.~J. Yang, S.~Y. Lim, H.~J. Park, and N.~H. Park, ``Solving the data overload:
  {Device-to-device} bearer control architecture for cellular data
  offloading,'' \emph{IEEE Veh. Technol. Mag.}, vol.~8, no.~1, pp. 31--39, Mar.
  2013.

\bibitem{Ph2013survey}
P.~Phunchongharn, E.~Hossain, and D.~Kim, ``Resource allocation for
  device-to-device communications underlaying {LTE-advanced} networks,''
  \emph{IEEE Wireless Commun. Mag.}, vol.~20, no.~4, pp. 91--100, Aug. 2013.

\bibitem{D2D_CM}
D.-Q. Feng, L.~Lu, Y.~Yuan-Wu, G.~Y. Li, S.-Q. Li, and G.~Feng,
  ``Device-to-device communications in cellular networks,'' \emph{IEEE Commun.
  Mag.}, vol.~52, no.~4, pp. 49--55, Apr. 2014.

\bibitem{3GPP}
\BIBentryALTinterwordspacing
3GPP, ``3rd generation partnership project; technical specification group
  services and system aspects; study on architecture enhancements to support
  proximity-based services {(ProSe)},'' TR23.703 V0.4.1., Release 12, Dec.
  2013. [Online]. Available: \url{http://www.3gpp.org/DynaReport/23703.htm}
\BIBentrySTDinterwordspacing

\bibitem{fodor2011distributed}
G.~Fodor and N.~Reider, ``A distributed power control scheme for cellular
  network assisted {D2D} communications,'' in \emph{Proc. IEEE Global
  Telecommun. Conf. (GLOBECOM' 11)}, Dec. 2011, pp. 1--6.

\bibitem{yu2011resource}
C.-H. Yu, K.~Doppler, C.~Ribeiro, and O.~Tirkkonen, ``Resource sharing
  optimization for device-to-device communication underlaying cellular
  networks,'' \emph{{IEEE} Trans. Wireless Commun.}, vol.~10, no.~8, pp. 2752
  --2763, Aug. 2011.

\bibitem{kaufman2008cellular}
B.~Kaufman and B.~Aazhang, ``Cellular networks with an overlaid device to
  device network,'' in \emph{Proc. IEEE 42nd Asilomar Conf. on Signals, Syst.
  and Comput.}, Oct. 2008, pp. 1537--1541.

\bibitem{chen2010timehopping}
T.~Chen, G.~Charbit, and S.~Hakola, ``Time hopping for device-to-device
  communication in {LTE} cellular system,'' in \emph{Proc. IEEE Wireless
  Commun. and Networking Conf. (WCNC' 10)}, Apr. 2010, pp. 1 --6.

\bibitem{min2011reliability}
H.~Min, W.~Seo, J.~Lee, S.~Park, and D.~Hong, ``Reliability improvement using
  receive mode selection in the device-to-device uplink period underlaying
  cellular networks,'' \emph{{IEEE} Trans. Wireless Commun.}, vol.~10, no.~2,
  pp. 413--418, Feb. 2011.

\bibitem{zulhasnine2010efficient}
M.~Zulhasnine, C.~Huang, and A.~Srinivasan, ``Efficient resource allocation for
  device-to-device communication underlaying lte network,'' in \emph{Proc. IEEE
  6th Int. Conf. on Wireless and Mobile Computing, Networking and Commun.
  (WiMob' 10)}, Oct. 2010, pp. 368--375.

\bibitem{janis2009interferenceMIMO}
P.~J{\"a}nis, V.~Koivunen, C.~Ribeiro, K.~Doppler, and K.~Hugl,
  ``Interference-avoiding {MIMO} schemes for device-to-device radio underlaying
  cellular networks,'' in \emph{Proc. IEEE 20th Int. Symp. on Personal, Indoor
  and Mobile Radio Commun. (PIMRC' 09)}, Sept. 2009, pp. 2385--2389.

\bibitem{ZhanghuaMIMO}
D.~Zhu, W.~Xu, H.~Zhang, C.~Zhao, J.~C. Li, and M.~Lei, ``Rate-maximized
  transceiver optimization for multi-antenna device-to-device communications,''
  in \emph{Proc. IEEE Wireless Commun. and Networking Conf. (WCNC' 13)}, Apr.
  2013, pp. 4152--4157.

\bibitem{Tcom}
D.-Q. Feng, L.~Lu, Y.~Yuan-Wu, G.~Y. Li, G.~Feng, and S.-Q. Li,
  ``Device-to-device communications underlaying cellular networks,'' \emph{IEEE
  Trans. Commun.}, vol.~61, no.~8, pp. 3541--3551, Aug. 2013.

\bibitem{Wangame}
F.~Wang, C.~Xu, L.~Song, Q.~Zhao, X.~Wang, and Z.~Han, ``Energy-aware resource
  allocation for device-to-device underlay communication,'' in \emph{Proc. IEEE
  Conf. Commun. (ICC' 13)}, June 2013, pp. 6076--6080.

\bibitem{Xugame}
C.~Xu, L.~Song, Z.~Han, Q.~Zhao, X.~Wang, X.~Cheng, and B.~Jiao, ``Efficiency
  resource allocation for device-to-device underlay communication systems: A
  reverse iterative combinatorial auction based approach,'' \emph{IEEE J. Sel.
  Areas Commun.}, vol.~31, no.~9, pp. 348--358, Sept. 2013.

\bibitem{yinrui11}
R.~Yin, G.~Yu, H.~Zhang, Z.~Zhang, and G.~Y. Li, ``Pricing-based interference
  coordination for {D2D} communications in cellular networks,'' \emph{IEEE
  Trans. Wireless Commun.}, vol.~14, no.~3, pp. 1519--1532, Mar. 2015.

\bibitem{yinrui22}
R.~Yin, C.~Zhong, G.~Yu, Z.~Zhang, K.-K. Wong, and X.~Chen, ``Joint spectrum
  and power allocation for {D2D} communications underlaying cellular
  networks,'' \emph{IEEE Trans. Veh. Technol.}, vol.~PP, no.~99, pp. 1--1,
  2015.

\bibitem{fengse}
D.-Q. Feng, G.-D. Yu, Y.~Yuan-Wu, G.~Y. Li, G.~Feng, and S.-Q. Li, ``Mode
  switching for device-to-device communications in cellular networks,'' in
  \emph{Proc. IEEE Global Conf. on Signal and Inform. Process. (GlobalSIP'14)},
  Dec. 2014.

\bibitem{fengee}
D.-Q. Feng, G.-D. Yu, C.~Xiong, Y.~Yuan-Wu, G.~Y. Li, G.~Feng, and S.-Q. Li,
  ``Mode switching for energy-efficient device-to-device communications in
  cellular networks,'' \emph{IEEE Trans. Wireless Commun.}, vol.~PP, no.~99,
  pp. 1--1, 2015.

\bibitem{doppler2010mode}
K.~Doppler, C.-H. Yu, C.~Ribeiro, and P.~J{\"a}nis, ``Mode selection for
  device-to-device communication underlaying an {LTE-advanced} network,'' in
  \emph{Proc. IEEE Wireless Commun. and Networking Conf. (WCNC' 10)}, Apr.
  2010, pp. 1--6.

\bibitem{Simode}
S.~Wen, X.~Zhu, X.~Zhang, and D.~Yang, ``Qos-aware mode selection and resource
  allocation scheme for device-to-device {(D2D)} communication in cellular
  networks,'' in \emph{Proc. IEEE Conf. Commun. (ICC' 13)}, June 2013, pp.
  101--105.

\bibitem{yumode}
G.-D. Yu, L.-K. Xu, D.-Q. Feng, R.~Yin, G.~Y. Li, and Y.-H. Jiang, ``Joint mode
  selection and resource allocation for device-to-device communications,''
  \emph{IEEE Trans. Commun.}, vol.~62, no.~11, pp. 3814--3824, Nov. 2014.

\bibitem{YeAndd}
Q.~Ye, M.~Al-Shalash, C.~Caramanis, and J.~G. Andrews, ``Device-to-device
  modeling and analysis with a modified matern hardcore bs location model,'' in
  \emph{Proc. IEEE Global Telecommun. Conf. (GLOBECOM' 13)}, Dec. 2013, pp.
  1--6.

\bibitem{LinAndD2D}
X.~Lin and J.~G. Andrews, ``Optimal spectrum partition and mode selection in
  device-to-device overlaid cellular networks,'' in \emph{Proc. IEEE Global
  Telecommun. Conf. (GLOBECOM' 13)}, Dec. 2013, pp. 1--6.

\bibitem{Wang2013survey}
A.~Asadi, Q.~Wang, and V.~Mancuso, ``A survey on device-to-device communication
  in cellular networks,'' \emph{IEEE Commun. Surveys Tuts.}, vol.~16, no.~4,
  pp. 1801 -- 1819, Nov. 2014.

\bibitem{Miaooutage}
S.~Shalmashi, G.~Miao, and S.~Ben~Slimane, ``Interference management for
  multiple device-to-device communications underlaying cellular networks,'' in
  \emph{Proc. IEEE 24th Int. Symp. on Personal, Indoor and Mobile Radio Commun.
  (PIMRC' 13)}, Sept. 2013, pp. 223--227.

\bibitem{Maioutage}
A.~Abu Al~Haija and M.~{Vu}, ``{Spectral efficiency and outage performance for
  device-to-device cooperation in uplink cellular communication},'' \emph{IEEE
  Trans. Wireless Commun.}, vol.~14, no.~3, pp. 1183 -- 1198, Mar. 2015.

\bibitem{D2D_G}
D.-Q. Feng, L.~Lu, Y.~Yuan-Wu, G.~Y. Li, G.~Feng, and S.-Q. Li, ``Optimal
  resource allocation for device-to-device communications in fading channels,''
  in \emph{Proc. IEEE Global Telecommun. Conf. (GLOBECOM' 13)}, Dec. 2013, pp.
  1--5.

\bibitem{D2D_P}
------, ``User selection based on limited feedback in device-to-device
  communications,'' in \emph{Proc. IEEE 24th Int. Symp. on Personal, Indoor and
  Mobile Radio Commun. (PIMRC' 13)}, Sept. 2013, pp. 1--5.

\bibitem{Xin2011}
X.~Kang, R.~Zhang, Y.-C. Liang, and H.~K. Garg, ``Optimal power allocation
  strategies for fading cognitive radio channels with primary user outage
  constraint,'' \emph{IEEE J. Sel. Areas Commun.}, vol.~29, no.~2, pp.
  374--383, Feb. 2011.

\bibitem{1056848}
R.~McEliece and W.~Stark, ``Channels with block interference,'' \emph{IEEE
  Trans. Inform. Theory}, vol.~30, no.~1, pp. 44--53, Jan. 1984.

\bibitem{FodorCM2012}
G.~Fodor, E.~Dahlman, G.~Mildh, S.~Parkvall, N.~Reider, G.~Miklos, and
  Z.~Turanyi, ``Design aspects of network assisted device-to-device
  communications,'' \emph{IEEE Commun. Mag.}, vol.~50, no.~3, pp. 170--177,
  Mar. 2012.

\bibitem{qian2009mapel}
L.~P. Qian, Y.~J. Zhang, and J.~Huang, ``{MAPEL: Achieving global optimality
  for a non-convex wireless power control problem},'' \emph{IEEE Trans.
  Wireless Commun.}, vol.~8, no.~3, pp. 1553--1563, Mar. 2009.

\bibitem{tse2005fundamentals}
D.~Tse and P.~Viswanath, \emph{Fundamentals of wireless communication}.\hskip
  1em plus 0.5em minus 0.4em\relax Cambridge university press, 2005.

\bibitem{papoulis2002probability}
A.~Papoulis and S.~Pillai, \emph{Probability, random variables, and stochastic
  processes}, 4th~ed.\hskip 1em plus 0.5em minus 0.4em\relax NY: McGraw-Hill,
  2002.

\bibitem{stuber2012principles}
G.~L. St{\"u}ber, \emph{Principles of Mobile Communication}, 3rd~ed.\hskip 1em
  plus 0.5em minus 0.4em\relax New York, NY: Springer US, 2012.

\bibitem{Abdi}
A.~Abdi and M.~Kaveh, ``K distribution: an appropriate substitute for
  rayleigh-lognormal distribution in fading-shadowing wireless channels,''
  \emph{Electron. Lett.}, vol.~34, no.~9, pp. 851--852, Apr. 1998.

\bibitem{Theo2008}
P.~Theofilakos, A.~Kanatas, and G.~Efthymoglou, ``Performance of generalized
  selection combining receivers in {K} fading channels,'' \emph{IEEE Commun.
  Lett.}, vol.~12, no.~11, pp. 816--818, Nov. 2008.

\bibitem{west2001introduction}
D.~West \emph{et~al.}, \emph{Introduction to Graph Theory}.\hskip 1em plus
  0.5em minus 0.4em\relax Upper Saddle River, NJ: Prentice Hall, 2001.

\bibitem{janis2009interference}
P.~J{\"a}nis, V.~Koivunen, C.~Ribeiro, J.~Korhonen, K.~Doppler, and K.~Hugl,
  ``Interference-aware resource allocation for device-to-device radio
  underlaying cellular networks,'' in \emph{Proc. IEEE 69th Veh. Technol. Conf.
  (VTC Spring' 09)}, Apr. 2009, pp. 1--5.

\end{thebibliography}


% Generated by IEEEtran.bst, version: 1.14 (2015/08/26)
\begin{thebibliography}{10}
\providecommand{\url}[1]{#1}
\csname url@samestyle\endcsname
\providecommand{\newblock}{\relax}
\providecommand{\bibinfo}[2]{#2}
\providecommand{\BIBentrySTDinterwordspacing}{\spaceskip=0pt\relax}
\providecommand{\BIBentryALTinterwordstretchfactor}{4}
\providecommand{\BIBentryALTinterwordspacing}{\spaceskip=\fontdimen2\font plus
\BIBentryALTinterwordstretchfactor\fontdimen3\font minus
  \fontdimen4\font\relax}
\providecommand{\BIBforeignlanguage}[2]{{%
\expandafter\ifx\csname l@#1\endcsname\relax
\typeout{** WARNING: IEEEtran.bst: No hyphenation pattern has been}%
\typeout{** loaded for the language `#1'. Using the pattern for}%
\typeout{** the default language instead.}%
\else
\language=\csname l@#1\endcsname
\fi
#2}}
\providecommand{\BIBdecl}{\relax}
\BIBdecl

\bibitem{3GPP261}
3GPP, ``Service requirements for the {5G} system,'' TS 22.261 V19.5.0., Release
  19, Dec. 2023.

\bibitem{3GPP824}
{3GPP}, ``Study on physical layer enhancements for {NR} ultra-reliable and low
  latency case ({URLLC}),'' TR 38.824 V16.0.0., Release 16, Mar. 2019.

\bibitem{Wang2023On}
C.-X. Wang \emph{et~al.}, ``On the road to {6G}: visions, requirements, key
  technologies, and testbeds,'' \emph{IEEE Commun. Surveys \& Tutorials},
  vol.~25, no.~2, pp. 905--974, Secondquarter 2023.

\bibitem{You2021Towards}
X.~You \emph{et~al.}, ``Towards {6G} wireless communication networks: vision,
  enabling technologies, and new paradigm shifts,'' \emph{Sci. China Inf.
  Sci.}, vol.~64, pp. 1--74, Jan. 2021, doi:10.1007/s11432-020-2955-6.

\bibitem{she2021PIEEE}
C.~She \emph{et~al.}, ``A tutorial of ultra-reliable and low-latency
  communications in 6{G}: Integrating domain knowledge into deep learning,''
  \emph{Proc. IEEE}, vol. 109, no.~3, pp. 204--246, Mar. 2021.

\bibitem{3GPP912}
3GPP, ``Study on new radio ({NR}) access technology,'' TR 38.912 V17.0.0.,
  Release 17, Mar. 2022.

\bibitem{3GPP211}
{3GPP}, ``Physical channels and modulation,'' TS 38.211 V18.1.0., Release 18,
  Dec. 2023.

\bibitem{Feng2019toward}
D.~Feng \emph{et~al.}, ``{Towards Ultra-Reliable Low-Latency Communications:
  Typical Scenarios, Possible Solutions, and Open Issues},'' \emph{IEEE Veh.
  Technol. Mag.}, vol.~14, no.~2, pp. 94--102, June 2019.

\bibitem{Durisi2016Toward}
G.~Durisi, T.~Koch, and P.~Popovski, ``Toward massive, ultrareliable, and
  low-latency wireless communication with short packets,'' \emph{Proc. IEEE},
  vol. 104, no.~9, pp. 1711--1726, Sept. 2016.

\bibitem{Xu2019Sixty}
C.~Xu \emph{et~al.}, ``Sixty years of coherent versus non-coherent tradeoffs
  and the road from {5G} to wireless futures,'' \emph{IEEE Access}, vol.~7, pp.
  178\,246--178\,299, Dec. 2019.

\bibitem{Wang2012Dispensing}
L.~Wang and L.~Hanzo, ``Dispensing with channel estimation: Differentially
  modulated cooperative wireless communications,'' \emph{IEEE Commun. Surveys
  \& Tutorials}, vol.~14, no.~3, pp. 836--857, 2012.

\bibitem{Baeza2019Non}
V.~M. Baeza and A.~G. Armada, ``Non-coherent massive {SIMO} system based on
  {M-DPSK} for rician channels,'' \emph{IEEE Trans. on Veh. Technol.}, vol.~68,
  no.~3, pp. 2413--2426, Mar. 2019.

\bibitem{Avendi2014Performance}
M.~R. Avendi \emph{et~al.}, ``Performance of selection combining for
  differential amplify-and-forward relaying over time-varying channels,''
  \emph{IEEE Trans. on Wireless Commun.}, vol.~13, no.~8, pp. 4156--4166, Aug.
  2014.

\bibitem{Himsoon2008Differential}
T.~Himsoon \emph{et~al.}, ``Differential modulations for multinode cooperative
  communications,'' \emph{IEEE Trans. on Signal Process.}, vol.~56, no.~7, pp.
  2941--2956, July 2008.

\bibitem{Kun2004A}
K.~Zhong \emph{et~al.}, ``A general {SER} formula for an {OFDM} system with
  {MDPSK} in frequency domain over rayleigh fading channels,'' \emph{IEEE
  Trans. on Commun.}, vol.~52, no.~4, pp. 584--594, Apr. 2004.

\bibitem{Osaki2019Differentially}
S.~Osaki, M.~Nakao, T.~Ishihara, and S.~Sugiura, ``Differentially modulated
  spectrally efficient frequency-division multiplexing,'' \emph{IEEE Signal
  Process. Lett.}, vol.~26, no.~7, pp. 1046--1050, Jul. 2019.

\bibitem{Polyanskiy2011Scalar}
Y.~Polyanskiy and S.~Verdu, ``Scalar coherent fading channel: Dispersion
  analysis,'' in \emph{Proc. IEEE Int. Symp. on Inf. Theory}, 2011, pp.
  2959--2963.

\bibitem{Collins2019Coherent}
A.~Collins and Y.~Polyanskiy, ``Coherent multiple-antenna block-fading channels
  at finite blocklength,'' \emph{IEEE Trans. on Inf. Theory}, vol.~65, no.~1,
  pp. 380--405, Jan. 2019.

\bibitem{Yang2014Quasi}
W.~Yang, G.~Durisi, T.~Koch, and Y.~Polyanskiy, ``Quasi-static multiple-antenna
  fading channels at finite blocklength,'' \emph{IEEE Trans. on Inf. Theory},
  vol.~60, no.~7, pp. 4232--4265, July 2014.

\bibitem{Ferrante2018Pilot}
G.~C. Ferrante \emph{et~al.}, ``Pilot-assisted short-packet transmission over
  multiantenna fading channels: A {5G} case study,'' in \emph{Proc. IEEE
  Conf.on Inf. Sci. and Syst. (CISS)}, Mar. 2018, pp. 1--6.

\bibitem{Kislal2023Efficient}
A.~O. Kislal \emph{et~al.}, ``Efficient evaluation of the error probability for
  pilot-assisted {URLLC} with massive {MIMO},'' \emph{IEEE J. on Sel. Areas in
  Commun.}, vol.~41, no.~7, pp. 1969--1981, July 2023.

\bibitem{Johan2021URLLC}
J.~{\"O}stman, A.~Lancho, G.~Durisi, and L.~Sanguinetti, ``{URLLC} with massive
  {MIMO}: Analysis and design at finite blocklength,'' \emph{IEEE Trans. on
  Wireless Commun.}, vol.~20, no.~10, pp. 6387--6401, Oct. 2021.

\bibitem{Lancho2023Cell}
A.~Lancho, G.~Durisi, and L.~Sanguinetti, ``Cell-free massive {MIMO} for
  {URLLC}: A finite-blocklength analysis,'' \emph{IEEE Trans. on Wireless
  Commun.}, vol.~22, no.~12, pp. 8723--8735, Dec. 2023.

\bibitem{Mousaei2017Optimizing}
M.~Mousaei and B.~Smida, ``Optimizing pilot overhead for ultra-reliable
  short-packet transmission,'' in \emph{Proc. IEEE Conf. Commun. (ICC)}, May
  2017, pp. 1--5.

\bibitem{Cao2022Independent}
J.~Cao \emph{et~al.}, ``Independent pilots versus shared pilots: Short frame
  structure optimization for heterogeneous-traffic urllc networks,'' \emph{IEEE
  Trans. on Wireless Commun.}, vol.~21, no.~8, pp. 5755--5769, Aug. 2022.

\bibitem{Zhou2023Joint}
X.~Zhou \emph{et~al.}, ``Joint optimization of frame structure and power
  allocation for {URLLC} in short blocklength regime,'' \emph{IEEE Trans. on
  Commun.}, vol.~71, no.~12, pp. 7333--7346, Dec. 2023.

\bibitem{Lee2018Packet}
B.~Lee \emph{et~al.}, ``Packet structure and receiver design for low latency
  wireless communications with ultra-short packets,'' \emph{IEEE Trans. on
  Commun.}, vol.~66, no.~2, pp. 796--807, Feb. 2018.

\bibitem{Johan2019Short}
J.~{\"O}stman \emph{et~al.}, ``Short packets over block-memoryless fading
  channels: Pilot-assisted or noncoherent transmission?'' \emph{IEEE Trans. on
  Commun.}, vol.~67, no.~2, pp. 1521--1536, Feb. 2019.

\bibitem{Lancho2020On}
A.~Lancho, T.~Koch, and G.~Durisi, ``On single-antenna rayleigh block-fading
  channels at finite blocklength,'' \emph{IEEE Trans. on Inf. Theory}, vol.~66,
  no.~1, pp. 496--519, Jan. 2020.

\bibitem{Lancho2020Saddlepoint}
A.~Lancho \emph{et~al.}, ``Saddlepoint approximations for short-packet wireless
  communications,'' \emph{IEEE Trans. on Wireless Commun.}, vol.~19, no.~7, pp.
  4831--4846, Jul. 2020.

\bibitem{Qi2020A}
C.~Qi and T.~Koch, ``A high-{SNR} normal approximation for {MIMO} rayleigh
  block-fading channels,'' in \emph{Proc. IEEE Int. Symp. on Inf. Theory},
  2020, pp. 2314--2319.

\bibitem{Wu2020Pilot}
J.~Wu, W.~Kim, and B.~Shim, ``Pilot-less one-shot sparse coding for short
  packet-based machine-type communications,'' \emph{IEEE Trans. on Veh.
  Technol.}, vol.~69, no.~8, pp. 9117--9120, Aug. 2020.

\bibitem{Ji2019Pilot}
H.~Ji, W.~Kim, and B.~Shim, ``Pilot-less sparse vector coding for short packet
  transmission,'' \emph{IEEE Wireless Commun. Lett.}, vol.~8, no.~4, pp.
  1036--1039, Aug. 2019.

\bibitem{Liu2019Fast}
Y.~Liu \emph{et~al.}, ``Fast iterative semi-blind receiver for {URLLC} in
  short-frame full-duplex systems with {CFO},'' \emph{IEEE J. on Sel. Areas in
  Commun.}, vol.~37, no.~4, pp. 839--853, Apr. 2019.

\bibitem{Walk2019MOCZ}
P.~Walk, P.~Jung, and B.~Hassibi, ``{MOCZ} for blind short-packet
  communication: Basic principles,'' \emph{IEEE Trans. on Wireless Commun.},
  vol.~18, no.~11, pp. 5080--5097, Nov. 2019.

\bibitem{Siddiqui2024Spectrally}
A.~A. Siddiqui \emph{et~al.}, ``Spectrally-efficient modulation on
  conjugate-reciprocal zeros ({SE-MOCZ}) for non-coherent short packet
  communications,'' \emph{IEEE Trans. on Wireless Commun.}, vol.~23, no.~3, pp.
  2226--2240, Mar. 2024.

\bibitem{Choi2021Generalized}
J.~Choi and N.~Lee, ``Generalized differential index modulation for pilot-free
  communications,'' \emph{IEEE Internet of Things J.}, vol.~8, no.~10, pp.
  7973--7984, May 2021.

\bibitem{Jiang2019Packet}
X.~Jiang \emph{et~al.}, ``Packet detection by a single {OFDM} symbol in {URLLC}
  for critical industrial control: A realistic study,'' \emph{IEEE J. on Sel.
  Areas in Commun.}, vol.~37, no.~4, pp. 933--946, Apr. 2019.

\bibitem{Zheng2023Differential}
C.~Zheng \emph{et~al.}, ``Differential modulation for short packet transmission
  in {URLLC},'' \emph{IEEE Trans. on Wireless Commun.}, vol.~23, no.~7, pp.
  7230--7245, Jul. 2024.

\bibitem{Ding2006Performance}
D.-B. Lin \emph{et~al.}, ``Performance of noncoherent maximum-likelihood
  sequence detection for differential {OFDM} systems with diversity
  reception,'' \emph{IEEE Trans. on Broadcast.}, vol.~52, no.~1, pp. 62--70,
  Mar. 2006.

\bibitem{Liu2014Channel}
Y.~Liu \emph{et~al.}, ``Channel estimation for {OFDM},'' \emph{IEEE Commun.
  Surveys \& Tutorials}, vol.~16, no.~4, pp. 1891--1908, Fourthquarter 2014.

\bibitem{Kim2005Performance}
J.~Kim, J.~Park, and D.~Hong, ``Performance analysis of channel estimation in
  {OFDM} systems,'' \emph{IEEE Signal Process. Lett.}, vol.~12, no.~1, pp.
  60--62, Jan. 2005.

\bibitem{Polyanskiy2010Channel}
Y.~{Polyanskiy}, H.~V. {Poor}, and S.~{Verdu}, ``Channel coding rate in the
  finite blocklength regime,'' \emph{IEEE Trans. on Inf. Theory}, vol.~56,
  no.~5, pp. 2307--2359, May 2010.

\bibitem{MolavianJazi2015A}
E.~MolavianJazi and J.~N. Laneman, ``A second-order achievable rate region for
  gaussian multi-access channels via a central limit theorem for functions,''
  \emph{IEEE Trans. on Inf. Theory}, vol.~61, no.~12, pp. 6719--6733, Dec.
  2015.

\bibitem{Xiong2023Status}
Q.~Xiong \emph{et~al.}, ``Status prediction and data aggregation for
  {AoI}-oriented short-packet transmission in industrial {I}o{T},'' \emph{IEEE
  Trans. on Commun.}, vol.~71, no.~1, pp. 611--625, Jan. 2023.

\bibitem{Zheng2021Open}
C.~Zheng \emph{et~al.}, ``Open-loop communications for up-link {URLLC} under
  clustered user distribution,'' \emph{IEEE Trans. on Veh. Technol.}, vol.~70,
  no.~11, pp. 11\,509--11\,522, Nov. 2021.

\bibitem{A2014Jazi}
E.~MolavianJazi, \emph{A Unified Approach to Gaussian Channels with Finite
  Blocklength}.\hskip 1em plus 0.5em minus 0.4em\relax Ph.D. dissertation,
  Dept. Elect. Eng., Univ. Notre Dame, Notre Dame, IN, USA, July 2014.

\bibitem{Rao2018Adaptive}
R.~M. Rao, V.~Marojevic, and J.~H. Reed, ``Adaptive pilot patterns for
  {CA-OFDM} systems in nonstationary wireless channels,'' \emph{IEEE Trans. on
  Veh. Technol.}, vol.~67, no.~2, pp. 1231--1244, Feb. 2018.

\bibitem{Simko2013Adaptive}
M.~{\v{S}}imko \emph{et~al.}, ``Adaptive pilot-symbol patterns for {MIMO}
  {OFDM} systems,'' \emph{IEEE Trans. on Wireless Commun.}, vol.~12, no.~9, pp.
  4705--4715, Sep. 2013.

\bibitem{Edfors1998OFDM}
O.~Edfors \emph{et~al.}, ``{OFDM} channel estimation by singular value
  decomposition,'' \emph{IEEE Trans. on Commun.}, vol.~46, no.~7, pp. 931--939,
  Jul. 1998.

\bibitem{Yuan2021Polar}
P.~Yuan \emph{et~al.}, ``Polar-coded non-coherent communication,'' \emph{IEEE
  Commun. Lett.}, vol.~25, no.~6, pp. 1786--1790, Feb. 2021.

\end{thebibliography}

%\end{IEEEbiography}
%\begin{IEEEbiography}{Michael Shell}
%Biography text here.
%\end{IEEEbiography}
%
%% if you will not have a photo at all:
%\begin{IEEEbiographynophoto}{John Doe}
%Biography text here.
%\end{IEEEbiographynophoto}
%
%% insert where needed to balance the two columns on the last page with
%% biographies
%%\newpage
%
%\begin{IEEEbiographynophoto}{Jane Doe}
%Biography text here.
%\end{IEEEbiographynophoto}

\end{document}